\documentstyle[12pt]{article}
\input{psfig}
\begin{document}
\hfill hep-th/0311186

\hfill HU-EP 03/181
\catcode`@=11
\def\seceqaa{\@addtoreset{equation}{section}
\def\theequation{A\arabic{equation}}}
\def\seceqbb{\@addtoreset{equation}{section}
\def\theequation{B\arabic{equation}}}
\catcode`@=12
\textheight 20.5cm
\textwidth 20.5cm
\hoffset -1.5cm
\headsep 0.1cm
\parindent 1.0em
\begin{center}
\Large 
{\bf
Type IIA on a Compact Calabi-Yau and $D=11$ Supergravity Uplift of its Orientifold} 
\vskip 0.1in
{Aalok Misra\footnote{Address from Jan 3, 2004: Indian Institute of Technology, Roorkee, India}
\\
Institut f\"{u}r Physik, Humboldt Universit\"{a}t, Berlin, Germany}
\end{center}
\begin{abstract}
Using the prescription of \cite{HV} for defining period integrals in the Landau-Ginsburg
theory for compact Calabi-Yau's, we obtain the Picard-Fuchs equation and
the Meijer basis of solutions for the compact Calabi-Yau $CY_3(3,243)$ expressed
as a degree-24 Fermat hypersurface {\it after} resolution of the orbifold  singularities.
This is similar in spirit to the method of obtaining Meijer basis of solutions in \cite{AM2}
for the case wherein one is away from the orbifold singularities, and one is considering
the large-base limit of the Calabi-Yau. The importance of the method
lies in the ease with which one can consider the large {\it and} small complex structure
limits, 
as well as the ability to get the "ln"-terms in the periods without having to 
parametrically differentiate infinite series. 
We consider in detail the evaluation of the monodromy matrix in the large and small complex
structure limits.
We also consider the action of the freely acting antiholomorphic involution of \cite{AM1,AM2}
on $D=11$ supergravity compactified on $CY_3(3,243)\times S^1$ \cite{BCF} 
and obtain the K\"{a}hler
potential for the same in the limit of large volume of the Calabi-Yau. As a by-product,
we also give a conjecture for the action of the orientation-reversing antiholomorphic
involution on the periods, given its action on the cohomology, using a  canonical
(co)homology basis.
Finally, we also consider showing a null superpotential 
on the orientifold of type IIA on $CY_3(3,243)$, having taken care of the orbifold
singularities, thereby completing the argument initiated in \cite{AM2}.
\end{abstract}
\clearpage
\section{Introduction}

The periods are the building blocks, e.g., for getting the prepotential
(and hence the K\"{a}hler potential and the Yukawa coupling) in ${\cal N}=2$
type $II$ theories compactified on a Calabi-Yau, and the superpotential
for ${\cal N}=1$ type $II$ compactifications in the rpesence of (RR) fluxes.
It is in this regard that the Picard-Fuchs equation satisfied by the periods, become quite 
important. Also, traversing non-trivial loops in the complex structure 
moduli space of type $IIB$ on a Calabi-Yau mirror to the one on the type 
$IIA$ side, corresponds to shifting of the K\"{a}hler moduli in the 
K\"{a}hler moduli space on the type $IIA$ side. This results in mixing of 
flux numbers corresponding to $RR$ fluxes on the type $IIA$ side, implying
thereby that dimensions of cycles on the type $IIA$ side, loose 
their meaning. The mixing matrix for the flux numbers is given by the 
monodromy matrix. The mixing matrix for the flux numbers if given by the
monodromy matrix. It hence becomes important to evaluate the same. 
Non-compact Calabi-Yau's have typically been the ones that have been
extensively studied in this context as well as topological strings with
the same as target spaces. Compact Calabi-Yau's, in addition to the quintic,
thus become quite important to be studied. Additionally, mirror symmetry
becomes quite indispensible when working with compact Calabi-Yau's with
large value of $h^{2,1}$ and small $h^{1,1}$. One such compact Calabi-Yau
that we will study is $CY_3(3,243)$, i.e., one for which $h^{1,1}=3$, and
$h^{2,1}=243$. This compact Calabi-Yau plays a central role in ${\cal N}=2,1$
 type $IIA$/Heterotic dual paris in four dimensions in \cite{VW,KV}.

In \cite{AM2}, we addressed the issue of deriving the Picard-Fuchs equation
on the mirror Landau-Ginsburg side corresponding to the gauged linear sigma model for
a compact Calabi-Yau $CY_3(3,243)$, expressed as a degree-24 Fermat hypersurface in a 
suitableweighted complex projective space, but staying away from the orbifold singularities
by taking the large-base limit of the compact Calabi-Yau. Even though, one ended up
with more than the required number of solutions, but the essential idea that was highlighted
was the ease with which, both the large and small complex structure limits could be
addressed, and the fact that the nonanalytic $ln$-terms in the periods, could  be
easily obtained without having  to resort to parametric differentiations of infinite
series. In this paper, we address the problem of getting the right  number of the
right kind of solutions on the mirror Landau-Ginsburg side, but this time after having
resolved  the orbifold singularities. We also address the problems of showing that
unoriented instantons do not generate a superpotential on the type $IIA$ side 
in the  ${\cal N}=1$ Heterotic/type $IIA$ dual pair of \cite{VW}, whose $M$ and
$F$ theory uplifts were discussed in \cite{AM1}. It was  shown in \cite{AM2}, using
mirror symmetry, that as expected from the Heterotic and $F$ theory duals, there is
indeed no superpotential generated from ${\bf RP}^2$-instantons in the type $IIA$ side
in the large-base limit of $CY_3(3,243)$, away from the aforementioned orbifold
singularities of the relevant Fermat hypersurface. In this paper, we show that the
same remains true even after the resolution of  the orbifold singularities. Further,
we discuss the supergravity uplift of the type $IIA$ orientifold that figures in
the abovementioned ${\cal N}=1$ Heterotic/type $IIA$ dual pair, to $D=11$ supergravity.
We evaluate the K\"{a}hler potential in the large volume limit of $CY_3(3,243)$.
As an interesting aside, we give a conjecture about the action of the antiholomorphic
involution that figures in the definition  of  the type  $IIA$ orientifold, on the
periods, given its action on the cohomology of $CY_3(3,243)$,  using a canonical
(co)homology basis to expand  the holomorphic 3-form. We verify the conjecture for
$T^6$ and (partly) for the mirror  to the  quintic.  

The plan of the paper is the following. In section {\bf 2}, after a brief discussion of
the orbifold singularities' resolution corresponding to the Fermat hypersurface, we define
the period integral as per the prescription of Hori and Vafa, for a compact Calabi-Yau.
We then obtain the degree-8 Picard-Fuchs satisfied by the period integral, and then
discuss the guiding principle in obtaining  the right number of the right kind of solutions,
and obtain a set of solutions to the equation, and evaluate the relevant contour integrals.
We also discuss the evaluation of the monodromy matrix in the large and small complex 
structure limits.  In section {\bf 3}, we then discuss
showing that there is no superpotential generated (by the only possible source
- unoriented instantons) for type $IIA$ on an orientifold of $CY_3(3,243)$, {\it after}
resolution  of the aforementioned orbifold  singularities, thereby strengthening
the conjectured ${\cal N}=1$ triality in \cite{AM1}.  In section {\bf 4}, we discuss
the $D=11$ supergravity uplift of the aforementioned type $IIA$ orientifold on a
`barely' $G_2$ manifold \cite{AM1}, and also evaluate the K\"{a}hler potential in the
limit of the large volume of the Calabi-Yau. Finally, in section {\bf 5}, we 
discuss a conjecture related to the action of antiholomorphic involution on the periods
of $CY_3(3,243)$, given its action  on its cohomology, using a canonical (co)homology
basis. Section {\bf 6} has the conclusions and discussion. There is one appendix on
the evaluation of the monodromy matrix.

\section{Landau-Ginsburg PF equation for type $IIA$ on $CY_3(3,243)$}

 By following the alternative formulation of Hori and Vafa \cite{HV} for 
deriving the Picard-Fuchs equation for a definition of period integral in the mirror
Landau-Ginsburg model,
we obtain solutions valid in the large {\it and} small complex structure limits,
and get the ln terms as naturally as the analytic terms
(i.e. without using
parametric differentiation of infinite series). We also study in detail, the monodromy
matrix in the large and small complex structure limits.

Consider the Calabi-Yau 3-fold given as a degree-24 Fermat 
hypersurface in the weighted projective
space ${\bf WCP}^4[1,1,2,8,12]$:
\begin{equation}
\label{eq:Fermatdieuf}
P=z_1^{24}+z_2^{24}+z_3^{12}+z_4^{3}+z_5^2=0.
\end{equation}
It has a ${\bf Z}_2$-singularity curve and a ${\bf Z}_4$-singularity point. 
${\bf Z}_2$ and ${\bf Z}_4$ singularity resolution $\leftrightarrow$
The two new chiral superfields
 needed to be introduced as a consequence of singularity resolution,
correspond to the two ${\bf CP}^1$'s
that required to be introduced in blowing up the singularities. One then has to consider three instead
of a single $C^*$ action, and the $CY_3(3,243)$
\footnote{The $CY_3(3,243)$ considered in this paper will be
 an elliptic fibration over the Hirzebruch surface $F_2$.}
can be expressed as a suitable
holomorphic quotient corresponding to a smooth toric variety. To be more specific, one
considers the resolved Calabi-Yau $CY_3(3,243)$ as the holomorphic quotient:
${C^7-F\over(C^*)^3}|_{\rm
hyp\ constraint}$, where the diagonal $(C^*)^3$ actions
on the seven coordinates of $C^7$ are given by:
\begin{equation}
\label{eq:3Q's}
x^j\sim \lambda^{iQ^a_j}x^j,\ {\rm no\ sum\ over}\ j;\ a=1,2,3,
\end{equation}
where the three sets of charges $\{Q^{a=1,2,3}_{i=(0,),1,...,7}\}$ (the "0" being for
the extra chiral superfield with $Q^0_i=-\sum_{i=1}^7Q^a_i$\cite{W}) are give by the following:
\begin{equation}
\label{eq:3C*'s}
\begin{array}{ccccccccc}\\
& {\cal X}_0 &{\cal X}_1 & {\cal  X}_2 & {\cal  X}_3 & {\cal X}_4 & {\cal X}_5
& {\cal X}_6 & {\cal X}_7\\ \hline
Q^{(1)}_i:& 0 & 1 & 1 & -2
& 0 & 0 & 0 & 0 \\
Q^{(2)}_i:& 0 & 0 & 0 & 1 & 1 & 0 & 0 & -2 \\
Q^{(3)}_i: & -6 & 0 & 0 & 0 & 0 & 2 & 3 & 1 \\
\end{array}\end{equation}
where on noting:
\begin{equation}
\label{eq:linrel}
Q^{(1)}+2Q^{(2)}+4Q^{(3)}=\begin{array}{cccccccc}
-24&1&1&0&2&8&12&0
\end{array},
\end{equation}
one identifies ${\cal X}_{3,7}$ as the two extra chiral superfields introduced as a consequence of
singularity resolution.

The Landau-Ginsburg Period for the resolved $CY_3(3,243)$, as per the prescription of
Hori and Vafa, is given by: 
\begin{eqnarray}
\label{eq:PFcomp1}
& & \Pi(t_1,t_2,t_3)=\int\prod_{i=0}^7dY_i\prod_{a=1}^3dF^{(a)}
\sum_{a=1}^3d_{1a}F^{(a)}
e^{-\sum_{a=1}^3F^{(a)}(\sum_{i=1}^7Q^{(a)}_iY_i-Q^{(a)}_0Y_0-t_{(a)})-\sum_{i=0}^7e^{-Y_i}}
\nonumber\\
& & 
=\sum_{a=1}^3d_{1a}{\partial\over\partial t_{(a)}}\int\prod_{i=0}^7 dY_i\prod_{a=1}^3
\delta(\sum_{i=1}^7Q^{(a)}_iY_i-Q^{(a)}_0Y_0-t_{(a)})e^{-\sum_{i=0}^7e^{-Y_i}}\nonumber\\
& & 
\equiv\sum_{a=1}^3d_{1a}{\partial\over\partial t_{(a)}}\tilde{\Pi}(t_{1,2,3});
\end{eqnarray}
$d_{\alpha a}\equiv$ charge matrix,  $\alpha$ indexes the number of hypersurfaces
and $a$ indexes the number
of $U(1)$'s. For $CY_3(3,243)$, $\alpha=1$, $a=1,2,3$ with $d_{11}=d_{12}=0,d_{13}=6$.

Consider: 
\begin{equation}
\label{eq:dieufPi}
\tilde{\Pi}(t_{1,2,3},\{\mu_i\})\equiv\int\prod_{i=0}^7dY_i\prod_{a=1}^3
\delta(\sum_{i=1}^7Q^{(a)}_iY_i-Q^{(a)}_0Y_0-t_{(a)})e^{-\sum_{i=0}^7\mu_ie^{-Y_i}}.
\end{equation}
One can show that: 
\begin{equation}
\label{eq:redieufts}
\tilde{\Pi}(t_{1,2,3},\{\mu_i\})=\tilde{\Pi}(t_{1,2,3}^\prime,\{\mu_i=1\}),
\end{equation}
where 
\begin{equation}
\label{eq:t'sdieufs}
t_1^\prime\equiv t_1 + ln(\mu_3^2/\mu_1\mu_2),\ t_2^\prime=t_2+ln(\mu_7^2/\mu_3\mu_4),\
t_3^\prime=t_3+ln(\mu_0^6/\mu_7\mu_5^2\mu_6^3).
\end{equation}
Eliminating $Y_{0,3,7}$
 gives a order-24
Picard-Fuchs equation: 
\begin{equation}
\label{eq:PF}
{\partial^{24}\over\partial\mu_1\partial\mu_2\partial\mu_4^2
\partial\mu_5^8
\partial\mu_6^{12}}
\tilde{\Pi}(t_{1,3,4})=e^{-t_1+2t_2+4t_3}{\partial^{24}\over\partial\mu_0^{24}}
\tilde{\Pi} (t_{1,2,3}),
\end{equation}
which is the same as the PF equation for the unresolved hypersurface away from the
orbifold singularities. {\bf This overcounts the number of solutions}.

The right number of solutions must be $2h^{2,1}({\rm Mirror})+2=2.3+2=8$. To get this 
number, one notes that by adding the three constraints: 
\begin{equation}
\label{eq:consts}
Y_1+Y_2-2Y_3=t_1;\ Y_3+Y_4-2Y_7=t_2;\
-6Y_0+2Y_5+3Y_6+Y_7=t_3, 
\end{equation}
one gets: 
\begin{equation}
\label{eq:sumconsts}
-6Y_0-Y_3-Y_7+Y_1+Y_2+Y_4+2Y_5+3Y_6=t_1+t_2+t_3,
\end{equation}
which allows one to write the following order-8 PF equation:
\begin{equation}
\label{eq:PFcorrect1}
{\partial^8\over\partial\mu_1\partial\mu_2\partial\mu_4\partial\mu_5^2\partial\mu_6^3}
\tilde{\Pi}(t_{1,2,3})
=e^{-(t_1+t_2+t_3)}{\partial^8\over\partial\mu_0^6\partial\mu_3\partial\mu_7}
\tilde{\Pi}(t_{1,2,3}).
\end{equation}

If $\Theta_i\equiv{\partial\over\partial t_i^\prime}$, then one gets:
\begin{eqnarray}
\label{eq:PFcorrect2}
& & \biggl[\Theta_1^2\Theta_2\prod_{l=2}^3\prod_{k=0}^{l-1}(-l\Theta_3-k)-e^{-(t_1^\prime
+t_2^\prime+t_3^\prime)}(2\Theta_2-\Theta_3)(2\Theta_1-\Theta_2)\prod_{j=0}^5(6\Theta_3-j)
\biggr]\tilde{\Pi}=0\nonumber\\
& & {\rm with}\ z\equiv e^{-(t_1^\prime
+t_2^\prime+t_3^\prime)};\ z{d\over dz}\equiv\Delta_z,\ {\rm and\ rescaling}:
\nonumber\\
& & \biggl[\Delta_z^4(\Delta_z-{1\over2})\Delta_z(\Delta_z-{1\over3})(\Delta_z-{2\over3})
+z\prod_{j=0}^5(\Delta_z+{j\over6})\Delta_z^2\biggr]\tilde{\Pi}=0.
\end{eqnarray}

One solution to the above equation is: $\ _8F_7\left(\begin{array}{cccccccc}
0&{1\over6}&{2\over6}&{3\over6}&{4\over6}&{5\over6}&0&0\\
1&1&1&{1\over2}&1&{2\over3}&{1\over3}&\\
\end{array}\right)(-z)$

$\Biggl[$
For $e^{-t^\prime}\equiv z$, $t^\prime\equiv t_1^\prime+2t_2^\prime+4t_3^\prime$ and
suitable rescaling of $z$, the relevant order-24 PF equation {\bf for the unresolved
hypersurface} is:
\begin{equation}
\label{eq:unresPF}
\Delta^2_z\Delta_z(\Delta_z-{1\over2})\prod_{j=1}^8(\Delta_z-{j-1\over8})\prod_{j=1}^{12}
(\Delta_z-{j-1\over12})\tilde{\Pi}
=z\prod_{j=1}^{24}(\Delta_z+{j-1\over24})\tilde{\Pi}.
\end{equation}
One solution can be written in terms of the following generalized
hypergeometric function
\begin{equation}
\label{eq:PFsol1}
\ _{24}F_{23}\left(\begin{array}{ccccccc}
0 & {1\over24} & {2\over24} & {3\over24} & {4\over24} & {5\over24} & ....\ {23\over24}\\
1 & 1 & {1\over2} & {5\over8} & ...\ -{2\over8} & {1\over12} & ...\ -{10\over12}\\
\end{array}\right).
\end{equation}
$\Biggr]$

From the above solution, Meijer basis obtained using properties of 
$\ _pF_q$ and the Meijer function $I$:
\begin{eqnarray}
\label{eq:IpFqprops}
& & \ _pF_q\left(\begin{array}{cccc}
\alpha_1 & \alpha_2 & \alpha_3 & ....\ \alpha_p\\
\beta_1 & \beta_2 & \beta_3 & ....\ \beta_q\\
\end{array}\right)(z)={\prod_{i=1}^p\Gamma(\beta_i)\over\prod_{j=1}^q\Gamma(\alpha_j)}
I\left(\begin{array}{c|c}
0 & \alpha_1...\alpha_p\\
\hline
. & \beta_1...\beta_q\\
\end{array}\right)(-z)\ {\rm where}\nonumber\\
& & I\left(\begin{array}{c|c}
a_1...a_A & b_1...b_B\\
\hline
c_1...c_C & d_1...d_D\\
\end{array}\right)(z), I\left(\begin{array}{c|c}
a_1...(1-d_l)...a_A & b_1...b_B\\
\hline
c_1...c_C & d_1...\hat{d}_l...d_D\\
\end{array}\right)(-z)\nonumber\\
& & I\left(\begin{array}{c|c}
a_1...a_A & b_1..\hat{b}_j...b_B\\
\hline
c_1..(1-b_j)..c_C & d_1...d_D\\
\end{array}\right)(-z)
\end{eqnarray}
satisfy the same equation.

Now,
$z\equiv e^{-(t_1^\prime+t_2^\prime+t_3^\prime)}=e^{-(t_1+t_2+t_3)}
{\mu_1\mu_2\mu_4\mu_5^2\mu_6^3\over
\mu_0^{6}\mu_3\mu_7}.$
Hence, one can solve for large ($\equiv|z|<<1$) and small complex structure ($\equiv|z|>>1$)
limits, as well as large-size-Calabi-Yau
limit($\equiv t_i\rightarrow\infty\Leftrightarrow|z|<<1$)
on the mirror Landau-Ginsburg side with equal ease using 
Mellin-Barnes integral represention for the Meijer's function $I$,
as in \cite{GL} and in (\ref{eq:MBdieuf}) below.

Now, to get an infinite series expansion in $z$ for $|z|<1$ as well as $|z|>1$, 
one uses the following
\begin{equation}
\label{eq:MBdieuf}
I\left(\begin{array}{c|c}
a_1...a_A & b_1...b_B\\ \hline
c_1...c_C & d_1...d_D\\
\end{array}\right)(z)={1\over2\pi i}\int_\gamma ds {\prod_{i=1}^A\Gamma(a_i-s)\prod_{j=1}^B
\Gamma(b_j+s)\over\prod_{k=1}^C\Gamma(c_k-s)\prod_{l=1}^D\Gamma(d_l+s)}z^s,
\end{equation}
where the contour $\gamma$ lies to the right 
of:$s+b_j=-m\in{\bf Z}^-\cup\{0\}$ and
to the left of: $a_i-s=-m\in{\bf Z}^-\cup\{0\}$. 

\begin{figure}[htbp]
\vskip -2in
\centerline{{\psfig{file=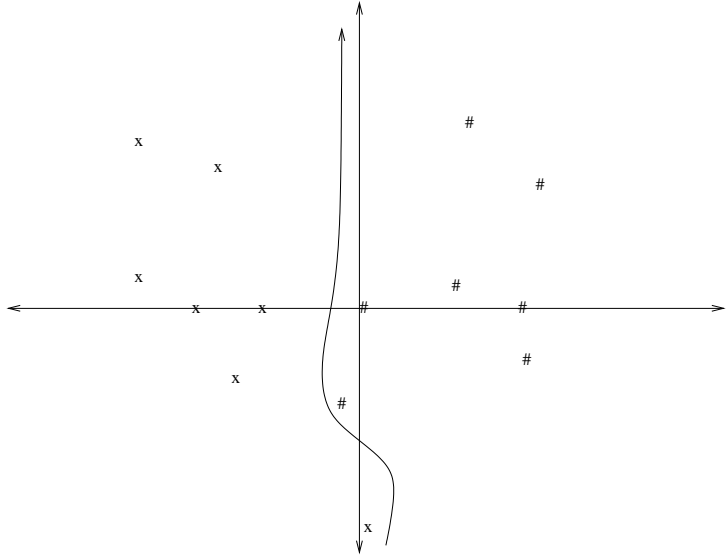,width=0.6\textwidth}}}
\vskip -1.8in
\caption{The contour $\gamma$ for $I$: The poles $s=a_i+m$ are denoted by $\#$ and the
poles $s=-m-b_j$ are denoted by $x$}
  \end{figure}

\begin{figure}[htbp]
\vskip -2in
\centerline{{\psfig{file=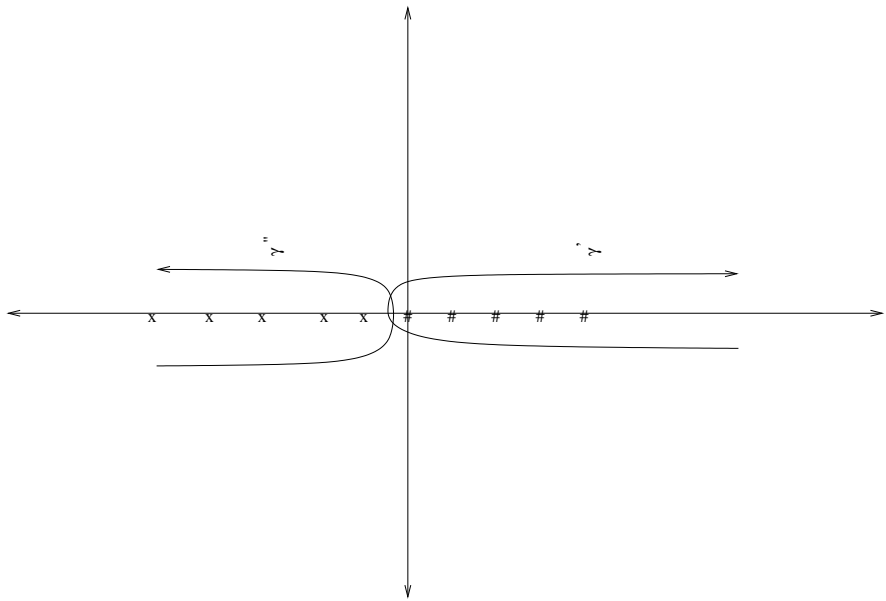,width=0.6\textwidth}}}
\vskip -1in
\caption{The deformed contour $\gamma^\prime$ valid for $|z|<1$, and $\gamma^{\prime\prime}$
valid for $|z|>>1$}
  \end{figure}
This, $|z|<<1$ and $|z|>>1$ can be dealt with equal ease by suitable deformations of the
contour $\gamma$ (see Fig. 1)
to $\gamma^\prime$ and $\gamma^{\prime\prime}$ respectively (See Fig. 2). 
Additionally, instead of performing parametric differentiation of 
infinite series to get the  $ln$-terms, one get the same (for the large complex structure
limit: $|z|<1$) by evaluation of the residue at $s=0$ in the Mellin-Barnes contour integral
in (\ref{eq:MBdieuf}) as is done explicitly to evaluate the eight integrals in (\ref{eq:sols}).
\footnote{From \cite{AM2}, one gets, for the case of large-base Calabi-Yau away from the
orbifold singularities: 
\begin{eqnarray}
\label{eq:sol1fin}
& & I\left(\begin{array}{c|c}
0& 0 {1\over24}...{23\over24}\\ 
&\\ \hline
&\\
. & 1 1 {1\over2} {5\over8}...-{2\over8} {1\over12}...-{10\over12}\\
\end{array}\right)(-z)
=\Theta(1-|z|)\sum_{m=1}^\infty{\prod_{j=1}^{23}\Gamma(m+{j\over24})z^m(-)^{m+1}\over\Gamma(m+{1\over2})
\prod_{k=-2}^5\Gamma(m+{k\over8})\prod_{l=-10}^1\Gamma(m+{l\over10})m^2m!(m-1)!}\nonumber\\
& & \times\biggl[\sum_{j=1}^{23}\Psi(m+{j\over24})+ln(-z)-1-\Psi({1\over2}+m)
-\sum_{k=-2}^5\Psi(m+{k\over8})-\sum_{l=10}^2\Psi({l\over12}+m)-{1\over m}
-\sum_{j=1}^m{1\over j}-\Psi(m+1)\biggr]\nonumber\\
& &
+\Theta(|z|-1)\sum_{m=0}^\infty\sum_{j=1}^{23}{-\Gamma({j\over24}+m)(-)^m\prod_{j\neq k=1}^{23}
\Gamma({k-j\over24}-m)\over m!(m+{j\over24})\Gamma(-m-{j\over24}+1)\prod_{k=-2}^8\Gamma
({k\over8}-m-{j\over24})\prod_{l=10}^1\Gamma({l\over12}-m-{j\over24})}(-z)^{-m-{j\over24}}\nonumber\\
& & =\Theta(1-|z|)\sum_{m=1}^\infty(2\pi)^{{7\over2}}
 {24^{{1\over2}-24m}\over 8^{{1\over2}-8[m-{1\over4}]}.10^{{1\over2}-10[m-1]}}
{\Gamma(24m)z^m(-)^{m+1}\over\Gamma(m+{1\over2})\Gamma(m-{1\over10})\Gamma(10m)m^2(m-1)!}
\nonumber\\
& & \times
\Biggl[-\Psi(m)+24\Psi(24m)+ln\biggl({12^{12}.8^8\over24^{24}}\biggr)+ln(-z)-1
-\Psi({1\over2}+m)
-8\Psi(8[m-{1\over4}])-12\Psi(12[m-{5\over6}])
\nonumber\\
& & 
-{1\over m}-\sum_{j=1}^m{1\over j}
-\Psi(m+1)\Biggr]\nonumber\\
& & -\Theta(|z|-1)\nonumber\\
& & \times\sum_{m=0}^\infty\sum_{j=1}^{23}{\Gamma({j\over24}+m)
\Gamma(m+{j\over24}+{1\over4})\Gamma(m+{j\over24}+{5\over6})(-)^m(-z)^{-m-{j\over24}}
\prod_{j\neq k}^{23}\Gamma({k-j\over24}-m)\over m!(m+{j\over24})\Gamma(-m-{j\over24}+1).
(2\pi)^9.8^{{1\over2}-8[m+{j\over24}+{1\over4}]}.12^{{1\over2}-12[m+{j\over24}+{5\over6}]}\Gamma(8[m+{j\over24}+{1\over4}])\Gamma(12[m+{j\over24}+{5\over6}])}
\nonumber\\
& & 
\end{eqnarray}}

The guiding principle is that of the eight solutions to $\tilde{\Pi}$, one should 
generate solutions in which one gets $(ln z)^P$, $P=1,...,4$ so that one gets 
$(ln z)^{P-1}$ for $\Pi$, and one can then identify terms independent of $ln z$ with
$Z^0$, three $(ln z)$ terms with $Z^{1,2,3}$, three $(ln z)^{P\leq2}$ 
terms with $F_{1,2,3}\equiv
{\partial F\over\partial Z^{1,2,3}}$,
and finally $(ln z)^{P\leq3}$ term with $F_0\equiv{\partial F\over\partial Z^0}$.

One (non-unique) choice of solutions for $\tilde{\Pi}(z)$ is given below:
\begin{eqnarray}
\label{eq:sols}
z{d\over dz}\left[\begin{array}{c}
I\left(\begin{array}{c|c}
0\ 0 & 0\ {1\over6}\ {2\over6}\ {3\over6}\ {4\over6}\ {5\over6}\ 0\ 0 \\
& \\ \hline
&\\
. & 1\ 1\ {1\over2}\ 1\ {2\over3}\ {1\over3}\\
\end{array}\right)(-z)\\ 
\\
I\left(\begin{array}{c|c}
0\ 0 {1\over2}\ {1\over3}\ {2\over3} & 
{1\over6}\ {2\over6}\ {3\over6}\ {4\over6}\ {5\over6}\\
& \\ \hline
&\\
1\ 1\ 1 & 1\ 1\ 1\\
\end{array}\right)(-z)\\
\\
I\left(\begin{array}{c|c}
0\ 0 \ {1\over2}& 0\ {1\over6}\ {2\over6}\ {3\over6}\ {4\over6}\ {5\over6}\ 0\ \\
& \\ \hline
&\\
1 & 1\ 1\ 1\ {2\over3}\ {1\over3}\\
\end{array}\right)(-z)\\
\\
I\left(\begin{array}{c|c}
0\ 0 \ {1\over3}& 0\ {1\over6}\ {2\over6}\ {3\over6}\ {4\over6}\ {5\over6}\ 0\ \\
& \\ \hline
&\\
1 & 1\ 1\ {1\over2}\ 1\ {1\over3}\\
\end{array}\right)(z)\\
\\
I\left(\begin{array}{c|c}
0\ 0 \ {2\over3}& 0\ {1\over6}\ {2\over6}\ {3\over6}\ {4\over6}\ {5\over6}\ 0\ \\
& \\ \hline
&\\
1 & 1\ 1\ {1\over2}\ 1\ {2\over3}\\
\end{array}\right)(-z)\\
\\
I\left(\begin{array}{c|c}
0\ 0 \ {1\over2}& {1\over6}\ {2\over6}\ {3\over6}\ {4\over6}\ {5\over6}\ 0\ \\
& \\ \hline
&\\
1\ 1 & 1\ 1\ 1\ {2\over3}\ {1\over3}\\
\end{array}\right)(z)\\
\\
I\left(\begin{array}{c|c}
0\ 0 \ {1\over3}& {1\over6}\ {2\over6}\ {3\over6}\ {4\over6}\ {5\over6}\ 0\ \\
& \\ \hline
&\\
1\ 1 & 1\ 1\ {1\over2}\ {2\over3}\ {1\over3}\\
\end{array}\right)(z)\\
\\
I\left(\begin{array}{c|c}
0\ 0 \ {2\over3}& {1\over6}\ {2\over6}\ {3\over6}\ {4\over6}\ {5\over6}\ 0\ \\
& \\ \hline
&\\
1\ 1 & 1\ 1\ {1\over2}\ {2\over3}\\
\end{array}\right)(z)\\
\end{array}\right]\sim\left[\begin{array}{c}
F_0\\
Z^0\\
F_1\\
F_2\\
F_3\\
Z^1\\
Z^2\\
Z^3\\
\end{array}\right]
\end{eqnarray}

(a)
\begin{eqnarray}
\label{eq:reltoF_0}
& & I\left(\begin{array}{c|c}
0\ 0 & 0\ {1\over6}\ {2\over6}\ {3\over6}\ {4\over6}\ {5\over6}\ 0\ 0 \\
& \\ \hline
& \\
. & 1\ 1\ {1\over2}\ 1\ {2\over3}\ {1\over3}\\
\end{array}\right)(-z)={1\over2\pi i}\int_\gamma ds{[\Gamma(-s)]^2[\Gamma(s)]^3\prod_{j=1}^5
\Gamma(s+{j\over6})\over[\Gamma(s+1)]^3\Gamma({1\over2}+s)\Gamma({2\over3}+
s)\Gamma({1\over3}+s)} (-z)^s
\nonumber\\
& & =\theta(1-|z|)\Biggl[{2(2\pi)^{{3\over2}}\over\sqrt{\pi}}\biggl[\biggl(ln\biggl({2^23^3\over6^6}\biggr) + ln(-z)\biggr)^4+{65\over3}
\biggl(ln\biggl({2^23^3\over6^6}\biggr) + ln(-z)\biggr)^2\nonumber\\
& & +{169\pi^4\over3}-1440\zeta(3)\biggl(
ln\biggl({2^23^3\over6^6}\biggr)+ln(-z)\biggr)+402\zeta(4)\biggr]
+\sum_{m=1}^\infty{(2\pi)^{{3\over2}}2^{{1\over2}-6m}\Gamma(6m)\over m^3(m!)^23^{3m}
\Gamma({1\over2}+m)\Gamma(3m)}\nonumber\\
& & \times\biggl[2\gamma+6\Psi(6m)-2\Psi(2m)-3\Psi(m)-3m^2+
ln\biggl({2^23^3\over6^6}\biggr)\biggr](-z)^m\Biggr]\nonumber\\
& & -\theta(|z|-1)\Biggl[ \sum_{m=0}^\infty{(-)^m
(\Gamma(m+{1\over6})^2\Gamma(-m+{2\over3})\over(m+{1\over6})^3m!}(-z)^{-m-{1\over6}}+
\nonumber\\
& & 
\sum_{m=0}^\infty{(-)^m(\Gamma(m+{5\over6})^2\Gamma(-m-{2\over3})\over(m+{5\over6})^3m!}(-z)^{-m-{5\over6}}\Biggr]
\end{eqnarray}

(b)

\begin{eqnarray}
\label{eq:reltoZ^0}
& & I\left(\begin{array}{c|c}
0\ 0 {1\over2}\ {1\over3}\ {2\over3} & 
{1\over6}\ {2\over6}\ {3\over6}\ {4\over6}\ {5\over6}\\
& \\ \hline
&\\
1\ 1\ 1 & 1\ 1\ 1\\
\end{array}\right)(-z)\nonumber\\
& & ={1\over2\pi i}\int_\gamma ds{[\Gamma(-s)]^2\Gamma({1\over2}-s)\Gamma(
{1\over3}-s)\Gamma({2\over3}-s)\prod_{j=1}^5\Gamma(s+{j\over6})\over[\Gamma(1-s)]^3[\Gamma
(1+s)]^2}(-z)^s
\nonumber\\
& & =\Theta(1-|z|)\biggl[{\sqrt{\pi}.2\pi.(2\pi)^{{5\over2}}\over\sqrt{18}}\biggl(
ln[{2^23^3\over6^6}]+ln(-z)\biggr)\nonumber\\
& & +\sum_{m=0}^\infty
{(-)^{m+1}(2\pi)^{{5\over2}}6^{{1\over2}-6[m+{1\over2}]}
\over m!(m+{1\over2})^2}
{\Gamma({1\over3}-[{1\over2}+m])\Gamma({2\over3} -[{1\over2}+m])\Gamma(6[m+{1\over2}])\over
[\Gamma({3\over2}-m)]^2\Gamma({1\over2}-m)\Gamma(m+{1\over2})}(-z)^{m+{1\over2}}
\nonumber\\
& & +\sum_{m=0}^\infty
{(-)^{m+1}(2\pi)^{{5\over2}}6^{{1\over2}-6[m+{1\over3}]}
\over m!(m+{1\over3})^2}
{\Gamma({1\over2}-[{1\over3}+m])\Gamma({2\over3} -[{1\over3}+m])\Gamma(6[m+{1\over3}])\over
[\Gamma({2\over3}-m)][\Gamma({2\over3}+m)]^2\Gamma(m+{1\over3})}(-z)^{m+{1\over3}}
\nonumber\\
& & +\sum_{m=0}^\infty
{(-)^{m+1}(2\pi)^{{5\over2}}6^{{1\over2}-6[m+{2\over3}]}
\over m!(m+{2\over3})^2}
{\Gamma({1\over2}-[{2\over3}+m])\Gamma({1\over3} -[{2\over3}+m])\Gamma(6[m+{2\over3}])\over
[\Gamma({1\over3}-m)][\Gamma({5\over3}+m)]^2\Gamma(m+{2\over3})}(-z)^{m+{2\over3}}\biggr]
\nonumber\\
& & +\Theta(|z|-1)\nonumber\\
& & \times\biggl[\sum_{m=0}^\infty\sum_{j=1}^5{(-)^{m+1}\over m!(m+{j\over6})^2}
{\Gamma({1\over2}+m+{j\over6}\Gamma({1\over3}+m+{j\over6})\Gamma({2\over3}+m+{j\over6})
\prod_{k\neq j}\Gamma(-m+{k-j\over6})\over
\Gamma(1+m+{j\over6})[\Gamma(1-m-{j\over6})]^2}(-z)^{-m-{j\over6}}\Biggr]\nonumber\\
& & 
\end{eqnarray}

(c) 

\begin{eqnarray}
\label{eq:reltoZ^1}
& & I\left(\begin{array}{c|c}
0\ 0 {1\over2} &
{1\over6}\ {2\over6}\ {3\over6}\ {4\over6}\ {5\over6}\ 0\\
& \\ \hline
&\\
1\ 1 & 1\ 1\ 1\ {2\over3}\ {1\over3}\\
\end{array}\right)(z)\nonumber\\
& & ={1\over2\pi i}\int_\gamma ds{[\Gamma(-s)]^2\Gamma({1\over2}-s)\Gamma(s)
\prod_{j=1}^5\Gamma(s+{j\over6})\over[\Gamma(1-s)]^2[\Gamma
(1+s)]^3\Gamma({2\over3}+s)\Gamma({1\over3}+s)}(z)^s
\nonumber\\
& & \Theta(1-|z|)\Biggl[2\pi^2\biggl[(ln z+ln\biggl({2^2.3^3\over6^6}\biggr)^2+{14\pi^2\over3}
\biggr]\nonumber\\
& & +\sum_{m=0}^\infty
{(-)^{m+1}z^{m+{1\over2}}\over m!}{(2\pi)^{{5\over2}}.6^{{1\over2}
-6[m+{1\over2}]}\Gamma(6[m+{1\over2}])\over(m+{1\over2})^3[\Gamma({3\over2}+m)]^2.2\pi.
3^{{1\over2}-3[m+{1\over2}]}\Gamma(3[m+{1\over2}])}
[\Gamma({3\over2}+m)]^2\Biggr]\nonumber\\
& & -\Theta(|z|-1)\Biggl[\sum_{m=0}^\infty{(-)^{m+1}.z^{-m-{j\over6}}\over m!}{
\Gamma({1\over2}+m+{j\over6})\prod_{k\neq j}\Gamma({k-j\over6}-m)\over(m+{j\over6})^3
[\Gamma(1-m-{j\over6})]^2\Gamma({2\over3}-m-{j\over6})\Gamma({1\over3}-m-{j\over6})}
\Biggr]
\nonumber\\
& & 
\end{eqnarray}

(d)

\begin{eqnarray}
\label{eq:eq:reltoZ^2}
& & I\left(\begin{array}{c|c}
0\ 0 {1\over3} &
{1\over6}\ {2\over6}\ {3\over6}\ {4\over6}\ {5\over6}\ 0\\
& \\ \hline
&\\
1\ 1 & 1\ 1\ {1\over2}\ 1\ {1\over3}\\
\end{array}\right)(z)\nonumber\\
& & ={1\over2\pi i}\int_\gamma ds{[\Gamma(-s)]^2\Gamma({1\over3}-s)\Gamma(s)
\prod_{j=1}^5\Gamma(s+{j\over6})\over[\Gamma(1-s)]^2[\Gamma
(1+s)]^3\Gamma({1\over2}+s)\Gamma({1\over3}+s)}(z)^s
\nonumber\\
& & =\Theta(1-|z|)\Biggl[\sqrt{16\over3}.\pi^2\biggl[\biggl(ln\biggl[{2^2.3^3\over6^6}\biggr]
+ln z +{\pi\over\sqrt{3}}\biggr)^2+5\pi^2\biggr]\nonumber\\
& & +\sum_{m=0}^\infty{(-)^{m+1}\over m!}{(2\pi)^{{5\over2}}.6^{{1\over2}-6[{1\over3}+m]}
\Gamma(6[{1\over3}+m])\over(m+{1\over3})^3\Gamma({1\over3}+m)[\Gamma({4\over3}+m)]^2
\Gamma({5\over6}+m)\Gamma({2\over3}+m)}\Biggr]\nonumber\\ 
& & -\Theta(|z|-1)\sum_{m=0}^\infty\sum_{j=1}^5{(-)^{m+1}\over m!}
{\Gamma({1\over3}+m+{j\over6})\prod_{k\neq j}^5\Gamma({k-j\over6}-m)
\over({j\over6}+m)^3[\Gamma(1-{j\over6}-m)]^2
\Gamma({1\over2}-{j\over6}-m)\Gamma({1\over3}-{j\over6}-m)}z^{-{j\over6}-m}
\nonumber\\
& & 
\end{eqnarray}

(e)

\begin{eqnarray}
\label{eq:eq:reltoZ^3}
& & I\left(\begin{array}{c|c}
0\ 0 {2\over3} &
{1\over6}\ {2\over6}\ {3\over6}\ {4\over6}\ {5\over6}\ 0\\
& \\ \hline
&\\
1\ 1 & 1\ 1\ {1\over2}\ 1\ {2\over3}\\
\end{array}\right)(z)\nonumber\\
& & ={1\over2\pi i}\int_\gamma ds{[\Gamma(-s)]^2\Gamma({2\over3}-s)\Gamma(s)
\prod_{j=1}^5\Gamma(s+{j\over6})\over[\Gamma(1-s)]^2[\Gamma
(1+s)]^3\Gamma({1\over2}+s)\Gamma({2\over3}+s)}(z)^s
\nonumber\\
& & =\Theta(1-|z|)\Biggl[\sqrt{16\over3}.\pi^2\biggl[\biggl(ln\biggl[{2^2.3^3\over6^6}\biggr]
+ln z -{\pi\over\sqrt{3}}\biggr)^2+5\pi^2\biggr]\nonumber\\
& & +\sum_{m=0}^\infty{(-)^{m+1}\over m!}{(2\pi)^{{5\over2}}.
6^{{1\over2}-6[{2\over3}+m]}
\Gamma(6[{2\over3}+m])\over(m+{1\over3})^3\Gamma({2\over3}+m)[\Gamma({4\over3}+m)]^2
\Gamma({7\over6}+m)\Gamma({4\over3}+m)}\Biggr]\nonumber\\ 
& & -\Theta(|z|-1)\sum_{m=0}^\infty\sum_{j=1}^5{(-)^{m+1}\over m!}{
\Gamma({2\over3}+m+{j\over6})\prod_{k\neq j}^5\Gamma({k-j\over6}-m)
\over({j\over6}+m)^3[\Gamma(1-{j\over6}-m)]^2
\Gamma({1\over2}-{j\over6}-m)\Gamma({2\over3}-{j\over6}-m)}z^{-{j\over6}-m}
\nonumber\\
& & 
\end{eqnarray}

(f)

\begin{eqnarray}
\label{eq:reltoF_1}
& & I\left(\begin{array}{c|c}
0\ 0 \ {1\over2}& 0\ {1\over6}\ {2\over6}\ {3\over6}\ {4\over6}\ {5\over6}\ 0\ \\
& \\ \hline
&\\
1 & 1\ 1\ 1\ {2\over3}\ {1\over3}\\
\end{array}\right)(-z)\nonumber\\
& & 
={1\over2\pi i}\int_\gamma{[\Gamma(-s)]^2[\Gamma(s)]^2\Gamma({1\over2}-s)
\prod_{j=1}^5\Gamma(s+{j\over6})\over\Gamma(1-s)[\Gamma(1+s)]^3\Gamma(s+{2\over3})
\Gamma(s+{1\over3})}(-z)^s
\nonumber\\
& & =\Theta(1-|z|)\Biggl[2\pi^2\Biggl(\biggl(ln(-z)+ln\biggl[{2^3.3^3\over6^6}\biggr]\biggr)^3
+15\pi^2\biggl(ln(-z)+ln\biggl[{2^2.3^3\over6^6}\biggr]\biggr)-356\zeta(3)\Biggr)\nonumber\\
& & 
+\sum_{m=1}^\infty{(-)^{m}(2\pi)^{{3\over2}}\over (m!)^2m^3}{6^{{1\over2}-6m}\over3^{{1\over2}
-3m}}{\Gamma(6m)\over\Gamma(3m)}\Gamma({1\over2}-m)(-z)^m\nonumber\\
& & +\sum_{m=1}^\infty{(-)^m\over(m!)^2(m+{1\over2})^3}.{(2\pi)^{{3\over2}}.6^{{1\over2}
-6[m+{1\over2}]}\over3^{{1\over2}-3[m+{1\over2}]}}
{\Gamma(6[m+{1\over2}])\over\Gamma(3[m+{1\over2}])}(-z)^{m+{1\over2}}
\Biggr]
\nonumber\\
& & +\Theta(|z|-1)\sum_{m=0}^\infty\sum_{j=1}^5{(-)^m\over 2\pi 
3^{{1\over2}+3[{j\over6}+m)]}m!({j\over6}+m)^3}{\Gamma({j\over6}
+m)\over\Gamma(1-{j\over6}-m)}.\Gamma({1\over2}+{j\over6}+m).{\Gamma(-[{j\over6}+m])\over\Gamma(-3[{j\over6}+m])}\nonumber\\
& & 
\times\prod_{k\neq j}
\Gamma({k-j\over6}-m).z^{-{j\over6}-m}
\end{eqnarray}

(g)
\begin{eqnarray}
\label{eq:reltoF_2}
& & I\left(\begin{array}{c|c}
0\ 0 \ {1\over3}& 0\ {1\over6}\ {2\over6}\ {3\over6}\ {4\over6}\ {5\over6}\ 0\ \\
& \\ \hline
&\\
1 & 1\ 1\ {1\over2}\ 1\ {1\over3}\\
\end{array}\right)(-z)\nonumber\\
& & ={1\over2\pi i}\int_\gamma{[\Gamma(-s)]^2[\Gamma(s)]^2\Gamma({1\over3}-s)
\prod_{j=1}^5\Gamma(s+{j\over6})\over\Gamma(1-s)[\Gamma(1+s)]^3
\Gamma({1\over2}+
s)\Gamma({1\over3}+s)}(-z)^s\nonumber\\
& & =\Theta(1-|z|)\Biggl[\pi^2\sqrt{{16\over3}}\biggl(\biggl[ln\biggl({2^2.3^3\over
6^6}\biggr)+ln(-z)+{\pi\over\sqrt{3}}\biggr)^3\nonumber\\
& & +16\pi^2\biggl[ln\biggl({2^2.3^3\over6^6}\biggr)+ln(-z)+{\pi\over\sqrt{3}}\biggr]
-360\zeta(3)-{8\pi^3\over 3^{{3\over2}}}\biggr)\nonumber\\
& & +\sum_{m=1}^\infty{(-)^{m+1}\over (m!)^2m^3}{\Gamma({1\over3}-m)\over
\Gamma({1\over3}+m)} (2\pi)^{{5\over2}}6^{{1\over2}-6m}{\Gamma(6m)\over\Gamma(m)
\Gamma({1\over2}+m)}(-z)^m\nonumber\\
& & 
+\sum_{m=0}^\infty{(-)^{m+1}\over m!({1\over3}+m)^4}{\Gamma({2\over3}-m)\over
\Gamma({4\over3}+m)}{(2\pi)^{{5\over2}}6^{{1\over2}-6[{1\over3}+m]}
\Gamma(6[{1\over3}+m])\over\Gamma({1\over3}+m)}{(-z)^{m+{1\over3}}\over
\Gamma({5\over6}+m)\Gamma({2\over3}+m)}\Biggr]
\nonumber\\
& & +\Theta(|z|-1)
\sum_{m=0}^\infty\sum_{j=1}^5{(-)^{m+1}\over m!({j\over6}+m)^4}
{\Gamma({1\over3}+{j\over6}+m)\Gamma(1+{j\over6}+m)
\over\Gamma(1-{j\over6}-m)\Gamma({1\over2}-{j\over6}-m)}{\prod_{k\neq j}
\Gamma({k-j\over6}-m)\over\Gamma({1\over3}-{j\over6}-m)}(-z)^{-{j\over6}-m}
\nonumber\\
& & 
\end{eqnarray}

(h)
\begin{eqnarray}
\label{eq:reltoF_3}
& & I\left(\begin{array}{c|c}
0\ 0 \ {2\over3}& 0\ {1\over6}\ {2\over6}\ {3\over6}\ {4\over6}\ {5\over6}\ 0\ \\
& \\ \hline
&\\
1 & 1\ 1\ {1\over2}\ 1\ {2\over3}\\
\end{array}\right)(-z)\nonumber\\
& & ={1\over2\pi i}\int_\gamma{[\Gamma(-s)]^2[\Gamma(s)]^2\Gamma({2\over3}-s)
\prod_{j=1}^5\Gamma(s+{j\over6})\over\Gamma(1-s)[\Gamma(1+s)]^3
\Gamma({1\over2}+
s)\Gamma({2\over3}+s)}(-z)^s\nonumber\\
& & =\Theta(1-|z|)\Biggl[\pi^2\sqrt{{16\over3}}\biggl(\biggl[ln\biggl({2^2.3^3\over
6^6}\biggr)+ln(-z)-{\pi\over\sqrt{3}}\biggr)^3\nonumber\\
& & +16\pi^2\biggl[ln\biggl({2^2.3^3\over6^6}\biggr)+ln(-z)+{\pi\over\sqrt{3}}\biggr]
-360\zeta(3)+{8\pi^3\over 3^{{3\over2}}}\biggr)\nonumber\\
& & +\sum_{m=1}^\infty{(-)^{m+1}\over (m!)^2m^3}{\Gamma({2\over3}-m)\over
\Gamma({1\over3}+m)} (2\pi)^{{5\over2}}6^{{1\over2}-6m}{\Gamma(6m)\over\Gamma(m)
\Gamma({1\over2}+m)}(-z)^m\nonumber\\
& & +\sum_{m=0}^\infty{(-)^{m+1}\over m!({2\over3}+m)^4}{\Gamma({1\over3}-m)\over
\Gamma({5\over3}+m)}{(2\pi)^{{5\over2}}6^{{1\over2}-6[{2\over3}+m]}
\Gamma(6[{2\over3}+m])\over\Gamma({2\over3}+m)}{(-z)^{m+{2\over3}}\over
\Gamma({7\over6}+m)\Gamma({4\over3}+m)}\Biggr]\nonumber\\
& & +\Theta(|z|-1)
\sum_{m=0}^\infty\sum_{j=1}^5{(-)^{m+1}\over m!({j\over6}+m)^4}
{\Gamma({2\over3}+{j\over6}+m)\Gamma(1+{j\over6}+m)
\over\Gamma(1-{j\over6}-m)\Gamma({1\over2}-{j\over6}-m)}{\prod_{k\neq j}
\Gamma({k-j\over6}-m)\over\Gamma({2\over3}-{j\over6}-m)}(-z)^{-{j\over6}-m}
\nonumber\\
& & 
\end{eqnarray}\footnote{In equations (\ref{eq:reltoF_0}) - (\ref{eq:reltoF_3}), use 
has been made of the following identities:
\begin{center}
$\prod_{l=0}^{n-1}\Gamma(x+{l\over n})=(2\pi)^{{n-1\over2}}n^{{1\over2}-nx}\Gamma(nx)$,

$\sum_{l=0}^{n-1}\psi(x+{l\over n})-n\psi(nx)=-n ln n$,

$\psi(1)=-\gamma,\ 
\psi^{(n)}=(-)^{n-1}n!\zeta(n+1)$.
\end{center}

Using these, one arrives at:
\begin{center}
$\psi({1\over3})+\psi({2\over3})=-2\gamma-ln 3^3$,

$\psi({1\over3})-\psi({2\over3})=-{\pi\over\sqrt{3}}$,

$\sum_{j=1}^5\psi({j\over6})=-5\gamma-ln 6^6$

$\psi^\prime({1\over3})+\psi^\prime({2\over3})=8\zeta(2)={4\over3}\pi^2,$

$\psi^{\prime\prime}({1\over3})+\psi^{\prime\prime}({2\over3})=-52\zeta(3)$,

$\psi^{\prime\prime\prime}({1\over3})+\psi^{\prime\prime\prime}({2\over3})=480\zeta(4)=
{16\pi^4\over3}$,

$\sum_{j=1}^5\psi^\prime({j\over6})=35\zeta(2)={35\over6}\pi^2$,

$\sum_{j=1}^5\psi^{\prime\prime}({j\over6})=-430\zeta(3)$,

$\sum_{j=1}^5\psi^{\prime\prime\prime}({j\over6})=7770\zeta(4)={259\pi^4\over3}$.
\end{center} }

The connection between (\ref{eq:sols}) 
that {\it effectively depends
only on one complex structure parameter
$z=e^{-(t_1+t_2+t_3)}{1\over z_1z_2z_3}$}, and the solutions
 given in the literature \cite{HKTY}of the form:
\begin{equation}
\label{eq:solklemm}
\partial_{\rho_m}^{s_m}\partial_{\rho_n}^{s_n}\partial_{\rho_p}^{s_p}
\sum_{m,n,p}
c(m,n,p;\rho_m,\rho_n,\rho_p)
z_1^{m+\rho_m}z_2^{n+\rho_n}z_3^{p+\rho_p}|_{\rho_m=\rho_n=\rho_p=0},
\end{equation}
with $s_m+s_n+s_p\leq3$,
and $z_1\equiv{\mu_1\mu_2\over\mu_3^2},\ z_2\equiv{\mu_3\mu_4\over\mu_7^2},\ z_3\equiv
{\mu_7\mu_5^2\mu_6^3\over\mu_0^6}$, needs to be understood.
The appearance of $\partial_{\rho_m}^{s_m}\partial_{\rho_n}^{s_n}\partial_{\rho_p}^{s_p}
\sum_{m,n,p}$ in (\ref{eq:solklemm}) is what was referred to earlier on as
parametric differentiation of infinite series, something which, as we have explicitly shown
above, is not needed in the approach followed in this work.

The Picard-Fuchs equation can be written in the form\cite{M}:
\begin{equation}
\label{eq:PFform}
\biggl(\Delta_z^{8}+\sum_{i=1}^{7}{\bf B}_i(z)\Delta_z^i\biggr)\tilde{\Pi}(z)
=0.
\end{equation}
The Picard-Fuchs
 equation in the form written in (\ref{eq:PFform}) can alternatively be expressed as
the following system of eight linear differential equations:
\begin{eqnarray}
\label{eq:PFdiffeq3}
& & \Delta_z\left(\begin{array}{c}\\
\tilde{\Pi}(z)\\
\Delta_z
\tilde{\Pi}(z)\\
(\Delta_z)^2
\tilde{\Pi}(z)\\
...\\
(\Delta_z)^7
\tilde{\Pi}(z)\\
\end{array}\right)
=\nonumber\\
& & \left(\begin{array}{ccccc}\\
0 & 1 & 0 & ... 0 & 0\\
0 & 0 & 1 & ... 0 & 0\\
. & . & . & ... . & . \\
0 & 0 & 0 & ... 0 & 1 \\
0 & -{\bf B}_1(z) & -{\bf B}_2(z) & ... -{\bf B}_6(z) & -{\bf B}_7(z)\\
\end{array}\right)
\left(\begin{array}{c}\\
\tilde{\Pi}(z)\\
\Delta_z
\tilde{\Pi}(z)\\
(\Delta_z)^2
\tilde{\Pi}(z)\\
...\\
(\Delta_z)^7
\tilde{\Pi}(z)\\
\end{array}\right)
\end{eqnarray}
The matrix on the RHS of (\ref{eq:PFdiffeq3}) is usually denoted by $A(z)$.
 
If the eight solutions, $\{\tilde{\Pi}_{I=1,...,8}\}$,
 are collected as a column vector $\tilde{\Pi}(z)$, then the {\it constant}\footnote{This 
thus implies that both $\tilde{\Pi}$ and $\Pi$, have the {\it same} monodromy matrix.}
monodromy matrix $T$ for $|z|<<1$ is defined by:
\begin{equation}
\tilde{\Pi}(e^{2\pi i}z)=T\tilde{\Pi}(z).
\end{equation}
The basis for the space of solutions can be collected as the
columns of the ``fundamental matrix" $\Phi(z)$ given by:
\begin{equation}
\label{eq:PFdiffeqsol}
\Phi(z)=
S_8(z)z^{R_8},
\end{equation}
where $S_8(z)$ and $R_8$ are 8$\times$8 matrices that single and
multiple-valued respectively. Note that ${\bf B}_i(0)\neq0$, which influences
the monodromy properties. Also,
\begin{equation}
\label{eq:Phidieuf2}
\Phi(z)_{ij}=\left(\begin{array}{ccc}\\
\tilde{\Pi}_1(z) & ... & \tilde{\Pi}_8(z) \\
\Delta_z \tilde{\Pi}_1(z) & ... & \Delta \tilde{\Pi}_8(z) \\
\Delta_z^2 \tilde{\Pi}_2(z) & ... & \Delta^2 \tilde{\Pi}_8(z) \\
... & ... & ... \\
\Delta_z^7 \tilde{\Pi}_1(z) & ... & \Delta_z^7 \tilde{\Pi}_8(z)\\
\end{array}\right)_{ij},
\end{equation}
implying that
\begin{equation}
\label{eq:Tdieuf}
T=e^{2\pi i R^t}.
\end{equation}
Now, writing $z^R=e^{R ln z}=1 + R ln z + R^2(ln z)^2+...$, and further
noting that there are no terms of order higher than $(ln z)^4$ in
$\tilde{\Pi}(z)$
obtained above, implies that the matrix $R$ must satisfy the property:
$R^{m}=0,\ m=5,...\infty$. Hence, $T=e^{2\pi i R^t}=1 + 2\pi i R^t +
{(2\pi i)^2\over2}(R^t)^2+{(2\pi i)^3\over 6}(R^t)^3+{(2\pi i)^4\over24}(R^t)^4.$ 
Irrespective of whether or not the
distinct eigenvalues of $A(0)$ differ by integers, one has to evaluate
$e^{2\pi i A(0)}$. The eigenvalues of $A(0)$ of (\ref{eq:A(0)}), are $0^5,{1\over3},{1\over2},
{2\over3}$, and hence five of the eight eigenvalues differ by an integer (0). 

Now, the Picard-Fuchs equation (\ref{eq:PFcorrect2}) can be rewritten in the form
(\ref{eq:PFform}), with the following values of $B_i$'s:
\begin{eqnarray}
\label{eq:Bis}
& & {\bf B}_{1,2}=0\nonumber\\
& & {\bf B}_3={5z\over324(1+z)}\nonumber\\
& & {\bf B}_4={137\over648(1+z)}\nonumber\\
& & {\bf B}_5={({25\over24}z-{1\over9})\over(1+z)}\nonumber\\
& & {\bf B}_6={(26+85z)\over36(1+z)}\nonumber\\
& & {\bf B}_7=-{(3-5z)\over2(1+z)}.
\end{eqnarray}

Under the change of basis
$\tilde{\Pi}(z)\rightarrow \tilde{\Pi}^\prime(z)=M^{-1}\tilde{\Pi}(z)$, 
and writing $\tilde{\Pi}_j(z)=\sum_{i=0}^4(ln z)^iq_{ij}(z)$ (See \cite{GL} and the
appendix), one sees that 
\begin{eqnarray}
\label{eq:primed}
& & \tilde{\Pi}^\prime_j(z)=\sum_{i=0}^4(ln z)^iq^\prime_{ij}(z),\nonumber\\
& & q^\prime(z)=q(z)(M^{-1})^t,\nonumber\\
& &  \Phi^\prime(z)\Phi(z)(M^{-1})^t,\ S^\prime(z)=S(z)(M^{-1})^t,\ R^\prime=M^tR(M^{-1})^t.
\end{eqnarray}
By choosing $M$ such that
$S^\prime(0)={\bf 1}_{24}$, one gets 
\begin{equation}
\label{eq:monodromy1}
T(0)=M(e^{2i\pi A(0)})^tM^{-1}.
\end{equation}
We evaluate $(M^{-1})^t$ and $e^{2\pi i A(0)}$ in the appendix. From the
form of $(M^{-1})^t$, one gets:
\begin{eqnarray}
\label{eq:Mexpression}
& & M=\{ \{ 1,0,0,\frac{-6\,\left( -16 + 16\,q_{00} - 29499\,q_{10} \right) }{1617589},
\frac{5586 - 5586\,q_{00} + 795503\,q_{10}}{14558301},\nonumber\\
& & \frac{-4\,\left( -110 + 110\,q_{00} - 607\,q_{10} \right) }{4852767},
 \frac{-5\,\left( -2052 + 2052\,q_{00} + 665123\,q_{10} \right) }{131024709},\nonumber\\
& &    \frac{16212 - 16212\,q_{00} - 8527877\,q_{10}}{262049418}\} ,
  \{ 0,1,0,\frac{-2\,\left( 88497 + 48\,q_{01} - 88497\,q_{11} + 1030121\,q_{21} \right) }
{1617589},\nonumber\\
& &   \frac{-795503 - 5586\,q_{01} + 795503\,q_{11} - 6851300\,q_{21}}{14558301},
   \frac{-4\,\left( 607 + 110\,q_{01} - 607\,q_{11} + 1710\,q_{21} \right) }{4852767},
\nonumber\\
& & 
   \frac{3325615 - 10260\,q_{01} - 3325615\,q_{11} + 25280767\,q_{21}}{131024709},\nonumber\\
& & \frac{8527877 - 16212\,q_{01} - 8527877\,q_{11} + 62539841\,q_{21}}{262049418}\} ,
\nonumber\\
& & \{ 0,0,1,\frac{-96\,q_{02} + 176994\,q_{12} - 1030121\,\left( -1 
+ 2\,q_{22} \right) }{1617589},\nonumber\\
& &    \frac{-5586\,q_{02} + 795503\,q_{12} - 3425650\,\left( -1 + 2\,q_{22} \right) }{14558301},
   \frac{-4\,\left( 110\,q_{02} - 607\,q_{12} + 855\,\left( -1 + 2\,q_{22} \right)  \right) }{4852767},\nonumber\\
& & 
   \frac{-20520\,q_{02} - 6651230\,q_{12} + 25280767\,\left( -1 + 2\,q_{22} \right) }
{262049418},\nonumber\\
& &    \frac{-32424\,q_{02} - 17055754\,q_{12} + 62539841\,\left( -1 + 2\,q_{22} \right) }{524098836}\} ,\nonumber\\
& & 
  \{ 0,0,0,\frac{-2\,\left( 48\,q_{03} - 88497\,q_{13} + 1030121\,q_{23} \right) }{1617589},
\nonumber\\
& & 
   \frac{-5586\,q_{03} + 795503\,q_{13} - 6851300\,q_{23}}{14558301},\nonumber\\
& &    \frac{-4\,\left( 110\,q_{03} - 607\,q_{13} + 1710\,q_{23} \right) }{4852767},
   \frac{-10260\,q_{03} - 3325615\,q_{13} + 25280767\,q_{23}}{131024709},\nonumber\\
& & 
   \frac{-16212\,q_{03} - 8527877\,q_{13} + 62539841\,q_{23}}{262049418}\} ,\nonumber\\
& &   \{ 0,0,0,\frac{-2\,\left( 48\,q_{04} - 88497\,q_{14} + 1030121\,q_{24} - 4852767\,q_{34} \right) }{1617589},\nonumber\\
& & 
   \frac{-5586\,q_{04} + 795503\,q_{14} - 6851300\,q_{24} + 349399224\,q_{44}}{14558301},
\nonumber\\
& &    \frac{-4\,\left( 110\,q_{04} - 607\,q_{14} + 1710\,q_{24} \right) }{4852767},
   \frac{-10260\,q_{04} - 3325615\,q_{14} + 25280767\,q_{24}}{131024709},\nonumber\\
& &    \frac{-16212\,q_{04} - 8527877\,q_{14} + 62539841\,q_{24}}{262049418}\} ,\nonumber\\
& &   \{ 0,0,0,\frac{-2\,\left( 48\,q_{05} - 88497\,q_{15} + 1030121\,q_{25} - 4852767\,q_{35} \right) }{1617589},\nonumber\\
& & 
   \frac{-5586\,q_{05} + 795503\,q_{15} - 6851300\,q_{25}}{14558301},\nonumber\\
& &    \frac{-4\,\left( 110\,q_{05} - 607\,q_{15} + 1710\,q_{25} \right) }{4852767},
   \frac{-10260\,q_{05} - 3325615\,q_{15} + 25280767\,q_{25}}{131024709},\nonumber\\
& &    \frac{-16212\,q_{05} - 8527877\,q_{15} + 62539841\,q_{25}}{262049418}\} ,\nonumber\\
& &   \{ 0,0,0,\frac{-2\,\left( 48\,q_{06} - 88497\,q_{16} + 1030121\,q_{26} - 4852767\,q_{36} \right) }{1617589},\nonumber\\
& & 
   \frac{-5586\,q_{06} + 795503\,q_{16} - 6851300\,q_{26}}{14558301},\nonumber\\
& &    \frac{-4\,\left( 110\,q_{06} - 607\,q_{16} + 1710\,q_{26} \right) }{4852767},
   \frac{-10260\,q_{06} - 3325615\,q_{16} + 25280767\,q_{26}}{131024709},\nonumber\\
& &    \frac{-16212\,q_{06} - 8527877\,q_{16} + 62539841\,q_{26}}{262049418}\} ,\nonumber\\
& &   \{ 0,0,0,\frac{-2\,\left( 48\,q_{07} - 88497\,q_{17} + 1030121\,q_{27} - 4852767\,q_{37} \right) }{1617589},\nonumber\\
& & 
   \frac{-5586\,q_{07} + 795503\,q_{17} - 6851300\,q_{27}}{14558301},\nonumber\\
& &    \frac{-4\,\left( 110\,q_{07} - 607\,q_{17} + 1710\,q_{27} \right) }{4852767},
   \frac{-10260\,q_{07} - 3325615\,q_{17} + 25280767\,q_{27}}{131024709},\nonumber\\
& & 
   \frac{-16212\,q_{07} - 8527877\,q_{17} + 62539841\,q_{27}}{262049418}\} \}
\end{eqnarray}
The elements $q_{ij}$ of a matrix $q$, are as explained in the appendix. Using MATHEMATICA,
one can actually evaluate $T$, but the expression is extremely long and complicated
and  will not be given in this paper.

The monodromy around $z=\infty$ can be evaluated as follows(similar to the way
given in \cite{GL}). For $|z|>>1$, one can write:
\begin{equation}
\label{eq:infty1}
\tilde{\Pi}_a(z)=\sum_{j=1}^5A_{aj}(z)u_j(z),\ a=0,...,7,
\end{equation}
where $u_j(z)=e^{-{j\over6}}$. Now, as $z\rightarrow e^{2i\pi}z$, with obvious meanings
to the notation:
\begin{equation}
\label{eq:infty2}
T_u(\infty)=\left(\begin{array}{ccccc}
e^{-i{\pi\over3}}&0&0&0&0\\
0&e^{-{2i\pi\over3}}&0&0&0\\
0&0&e^{-i\pi}&0&0\\
0&0&0&e^{-{4i\pi\over3}}&0\\
0&0&0&0&e^{-{5i\pi\over3}}\\
\end{array}\right).
\end{equation}
Now, using
\begin{equation}
\label{eq:infty3}
\tilde{\Pi}(z\rightarrow e^{2i\pi}z)|_{z\rightarrow\infty}
={\cal A}(z\rightarrow e^{2i\pi}z)T_u(\infty)u(z)|_{z\rightarrow\infty}
\equiv T(\infty){\cal A}(z)u(z)
|_{z\rightarrow\infty}.
\end{equation}
So, equation (\ref{eq:infty3}) is the defining equation for the monodromy matrix 
around $z\rightarrow\infty$. Note, however, that from the point of view of computations,
given that the matrix $A$ is not a square matrix, (\ref{eq:infty3}) involves solving
40 equations in 64 variables. The matrix ${\cal A}$ is given in appendix B. Unfortunately,
MATHEMATICA is not able to handle such a computation, this time. However, it is clear
that it is in prinicple, a doable computation.

\section{Superpotential Calculation}

As done in \cite{Witten}, consider $F$-theory on an
elliptically fibered Calabi-Yau 4-fold $X_4$ with holomorphic map 
$\pi:X_4\rightarrow B_3$ and a 6-divisor $D_3$
as a section such that $\pi(D_3)=C_2\subset B_3$. Then for vanishing size of the the elliptic
fiber, it was argued in \cite{Witten} that 5-branes 
wrapped around $D_3$ in $M$-theory on the same $X_4$ obeying
the unit-arithmetic genus condition, $\chi(D_3,{\cal O}_{D_3})=1,$
correspond to 3-branes wrapped around $C_2$ in type $IIB$, or equivalently $F$-theory 3-branes wrapped around
$C_2\subset B_3$. It was shown in \cite{Witten} that 
only 3-branes contribute to the superpotential in $F$-theory. 
As there are no 3-branes in the $F$-theory dual \cite{AM1}, this implies that 
no superpotential is generated on the $F$-theory side.
As F-theory 3-branes correspond to Heterotic instantons, one again expects no
superpotential to
be generated in  Heterotic theory on the self-mirror $CY_3(11,11)$ based on the 
${\cal N}=2$ type IIA/Heterotic
dual of Ferrara et al 
where the same self-mirror Calabi-Yau figured on the type IIA side and the self-mirror nature was argued to show
that there are no world-sheet or space-time instanton corrections to the 
classical moduli space.

If the abovementioned triality is correct, then one 
must be able to show that there is no superpotential generated
on type $IIA$ side on the freely-acting antiholomorphic involution of $CY_3(3,243)$. 

On the mirror type $IIB$ side, the  $W$ is generated from domain-wall (
$\equiv D5$-branes wrapped around supersymmetric 3-cycles $\hookrightarrow$ $CY_3$'s)
tention.
$W_{IIB}=\int_{C:\partial C=\sum_i D_i}\Omega_3,$
$D_i$'s are 2-cycles corresponding to the positions of $D5$-branes or $O5$-planes, i.e., objects carrying
$D5$ brane charge. 
From the world-sheet point of view, the $D5$ branes correspond to disc amplitudes and
$O5$-planes correspond to ${\bf RP}^2$ amplitudes.
 As there are no branes in our theory, we need to consider only ${\bf RP}^2$
amplitudes. 
Now, type $IIA$ on a freely acting involution of a Calabi-Yau with no branes or fluxes can still
generate a superpotential because it is possible that free involution on type $IIA$ side corresponds to
orientifold planes in the mirror type $IIB$ side, which can generate a superpotential. 
The same can also be studied using localization techniques in 
unoriented closed string enumerative geometry \cite{DFM}.
Consider an orientation-reversing diffeomorphism $\sigma:\Sigma\rightarrow\Sigma$, 
an antiholomorhpic involution on the Calabi-Yau $X$ $I:X\rightarrow X$ and an
equivariant map $f:\Sigma\rightarrow X$ [satisfying $f\circ\sigma=I\circ f$]),
then the quotient spaces in $\tilde{f}:
{\Sigma\over<\sigma>}\rightarrow{X\over I}$ possesses a dianalytic structure. 

In the
unoriented theory, one then has to sum over holomorphic and antiholomorphic instantons.
For connected ${\Sigma\over<\sigma>}$, the two contributions are the same; hence 
sufficient to consider only equivariant holomorphic maps.
One constructs one-dimensional torus action, $T$, on $X$ compatible with $I$ with
isolated fixed points. The action $T$ induces an action on the moduli space of 
equivariant holomorphic maps, and one then evaluates
 the localized contributions from the fixed
points, using an equivariant version of the the Atiyah-Bott formula, 
much on the lines of Kontsevich's work.
For a Calabi-Yau 3-fold, the virtual cycle ``$[{\bar{M}}_{g,0}(X,\beta)]^{virt}$"
is zero-dimensional, and 
one has to evaluate $\int_{\Xi_s^{virt}}{1\over e_T(N_{\Xi_s}^{virt})}$,
where $\Xi_s\equiv$ is the fixed locus in the moduli space of symmetric holomorphic maps,
and one sees that one gets a match with similar calculations based on large $N$ dualities
and mirror symmetry 

For $\int d^2\theta W_{LG}$ to be invariant under $\Omega.\omega$,
given that the measure is reflected under $\Omega$, 
$\omega:W_{LG}\rightarrow -W_{LG}.$

\underline{away from the orbifold singularities}:
Promoting the action of $\omega$ to the one on the chiral superfields: 
\begin{equation}
\label{eq:ah1}
\omega:({\cal X}_1,{\cal X}_2,{\cal X}_4,{\cal X}_5,{\cal X}_6,{\cal X}_0)
\rightarrow({\bar {\cal X}}_2,-{\bar {\cal X}}_1,{\bar {\cal X}}_4,{\bar {\cal X}}_5,
{\bar {\cal X}}_6,{\bar {\cal X}}_0),
\end{equation}
and using $Re(Y_i)=|{\cal X}_i|^2$, one gets the following action of $\omega$ on the twisted chiral superfields $Y_i$'s:
\begin{equation}
\label{eq:ahmirror}
\omega:Y_1\rightarrow Y_2+i\pi,\ Y_2\rightarrow Y_1+i\pi;\
Y_{0,4,5,6}\rightarrow Y_{0,4,5,6}+i\pi.
\end{equation}
The action of $\omega$ on $Y_{4,5,6,0}$ implies that $\omega$ acts without fixed points even on the twisted
chiral superfields, further implying that there are no orientifold fixed planes, and {\it hence no
superpotential is generated on the type $IIA$ side away from the orbifold
singularities}.

\underline{after singularity resolution}:
Writing $W=\prod_{i=0}^7a_{e_{0,...,7}}{\cal X}_i^{e_i}$ with the requirements that
$\vec{l}^{(a)}\cdot\vec{e}=0$ for $a=1,2,3$ and $e_i\leq1$\cite{VafaLG}\footnote{We thank
A.Klemm for bringing \cite{VafaLG} to our attention.}, one sees that
${\cal X}_0\prod_{i=1}^7{\cal X}_i$ is an allowed term in the superpotential.
A valid antiholomorphic involution this time can be:
\begin{equation}
\label{eq:ah2}
\omega:({\cal X}_0, {\cal X}_1, {\cal X}_2, {\cal X}_3, {\cal X}_4, {\cal X}_5, {\cal X}_6, 
{\cal X}_7)\rightarrow 
({\bar{\cal X}_0}, {\bar{\cal X}_2}, -{\bar{\cal X}_1}, -{\bar{\cal X}_3}, {\bar{\cal X}_4},
{\bar{\cal X}_5}, {\bar{\cal X}_6}, {\bar{\cal X}_7}).
\end{equation}
This on the mirror LG side again implies that one will have free actions w.r.t. 
$Y_{0,3,4,5,6,7}$ 
implying there can be no orientifold planes and {\it no superpotential (is likely to be) 
generated even after singularity resolution}.

\section{Action of the antiholomorphic involution on $M$ theory compactified on 
$CY_3(3,243)\times S^1$}

The $M$-theory uplift of the type $IIA$ side of the ${\cal N}=1$ Heterotic/type $IIA$
dual pair of \cite{VW}, as obtained in \cite{AM1}, involves the `barely $G_2$-manifold'
${CY_3(3,243)\times S^1\over{\bf Z}_2}$. In this section we consider the $D=11$
supergravity limit of $M$-theory and construct the ${\cal N}=1, D=4$ supergravity
action, and evaluate the K\"{a}hler potential for the same.

The effect of the ${\bf Z}_2$ involution that reflects the $S^1$, $H^{1,1}(CY_3)$
and takes $H^{p,q}(CY_3)$ to $H^{q,p}(CY_3)$ for $p+q=3$, where the $CY_3$ is the one 
that figures in ${CY_3\times S^1\over{\bf Z}_2}$,
at the level of $D=11$ supergravity can be obtained by first compactifying the same
on an $S^1$, then on $CY_3$ (following \cite{BCF}) 
and eventually modding out the action by the abovementioned
${\bf Z}_2$ action.

The $D=11$ supergravity action of Cremmer et al is:
\begin{equation}
\label{eq:L11}
{\cal L}_{11}=-{1\over2}e_{11}R_{11}-{1\over48}(G_{MNPQ})^2+{\sqrt{2}\over(12)^4}
\epsilon^{M_0...M_{10}}G_{M_0...M_3}G_{M_4...M_7}C^{11}_{M_8M_9M_{10}},
\end{equation}
which after dimensional reduction on an $S^1$, gives:
\begin{eqnarray}
\label{eq:L10}
& & {\cal L}_{10}=-{1\over2}e_{10}R_{10}-{1\over8}e_{10}\phi^{{9\over4}}F_{mn}^2-{9\over16}
e_{10}(\partial_m ln\phi)^2-{1\over48}e_{10}\phi^{{3\over4}}(F_{mnpq}+6F_{[mn}B_{pq]})^2
\nonumber\\
& & -{1\over12}e_{10}\phi^{-{3\over2}}H_{mnp}^2+{\sqrt{2}\over(48)^2}\epsilon^{m_0...m_9}
(F_{m_0...m_3}+6F_{m_0m_1}B_{m_2m_3})F_{m_4...m_7}B_{m_8m_9},
\end{eqnarray}
where 
\begin{eqnarray}
\label{eq:dieufs}
& & G_{MNPQ}=\partial_{[M}C_{NPQ]}\nonumber\\
& & F_{mnpq}=4\partial_{[m}C_{npq]};\ B_{mn}=C_{mn10};\ H_{mnp}=3\partial_{[m}B_{np]};\
F_{mn}=2\partial_{[m}A_{n]};\nonumber\\ 
& & C_{mnp}=A_{mnp}+3A_{[m}B_{np]},
\end{eqnarray}
and
\begin{eqnarray}
\label{eq:dieufs11to10}
& & e_{11}\ ^A_M=\left(\begin{array}{cc}
e_{10}\ ^a_m & \phi A_M \\
0 & \phi\\ 
\end{array}\right)\nonumber\\
& & A,M=0,...,10;\ a,m=0,...,9.
\end{eqnarray}
After compactifying on a $CY_3$, one gets the following Lagrangian density:
\begin{equation}
\label{eq:S1.CY31}
{\cal L}_4={\cal L}_4^{{\rm grav}+H^0}+{\cal L}_4^{H^2}+{\cal L}_4^{H^3},
\end{equation}
where after a suitable Weyl scaling:
\begin{equation}
\label{eq:grav}
{\cal L}_4^{{\rm grav}}=e\biggl[-{R\over2}-{9\over16}
\biggl({\partial_\mu\phi\over\phi}\biggr)^2+G_{\alpha\bar{\beta}}\partial_\mu Z_\alpha
\partial^\mu{\bar Z}^{\bar{\beta}}-\partial_\mu M^A\partial^\mu M^B\biggl({1\over2}G_{AB}
+{9\over4}{\int e^A\wedge J\wedge J\int e^B\wedge J\wedge J\over(\int J\wedge J\wedge J)^2}
\biggr)\biggr],
\end{equation}
where $A,B=1,...,h^{1,1}(CY_3)$, $\alpha,\beta=1,...,h^{2,1}(CY_3)$, 
\begin{eqnarray}
\label{eq:dieufs2}
& & G_{AB}={i\int e^A\wedge *e^B\over2{\cal V}},\nonumber\\
& &  {\cal V}={1\over3!}\int J\wedge J\wedge J,\nonumber\\
& & G_{\alpha\bar{\beta}}=-{i\int b_\alpha\wedge{\bar b}_\beta\over{\cal V}},
\end{eqnarray}
where
\begin{equation}
\label{eq:1,1dieuf}
\bar{b}_{\alpha\ ij}={i\over||\Omega||^2}\Omega_i^{\bar{l}\bar{k}}\bar{\Phi}_{\alpha\ \bar{l}
\bar{k}j},
\end{equation}
$\Phi$ being a (2,1) form, and the $h^{1,1}$ moduli $M^A$ entering in
the variation of the metric with mixed indices and the $h^{2,1}$ moduli $Z_\alpha$
entering in the variation of the metric with same indices. 
Writing (\ref{eq:grav}) in terms of $v^A$ defined via: $M^A=\sqrt{2}v^A\phi^{-{3\over4}}$,
one gets:
\begin{displaymath}
\label{eq:gravredieuf}
{\cal L}_4^{\rm grav}=e\biggl[-{R\over2}-{G_{AB}\over2}\partial_\mu v^A\partial^\mu v^B
-{1\over4}{\partial_\mu({\cal V}\phi^{-3})^2\over({\cal V}\phi^{-3})^2}
+G_{\alpha{\bar\beta}}\partial_\mu Z^\alpha\partial^\mu{\bar Z}^\beta\biggr],
\nonumber
\end{displaymath}
Under the freely-acting antiholomorphic involution, the $h^{1,1}$-moduli $M^A$/$v^A$ 
get projected out, $G_{AB}$ is even, and $A_\mu^A$ gets projected out.
Thus, one gets:
\begin{equation}
\label{eq:LgravZ2}
{\cal L}_4^{\rm grav}/{\bf Z}_2=e\biggl[
-{R\over2} -{1\over4}{(\partial_\mu({\cal V}\phi^{-3}))^2\over({\cal V}\phi^{-3})^2}
+G_{\alpha{\bar\beta}}\partial_\mu Z^\alpha\partial^\mu{\bar Z}^\beta\biggr].
\end{equation}
Additionally, after a Weyl scaling:
\begin{equation}
\label{eq:H0F}
{\cal L}_4^{H^0(A_\mu)}=-{\sqrt{2}^3\over8}e{\cal V}(v)F_{\mu\nu}^2
\end{equation}
\begin{equation}
\label{eq:H0A3}
{\cal L}_4^{H^0(A_{\mu\nu\rho})}=-{1\over48}e{\cal V}^3(v)\phi^{{3\over4}}(F_{\mu\nu\rho}+
6F_{[\mu\nu}B_{\rho\sigma]})^2,
\end{equation}
Thus, under the ${\bf Z}_2$-involution, ${\cal L}_4^{H^0(A_\mu)}$ and
${\cal L}_4^{H^0(A_{\mu\nu\rho})}$ get projected out.
The term ${\cal L}_4^{H^0(H_{\mu\nu\rho},\phi)}$ will be considered in conjunction with
${\cal L}_4^{H^3}$ later. 

Now, the $H^2$-sector Lagragian density is given by:
\begin{eqnarray}
\label{eq:H2}
& & {\cal L}_4^{H^2}=
e\biggl[-{1\over2}G_{AB}\partial_\mu W^A\partial^\mu W^B+{1\over\sqrt{2}}F_{\mu\nu}^2
\biggl(-{1\over12}\int J\wedge J\wedge J\nonumber\\
& & +{1\over2}\biggl[\int e^A\wedge e^B\wedge J-{3\over2}
{\int e^A\wedge J\wedge J\int e^B\wedge J\wedge J\over\int J\wedge J\wedge J}\biggr)
a^A a^B\biggr]\nonumber\\
& & +{1\over\sqrt{2}}\biggl(\int e^A\wedge e^B\wedge J-{3\over2}{\int e^A\wedge J\wedge J\int 
e^B\wedge J\wedge J\over\int J\wedge J\wedge J}f_{\mu\nu}^Aa^Bf^{\mu\nu}
\nonumber\\
& & +{\sqrt{2}\over4}
\biggl(\int e^A\wedge e^B\wedge J-{3\over2}
{\int e^A\wedge J\wedge J\int e^B\wedge J\wedge J\over\int J\wedge J\wedge J}\biggr)
f_{\mu\nu}^Af^{\mu\nu\ B}\biggr]\nonumber\\
& & +i\biggl[{\sqrt{2}\over12}a^A a^B a^C\tilde{F}_{\mu\nu}
F^{\mu\nu}\int e^A\wedge e^B\wedge e^C 
+{\sqrt{2}\over8}a^A a^B \tilde{F}_{\mu\nu}f^{C\ \mu\nu}\int e^A\wedge e^B\wedge e^C 
\nonumber\\
& & +{\sqrt{2}\over8} a^A a^B F_{\mu\nu}\tilde{F}^C_{\mu\nu}\int e^A\wedge e^B\wedge e^C 
+{\sqrt{2}\over4}a^A f_{\mu\nu}^B\tilde{f}^{C\ \mu\nu}\int e^A\wedge e^B\wedge e^C\biggr],
\end{eqnarray}
where $W^A\equiv a^A+iv^A$ and, e.g., $e\tilde{F}^{\mu\nu}\equiv {1\over2}\epsilon^{\mu\nu\rho\lambda}
F_{\rho\lambda}$. Thus, under the ${\bf Z}_2$-involution, only 
$-{1\over2}eG_{AB}\partial_\mu a^A\partial^\mu a^B$ in ${\cal L}_4^{H^2}$, survives, i.e.,
\begin{equation}
\label{eq:L2Z2}
{\cal L}_4^{H^2}/{\bf Z}_2=-{1\over2}eG_{AB}\partial_\mu a^A\partial^\mu a^B.
\end{equation}

Now, using the notations of \cite{BCF}, 
\begin{eqnarray}
\label{eq:H3+H0}
& & {\cal L}^{H^3}_4+{\cal L}_4^{H^0}=-e{1\over4\tilde{\phi}^2}|\partial_\mu\tilde{\phi}+i
\partial_\mu D-{1\over4}\partial_\mu[(\Psi+{\bar\Psi})R^{-1}(\Psi+{\bar\Psi})]+
(\Psi+{\bar\Psi})R^{-1}\partial_\mu\Psi\nonumber\\
& & -{1\over4}(\psi+{\bar\psi})R^{-1}\partial_\mu{\cal N} R^{-1}(\Psi+{\bar\Psi})|^2
\nonumber\\
& & =-{|\partial_\mu S+(\Psi+{\bar\Psi})R^{-1}\partial_\mu\Psi
-{1\over4}(\Psi+{\bar\Psi})R^{-1}\partial_\mu{\cal N}R^{-1}(\Psi+{\bar\Psi})|^2\over
[S+{\bar S}+{1\over2}(\Psi+{\bar\Psi})R^{-1}(\Psi+{\bar\Psi})]^2}\nonumber\\
& & +{[\partial_\mu\Psi-{1\over2}(\psi+{\bar\Psi})R^{-1}\partial_\mu{\cal N}]R^{-1}
[\partial_\mu{\bar\Psi}-{1\over2}\partial_\mu{\bar{\cal N}}R^{-1}(\Psi+{\bar\Psi})]\over
[(S+{\bar S})+{1\over2}(\Psi+{\bar\Psi})R^{-1}(\Psi+{\bar\Psi})]},
\end{eqnarray}
where $S\equiv\tilde{\phi}+iD-{1\over4}(\Psi+{\bar\Psi})R^{-1}(\psi+{\bar\Psi})$,
$D$ being a Lagrange multiplier, $\tilde{\phi}\equiv\sqrt{2}{\cal V}(v)\phi^{-3}$,
$\Psi_{I(\equiv0,1,...,h^{2,1})}$ 
appearing in the expansion of the real 3-form
$A_{mnp}$ in a canonical basis of $H^3$, and
\begin{eqnarray}
\label{eq:RcalNdieuf}
& & R_{IJ}\equiv Re[{\cal N}_{IJ}],\ {\cal N}_{IJ}\equiv {1\over4}{\bar F}_{IJ}
-{(NZ)_I(NZ)_J\over(ZNZ)},\nonumber\\
& & (R^{-1})^{IJ}=2\biggl(N^{-1}({\bf 1}-{\bar K}{\bar Z} - KZ)\biggr)^{IJ},
\end{eqnarray}
where $Z^I$ and $i F_I$ are the period integrals, $N_{IJ}={1\over4}(F_{IJ}+{\bar F}_{IJ})$,
 and $K_I\equiv{\int\Omega_I\wedge{\bar\Omega}\over
\int\Omega\wedge{\bar\Omega}}$, with $\Omega_I\equiv{\partial\Omega\over\partial Z^I}$.
Here, it is assumed that the holomorphic 3-form $\Omega$ is expaneded in a canonical
cohomology basis $(\alpha_I,\beta^I)$ satisfying
\begin{equation}
\label{eq:canbasdieufs}
\int_{A^J}\alpha_I=\int\alpha_I\wedge
\beta^J=-\int_{B_I}\beta^J=-\int\beta^J\wedge\alpha_I=\delta^J_I,
\end{equation}
with $(A^I,B_I)$ being
the dual homology basis. The period integrals are then defined to be:
$Z^I=\int_{A^I}\Omega$ and $iF_I=\int_{B_I}\Omega$. Hence,
\begin{equation}
\label{eq:omegaexpdieuf}
\Omega=Z^I\alpha_I + i F_I\beta^I.
\end{equation}

For the ${\cal N}=1$ case, we work in the large volume limit of the Calabi-Yau. In this
limit, one gets:
\begin{equation}
\label{eq:actionN=1}
{\cal L}^{{\rm grav}+H^0+H^2+H^3}/{\bf Z}_2=
e\biggl[-{R\over2}-G_{AB}\partial_\mu a^A\partial^\mu a^B+G_{\alpha\bar{\beta}}\partial_\mu 
Z^\alpha\partial^\mu Z^{\bar{\beta}}-{1\over2}{(\partial_\mu\tilde{\phi})^2\over\tilde{\phi}^2}
\biggr].
\end{equation}
Hence, one gets for the ${\cal N}=1$ K\"{a}hler potential $K_{{\cal N}=1}$:
\begin{equation}
\label{eq:KahN=1}
K_{{\cal N}=1}=K[a^A,Z^\alpha]+{1\over2}ln[\tilde{\phi}].
\end{equation}
At the ${\cal N}=2$ level, if there were decoupling between the fields of the $H^3$-sector
from the other fields of the other sectors, the K\"{a}hler potential would be consisting of 
$ln\biggl[S+{\bar S}+{1\over2} (\Psi+{\bar\Psi})R^{-1}(\Psi+{\bar\psi})\biggr],$\footnote{This
however assumes that ${\partial{\bar{\cal N}}_{IJ}\over\partial z^K}=0\leftrightarrow
{1\over4}F_{IJK}-{({1\over4}F_{IRK}{\bar z}^R(N{\bar Z})_J+{1\over4}(N{\bar Z}_I
F_{JLK}{\bar Z}^L)\over({\bar Z}N{\bar Z})}+{(N{\bar Z})_I(N{\bar Z})_J\over
(2{\bar Z}N{\bar Z})^2}{\bar Z}^P{\bar Z}^QF_{PQK}=0$} from the
$H^3$-sector. From the definitin of $S$ above, one sees that:
\begin{equation} 
\label{eq:S+cc}
S+{\bar S}+{1\over2} (\Psi+{\bar\Psi})R^{-1}(\Psi+{\bar\Psi})=2\tilde{\phi}.
\end{equation}
This partially explains the appearance of $ln[\tilde{\phi}]$ in $K_{{\cal N}=1}$.

\section{Action of Antiholomorphic Involution on the Periods - a Conjecture}

Given the action of an antiholomorphic involution on the cohomology, it is in general
quite non-trivial to figure out the action of the involution on the period integrals
using the canonical (co)homology basis of (\ref{eq:canbasdieufs}). We now discuss a 
reasonable guess for the same. From (\ref{eq:S+cc}), one sees that the RHS is reflected
under the antiholomorphic involution discussed towards the beginning of this section.
We now conjecture that on the LHS, this would imply that 
\begin{eqnarray}
\label{eq:conj1}
& & S\rightarrow-S,\
(\Psi+{\bar\Psi})^2\rightarrow(\Psi+{\bar\Psi})^2,\nonumber\\
& & R_{IJ}\rightarrow-R_{IJ}.
\end{eqnarray}
We further conjecture that $R_{IJ}\rightarrow-R_{IJ}$ is realized by
\begin{eqnarray}
\label{eq:conj2}
& & Z^I\rightarrow-{\bar Z}^I,\ F_I\rightarrow{\bar F}_I\nonumber\\
& & \alpha_I\rightarrow-\alpha_I,\ \beta^I\rightarrow\beta^I.
\end{eqnarray}
One should note that given that the antiholomorphic involution is orientation reversing,
the intersection form $\int\alpha_I\wedge\beta^J$ is also reflected.\footnote{We are
grateful to D.Joyce for pointing this out to us.} This fact, e.g., can be explicitly
seen in the real basis of $H^3(T^6,{\bf Z})$\cite{KST}:
\begin{eqnarray}
\label{eq:basis}
& & 
\alpha_0=dx^1\wedge dx^2\wedge dx^3;\ \alpha_{ij}={1\over2}\epsilon_{ilm}dx^l\wedge dx^m
\wedge dy^j(1\leq i,j\leq3),\nonumber\\
& & \beta_0=dy^1\wedge dy^2\wedge dy^3;\ \beta^{ij}=-{1\over2}\epsilon_{jlm}dy^l\wedge dy^m
\wedge dx^i(1\leq i,j\leq3).
\end{eqnarray}
where $x^i,y^i, i=1,2,3$ are the six real coordinates on the $T^6$, using which one 
can construct the holomorphic one-form: $dz^i=dx^i+\tau^{ij}dy^j$, $\tau^{ij}$ being
the period matrix for the torus. Hence, under the antiholomorphic involution:
$(x^i,y^i,\tau^{ij})\rightarrow(x^i,-y^i,-{\bar\tau}^{ij})$, the intersection matrix
changes sign because $\alpha_{ij}$ and $\beta_0$ change signs.\footnote{We thank
M.Schulz for pointing this out to us.} The holomorphic 3-form $\Omega$ is then given
by:
\begin{equation}
\label{eq:3form}
\Omega=dz^1\wedge dz^2\wedge dz^3=\alpha_0-\beta^{ij}(cof\tau)_{ij}+\alpha_{ij}\tau^{ij}
+\beta_0(det\tau),
\end{equation}
which on comparison with:
\begin{equation}
\label{eq:3-form}
\Omega=Z^I\alpha_I+iF_I\beta^I,
\end{equation}
for inhomogenous coordinates ($I=0,1,...,h^{2,1}$ and $h^{2,1}(T^6)=9$),
implies that 
\begin{eqnarray}
\label{eq:analogs}
& & (\alpha_0,\beta^{ij})\sim \beta^I,\ (\beta_0,\alpha_{ij})\sim\alpha_I;\nonumber\\
& & (1,-(cof\tau)_{ij})\sim F_I,\ (det\tau,\tau^{ij})\sim Z^I.
\end{eqnarray}
One thus sees that the conjecture (\ref{eq:conj2}) is satisfied.
Further, for a real canonical cohomology basis $(\alpha_I,\beta^I)$, ${\bar\Omega}$ will be
given by ${\bar\Omega}={\bar Z}^I\alpha_I - i{\bar F}_I\beta^I$. One sees that
(\ref{eq:conj2}) is consistent with complex conjugation of the holomorphic 3-form
$\Omega$.

Further, as a simple toy example, one can consider the mirror to the quintic,
for which $h^{2,1}=1$. If $(p,q)$ forms are denoted by $X^{p,q}$, then one choice of
real canonical cohomology basis satisfying (\ref{eq:canbasdieufs})
is 
\begin{eqnarray}
\label{eq:quintic}
& & a_0={i\over\sqrt{2}}(\Omega - {\bar\Omega}),\ a_1={i\over\sqrt{2}}(X^{2,1} - X^{1,2});
\nonumber\\
& & 
b^0={1\over\sqrt{2}}(\Omega + {\bar\Omega}),\ b^1={1\over\sqrt{2}}(X^{2,1}+X^{1,2}).
\end{eqnarray}
One sees that that indeed, $(a_0,a_1)$ under the aformentioned antiholomorphic involution
goes to $-(a_0,a_1)$ and $(b^0,b^1)$ goes to $(b^0,b^1)$
\section{Conclusion and Discussions}

We obtained the Meijer basis of solution to the Picard-Fuchs equation for the Landau-Ginsburg
model corresponding to $CY_3(3,243)$ {\it after} the resolution of the orbifold singularities
of the degree-24 Fermat hypersurface in ${\bf WCP}^4[1,1,2,8,12]$, 
in the large {\it and} small complex structure
limits, getting the $ln$-terms without resorting to parametric differentiations of
infinite series. 
We also discussed in detail the evaluation of the monodromy matrix in the large complex
structure limit.
We also considered the action of an antiholomorphic involution on 
$D=11$ supergravity compactified on $CY_3(3,243)\times S^1$, and evaluated the form of the
${\cal N}=1$ K\"{a}hler potential. In the process, we also gave a conjecture on the
action of the involution on the periods of $CY_3(3,243)$, given its action on the
cohomology of the same. We verified the conjecture for $T^6$ for the periods and
cohomology basis, and for the mirror  to the quintic for the cohomology basis. Finally, we 
showed that no superpotential is generated in type IIA and hence $M$-theory
sides using mirror symmetry, {\it after} the resolution of the orbifold singularities
associated with the Fermat hypersurface whose blow up gives $CY_3(3,243)$.

For future work, it will be nice to explicitly work out the K\"{a}hler potential, or 
equivalently the periods for the ``(12,12)" $E_8\times E_8$ Heterotic model, analogous
to the ``(14,10)" model of \cite{LSTY,CCLM}
and perhaps using the conjecture that both
models lie in the same moduli space\footnote{We are grateful to T.Mohaupt for pointing
this out to us and for bringing \cite{LSTY} to our attention}
 \cite{AFIQ}, so as to be able to compare with
the K\"{a}hler potential of type IIA on $CY_3(3,243)$ or equivalently $D=11$
supergravity on ${CY_3(3,243)\times S^1\over{\bf Z}_2}$. Further, it will be 
interesting to see whether or  not the conjectured ${\cal N}=1$ triality holds up
when one turns on fluxes\cite{fluxes}
 - issues of  the right choice of non-K\"{a}hler manifolds,
flux-induced superpotentials, etc will become relevant for  investigation.

\section*{Acknowledgements}

We are extremely grateful to A.Klemm and D.Joyce for very useful discussions and
clarifications, T.Mohaupt and M.Schulz for  useful communications,  
and to D.L\"{u}st for going through a preliminary version of the manuscript.
The research work is supported by the Alexander von Humboldt foundation.

\appendix
\setcounter{equation}{0}
\section{Monodromy around $z=0$}
\seceqaa
In 
this appendix, we give the details relevant to the evaluation of the matrix $M$
that is essential for the evaluation of the monodromy matrix $T$. MATHEMATICA has been
heavily used in this appendix. We first give the exponentiation of the relevant matrix $A(0)$
followed by the value of $z^{A(0)}$, again necessary for evaluation of $M$ and hence $T$.

The matrix $A(0)$ is given by:
\begin{equation}
\label{eq:A(0)}
A(0)=\left(\begin{array}{cccccccc}\\
0&1&0&0&0&0&0&0\\
0&0&1&0&0&0&0&0\\
0&0&0&1&0&0&0&0\\
0&0&0&0&1&0&0&0\\
0&0&0&0&0&1&0&0\\
0&0&0&0&0&0&1&0\\
0&0&0&0&0&0&0&1\\
0&0&0&0&0&{1\over9}&-{13\over18}&{3\over2}\\
\end{array}\right)
\end{equation}

Using MATHEMATICA, one then can evaluate the ``matrix exponent" involving $A(0)$.
The expression given below, has been written in the MATHEMATICA notebook format
(by cutting and pasting the ``TerXForm" of the output).

$e^{2\pi i A(0)}=\{ \{ 1,2\,i  \,\pi ,-2\,{\pi }^2,\frac{-4\,i  }{3}\,{\pi }^3,\frac{2\,{\pi }^4}{3},
   -\left( \frac{31001}{32} \right)  + \frac{729}{32\,e^{\frac{2\,i  }{3}\,\pi }} + 1458\,e^{\frac{2\,i  }{3}\,\pi } - \frac{5971\,i  }{8}\,\pi  + 
    \frac{865\,{\pi }^2}{4} + \frac{115\,i  }{3}\,{\pi }^3 - \frac{13\,{\pi }^4}{3},
   \frac{93213}{32} - \frac{3645}{32\,e^{\frac{2\,i  }{3}\,\pi }} - 5103\,e^{\frac{2\,i  }{3}\,\pi } + \frac{19215\,i  }{8}\,\pi  - 
    \frac{2637\,{\pi }^2}{4} - 105\,i  \,{\pi }^3 + 9\,{\pi }^4,
   -\left( \frac{35307}{16} \right)  + \frac{2187}{16\,e^{\frac{2\,i  }{3}\,\pi }} + 4374\,e^{\frac{2\,i  }{3}\,\pi } - \frac{7785\,i  }{4}\,\pi  + 
    \frac{1035\,{\pi }^2}{2} + 78\,i  \,{\pi }^3 - 6\,{\pi }^4\} ,\\
  \{ 0,1,2\,i  \,\pi ,-2\,{\pi }^2,\frac{-4\,i  }{3}\,{\pi }^3,
   128 + \frac{243}{16\,e^{\frac{2\,i  }{3}\,\pi }} + 486\,e^{\frac{2\,i  }{3}\,\pi } - 
    \frac{i  }{18}\,\left( \frac{-53739\,i  }{8} + \frac{7785\,\pi }{2} + 1035\,i  \,{\pi }^2 - 156\,{\pi }^3 \right) ,\\
   -576 - \frac{1215}{16\,e^{\frac{2\,i  }{3}\,\pi }} - 1701\,e^{\frac{2\,i  }{3}\,\pi } - 
    \frac{i  }{18}\,\left( \frac{172935\,i  }{8} - \frac{23733\,\pi }{2} - 2835\,i  \,{\pi }^2 + 324\,{\pi }^3 \right) ,
   576 + \frac{729}{8\,e^{\frac{2\,i  }{3}\,\pi }} + 1458\,e^{\frac{2\,i  }{3}\,\pi } - 
    \frac{i  }{18}\,\left( \frac{-70065\,i  }{4} + 9315\,\pi  + 2106\,i  \,{\pi }^2 - 216\,{\pi }^3 \right) \} ,\\
  \{ 0,0,1,2\,i  \,\pi ,-2\,{\pi }^2,64 + \frac{81}{8\,e^{\frac{2\,i  }{3}\,\pi }} + 162\,e^{\frac{2\,i  }{3}\,\pi } + 
    \frac{-\left( \frac{70065}{2} \right)  - 18630\,i  \,\pi  + 4212\,{\pi }^2}{324},
   -288 - \frac{405}{8\,e^{\frac{2\,i  }{3}\,\pi }} - 567\,e^{\frac{2\,i  }{3}\,\pi } + \frac{\frac{213597}{2} + 51030\,i  \,\pi  - 8748\,{\pi }^2}{324},
   288 + \frac{243}{4\,e^{\frac{2\,i  }{3}\,\pi }} + 486\,e^{\frac{2\,i  }{3}\,\pi } + \frac{-83835 - 37908\,i  \,\pi  + 5832\,{\pi }^2}{324}\} ,\\
  \{ 0,0,0,1,2\,i  \,\pi ,32 + \frac{27}{4\,e^{\frac{2\,i  }{3}\,\pi }} + 54\,e^{\frac{2\,i  }{3}\,\pi } + 
    \frac{i  }{5832}\,\left( 167670\,i   - 75816\,\pi  \right) ,
   -144 - \frac{135}{4\,e^{\frac{2\,i  }{3}\,\pi }} - 189\,e^{\frac{2\,i  }{3}\,\pi } + \frac{i  }{5832}\,\left( -459270\,i   + 157464\,\pi  \right) ,
   144 + \frac{81}{2\,e^{\frac{2\,i  }{3}\,\pi }} + 162\,e^{\frac{2\,i  }{3}\,\pi } + \frac{i  }{5832}\,\left( 341172\,i   - 104976\,\pi  \right) \} ,\\
  \{ 0,0,0,0,1,\frac{19}{2} + \frac{9}{2\,e^{\frac{2\,i  }{3}\,\pi }} + 18\,e^{\frac{2\,i  }{3}\,\pi },
   -\left( \frac{117}{2} \right)  - \frac{45}{2\,e^{\frac{2\,i  }{3}\,\pi }} - 63\,e^{\frac{2\,i  }{3}\,\pi },
   63 + \frac{27}{e^{\frac{2\,i  }{3}\,\pi }} + 54\,e^{\frac{2\,i  }{3}\,\pi }\} ,\\
  \{ 0,0,0,0,0,8 + \frac{3}{e^{\frac{2\,i  }{3}\,\pi }} + 6\,e^{\frac{2\,i  }{3}\,\pi },
   -36 - \frac{15}{e^{\frac{2\,i  }{3}\,\pi }} - 21\,e^{\frac{2\,i  }{3}\,\pi },
   36 + \frac{18}{e^{\frac{2\,i  }{3}\,\pi }} + 18\,e^{\frac{2\,i  }{3}\,\pi }\} ,\\
  \{ 0,0,0,0,0,4 + \frac{2}{e^{\frac{2\,i  }{3}\,\pi }} + 2\,e^{\frac{2\,i  }{3}\,\pi },
   -18 - \frac{10}{e^{\frac{2\,i  }{3}\,\pi }} - 7\,e^{\frac{2\,i  }{3}\,\pi },
   18 + \frac{12}{e^{\frac{2\,i  }{3}\,\pi }} + 6\,e^{\frac{2\,i  }{3}\,\pi }\} ,\\
  \{ 0,0,0,0,0,2 + \frac{4}{3\,e^{\frac{2\,i  }{3}\,\pi }} + \frac{2\,e^{\frac{2\,i  }{3}\,\pi }}{3},
   -9 - \frac{20}{3\,e^{\frac{2\,i  }{3}\,\pi }} - \frac{7\,e^{\frac{2\,i  }{3}\,\pi }}{3},
   9 + \frac{8}{e^{\frac{2\,i  }{3}\,\pi }} + 2\,e^{\frac{2\,i  }{3}\,\pi }\} \}$

Writing the solution vector $\tilde{\Pi}_i$ as $\tilde{\Pi}_i=\sum_{j=0}^4(ln z)^j q_{ji}$
(following the notation of \cite{GL}), one notes:
\begin{equation}
\label{eq:primedsol1}
(\Phi^\prime)_i=(\tilde{\Pi}^\prime)^t_i=\biggl(S^\prime z^{A(0)}\biggr)_{0i}=
(ln z)^j q^\prime_{ji}.
\end{equation}
From (\ref{eq:primedsol1}), one gets the following:
\begin{equation}
\label{eq:primedsol2}
(q^\prime(0))_{ji}={\delta_{ji}\over j!},\ 0\leq (i,j)\leq 4.
\end{equation}
For $5\leq i\leq 7$, consider. e.g., $i=5$. Then from the expression for $z^{A(0)}$ above,
\begin{eqnarray}
\label{eq:primedsol3}
& & \sum_{j=0}^4(q^\prime)_{j5}(ln z)^j=
(S^\prime)_{00}[f_{05}(z^{{1\over2}},z^{{1\over3}})
+\sum_{n=1}^4c_n^{(05}(ln z)^n]+
(S^\prime)_{01}[f_{15}(z^{{1\over2}},z^{{1\over3}})+\sum_{n=1}^3c_n^{(25)}(ln z)^n]\nonumber\\
& & 
(S^\prime)_{02}[f_{25}(z^{{1\over2}},z^{{1\over3}})
+\sum_{n=1}^2c_n^{(25}(ln z)^n]+
(S^\prime)_{03}[f_{35}(z^{{1\over2}},z^{{1\over3}})+c_1^{(35)}(ln z)]\nonumber\\
& & 
+(S^\prime)_{04}f_{45}(z^{{1\over2}},z^{{1\over3}})+(S^\prime)_{05}f_{55}(z^{{1\over3}})
+(S^\prime)_{06}f_{65}(z^{{1\over2}},z^{{1\over3}})+(S^\prime)_{07}f_{75}(z^{{1\over3}})
\end{eqnarray}
where the  $f_{ij}$'s and $c_n^{ij}$'s can be determined from the expression for
$z^{A(0}$ given below. From (\ref{eq:primedsol3}),
one gets:
\begin{eqnarray}
\label{eq:primedsol4}
& & (q^\prime)_{05}=\sum_{i=0}^7(S^\prime)_{0i}f_{i5}\nonumber\\
& & (q^\prime)_{15}=\sum_{i=0}^3(S^\prime)_{0i}c_1^{i5}\nonumber\\
& & (q^\prime)_{25}=\sum_{i=0}^2(S^\prime)_{0i}c_2^{i5}\nonumber\\
& & (q^\prime)_{35}=\sum_{i=0}^1(S^\prime)_{0i}c_3^{i5}\nonumber\\
& & (q^\prime)_{45}=(S^\prime)_{00}c_4^{05}.
\end{eqnarray}
From (\ref{eq:primedsol4}), one gets:
\begin{eqnarray}
\label{eq:primedsol6}
& & (q^\prime(0))_{0i}=f_{0i}(0),\nonumber\\
& & (q^\prime(0))_{ij}=c_i^{0j},\ 1\leq i\leq 4,\ 5\leq j\leq 7.
\end{eqnarray}

Again using the MATHEMATICA notebook format, the value of $z^{A(0)}$, as
evaluated by MATHEMATICA is given by:

$z^{A(0)}=e^{A(0) ln(z)}=
\{ \{ 1,\log (z),\frac{{\log (z)}^2}{2},\frac{{\log (z)}^3}{6},\frac{{\log (z)}^4}{24},\\
   \frac{3\,\left( -39193 + 46656\,z^{\frac{1}{3}} - 8192\,{\sqrt{z}} + 729\,z^{\frac{2}{3}} 
\right)  - 35826\,\log (z) - 5190\,{\log (z)}^2 - 460\,{\log (z)}^3 - 
      26\,{\log (z)}^4}{96},\\
\frac{-3\,\left( -43359 + 54432\,z^{\frac{1}{3}} - 12288\,{\sqrt{z}} + 1215\,z^{\frac{2}{3}} - 12810\,\log (z) - 1758\,{\log (z)}^2 - 
        140\,{\log (z)}^3 - 6\,{\log (z)}^4 \right) }{32},\\
\frac{3\,
      \left( -17913 + 23328\,z^{\frac{1}{3}} - 6144\,{\sqrt{z}} + 729\,z^{\frac{2}{3}} - 5190\,\log (z) - 690\,{\log (z)}^2 - 52\,{\log (z)}^3 - 2\,{\log (z)}^4
        \right) }{16}\} ,\\
\{ 0,1,\log (z),\frac{{\log (z)}^2}{2},\frac{{\log (z)}^3}{6},
   \frac{-17913 + 23328\,z^{\frac{1}{3}} - 6144\,{\sqrt{z}} + 729\,z^{\frac{2}{3}} 
- 5190\,\log (z) - 690\,{\log (z)}^2 - 52\,{\log (z)}^3}{48},\\
   \frac{-9\,\left( -2135 + 3024\,z^{\frac{1}{3}} - 1024\,{\sqrt{z}} 
+ 135\,z^{\frac{2}{3}} - 586\,\log (z) - 70\,{\log (z)}^2 - 4\,{\log (z)}^3 \right) }{16},\\
   \frac{3\,\left( 3\,\left( -865 + 1296\,z^{\frac{1}{3}} - 512\,{\sqrt{z}} + 81\,z^{\frac{2}{3}} \right)  - 690\,\log (z) - 78\,{\log (z)}^2 - 4\,{\log (z)}^3 \right)
        }{8}\},\\
\{ 0,0,1,\log (z),\frac{{\log (z)}^2}{2},\frac{-865 + 1296\,z^{\frac{1}{3}} 
- 512\,{\sqrt{z}} + 81\,z^{\frac{2}{3}} - 230\,\log (z) - 26\,{\log (z)}^2}
    {8},\\
\frac{-9\,\left( -293 + 504\,z^{\frac{1}{3}} 
- 256\,{\sqrt{z}} + 45\,z^{\frac{2}{3}} - 70\,\log (z) - 6\,{\log (z)}^2 \right) }{8},
   \frac{9\,\left( -115 + 216\,z^{\frac{1}{3}} - 128\,{\sqrt{z}} + 27\,z^{\frac{2}{3}} 
- 26\,\log (z) - 2\,{\log (z)}^2 \right) }{4}\} ,\\
  \{ 0,0,0,1,\log (z),\frac{-115 + 216\,z^{\frac{1}{3}} 
- 128\,{\sqrt{z}} + 27\,z^{\frac{2}{3}} - 26\,\log (z)}{4},
   \frac{-9\,\left( -35 + 84\,z^{\frac{1}{3}} - 64\,{\sqrt{z}} 
+ 15\,z^{\frac{2}{3}} - 6\,\log (z) \right) }{4},\\
   \frac{9\,\left( -13 + 36\,z^{\frac{1}{3}} - 32\,{\sqrt{z}} 
+ 9\,z^{\frac{2}{3}} - 2\,\log (z) \right) }{2}\} ,\
  \{ 0,0,0,0,1,\frac{-13 + 36\,z^{\frac{1}{3}} - 32\,{\sqrt{z}} + 9\,z^{\frac{2}{3}}}{2},
   \frac{-9\,\left( -3 + 14\,z^{\frac{1}{3}} - 16\,{\sqrt{z}} 
+ 5\,z^{\frac{2}{3}} \right) }{2},\\
   9\,{\left( -1 + z^{\frac{1}{6}} \right) }^3\,\left( 1 + 3\,z^{\frac{1}{6}} \right) \} ,\
  \{ 0,0,0,0,0,\left( 6 - 8\,z^{\frac{1}{6}} + 3\,z^{\frac{1}{3}} \right) \,z^{\frac{1}{3}},
   -3\,\left( 7 - 12\,z^{\frac{1}{6}} + 5\,z^{\frac{1}{3}} \right) \,z^{\frac{1}{3}},\\
18\,{\left( -1 + z^{\frac{1}{6}} \right) }^2\,z^{\frac{1}{3}}\} ,\
  \{ 0,0,0,0,0,2\,{\left( -1 + z^{\frac{1}{6}} \right) }^2
\,z^{\frac{1}{3}},
-7\,z^{\frac{1}{3}} + 18\,{\sqrt{z}} - 10\,z^{\frac{2}{3}},
   6\,\left( z^{\frac{1}{3}} - 3\,{\sqrt{z}} + 2\,z^{\frac{2}{3}} \right) \} ,\\
  \{ 0,0,0,0,0,\frac{2\,\left( z^{\frac{1}{3}} - 3\,{\sqrt{z}} 
+ 2\,z^{\frac{2}{3}} \right) }{3},
   \frac{-\left( \left( 7 - 27\,z^{\frac{1}{6}} + 20\,z^{\frac{1}{3}} 
\right) \,z^{\frac{1}{3}} \right) }{3},\
   \left( 2 - 9\,z^{\frac{1}{6}} + 8\,z^{\frac{1}{3}} \right) \,z^{\frac{1}{3}}\} \}$

The matrix $q^\prime$ introduced in (\ref{eq:primedsol1}) is given by:
\begin{equation}
\label{eq:q'dieuf}
q^\prime=\left(\begin{array}{cccccccc}
1 & 0 & 0 & 0 & 0 & -\left( \frac{391933}{32} \right)  & \frac{130077}{32} & 
-\left( \frac{53739}{16} \right)  \cr 0 & 1 & 0 & 0 & 0 & -\left( \frac{5971}
     {16} \right)  & \frac{19215}{16} & -\left( \frac{7785}{8} \right)  \cr 0 & 0 & \frac{1}{2} & 0 & 0 & -\left( \frac{865}{16} \right)  & \frac{2637}{16} & -\left
     ( \frac{1035}{8} \right)  \cr 0 & 0 & 0 & \frac{1}{6} & 0 & -\left( \frac{115}{24} \right)  & \frac{105}{8} & -\left( \frac{39}{4} \right)
      \cr 0 & 0 & 0 & 0 & \frac{1}{24} & -\left( \frac{13}{32} \right)  & \frac{9}{16} 
& -\left( \frac{3}{8} \right)  \cr 
\end{array}\right)
\end{equation} 

Further, the matrix $q$ is of the form:
\begin{equation}
\label{eq:qdieuf}
q=\left(\begin{array}{cccccccc} 
q_{00} & q_{01} & q_{02} & q_{03} & 
q_{04} & q_{05} & q_{06} & q_{07} \cr 
   q_{10} & q_{11} & q_{12} & q_{13} & q_{14} & q_{15} & q_{16} & q_{17} \cr 0 & q_{21} & q_{22} & q_{23} & q_{24} 
& q_{25} & q_{26} & q_{27} \cr 0 & 0 & 0 & 0 & q_{34} & 
q_{35} & q_{36} & q_{37} \cr 
0 & 0 & 0 & 0 & q_{44} & 0 & 0 & 0 \cr  
\end{array}\right)
\end{equation}

From the matrix equation $q^\prime=q(M^{-1})^t$, one 
sees that one has 40 equations in 64 variables, one has the freedom to (judiciously)
give arbitrary values to 24 variables, bearing in mind that from the forms of
the matrices $q^\prime$ and $q (M^{-1})^t$, the values of $(M^{-1})^t_{4i}$ are fixed.
We set: $M_{ij}=\delta_{ij}$ for $0\leq i\leq 3$ and $0\leq j\leq 7$.

Hence, one gets:

$q(M^{-1})^t=\{ \{ q_{00} + (M^{-1})^t_{30}\,q_{03} + (M^{-1})^t_{40}\,q_{04} 
+ (M^{-1})^t_{50}\,q_{05} + (M^{-1})^t_{60}\,q_{06} + 
    (M^{-1})^t_{70}\,q_{07},q_{01} + (M^{-1})^t_{31}\,q_{03} + (M^{-1})^t_{41}\,q_{04} + (M^{-1})^t_{51}\,q_{05} + 
    (M^{-1})^t_{61}\,q_{06} + (M^{-1})^t_{71}\,q_{07},
   q_{02} + (M^{-1})^t_{32}\,q_{03} + (M^{-1})^t_{42}\,q_{04} + (M^{-1})^t_{52}\,q_{05} + (M^{-1})^t_{62}\,q_{06} + 
    (M^{-1})^t_{72}\,q_{07},(M^{-1})^t_{33}\,q_{03} + (M^{-1})^t_{43}\,q_{04} + (M^{-1})^t_{53}\,q_{05} + 
    (M^{-1})^t_{63}\,q_{06} + (M^{-1})^t_{73}\,q_{07},
   (M^{-1})^t_{34}\,q_{03} + (M^{-1})^t_{44}\,q_{04} + (M^{-1})^t_{54}\,q_{05} + (M^{-1})^t_{64}\,q_{06} + 
    (M^{-1})^t_{74}\,q_{07},(M^{-1})^t_{35}\,q_{03} + (M^{-1})^t_{45}\,q_{04} + (M^{-1})^t_{55}\,q_{05} + 
    (M^{-1})^t_{65}\,q_{06} + (M^{-1})^t_{75}\,q_{07},
   (M^{-1})^t_{36}\,q_{03} + (M^{-1})^t_{46}\,q_{04} + (M^{-1})^t_{56}\,q_{05} + (M^{-1})^t_{66}\,q_{06} + 
    (M^{-1})^t_{76}\,q_{07},(M^{-1})^t_{37}\,q_{03} + (M^{-1})^t_{47}\,q_{04} + (M^{-1})^t_{57}\,q_{05} + 
    (M^{-1})^t_{67}\,q_{06} + (M^{-1})^t_{77}\,q_{07}\} ,
  \{ q_{10} + (M^{-1})^t_{30}\,q_{13} + (M^{-1})^t_{40}\,q_{14} + (M^{-1})^t_{50}\,q_{15} + (M^{-1})^t_{60}\,q_{16} + 
    (M^{-1})^t_{70}\,q_{17},q_{11} + (M^{-1})^t_{31}\,q_{13} + (M^{-1})^t_{41}\,q_{14} + (M^{-1})^t_{51}\,q_{15} + 
    (M^{-1})^t_{61}\,q_{16} + (M^{-1})^t_{71}\,q_{17},
   q_{12} + (M^{-1})^t_{32}\,q_{13} + (M^{-1})^t_{42}\,q_{14} + (M^{-1})^t_{52}\,q_{15} + (M^{-1})^t_{62}\,q_{16} + 
    (M^{-1})^t_{72}\,q_{17},(M^{-1})^t_{33}\,q_{13} + (M^{-1})^t_{43}\,q_{14} + (M^{-1})^t_{53}\,q_{15} + 
    (M^{-1})^t_{63}\,q_{16} + (M^{-1})^t_{73}\,q_{17},
   (M^{-1})^t_{34}\,q_{13} + (M^{-1})^t_{44}\,q_{14} + (M^{-1})^t_{54}\,q_{15} + (M^{-1})^t_{64}\,q_{16} + 
    (M^{-1})^t_{74}\,q_{17},(M^{-1})^t_{35}\,q_{13} + (M^{-1})^t_{45}\,q_{14} + (M^{-1})^t_{55}\,q_{15} + 
    (M^{-1})^t_{65}\,q_{16} + (M^{-1})^t_{75}\,q_{17},
   (M^{-1})^t_{36}\,q_{13} + (M^{-1})^t_{46}\,q_{14} + (M^{-1})^t_{56}\,q_{15} + (M^{-1})^t_{66}\,q_{16} + 
    (M^{-1})^t_{76}\,q_{17},(M^{-1})^t_{37}\,q_{13} + (M^{-1})^t_{47}\,q_{14} + (M^{-1})^t_{57}\,q_{15} + 
    (M^{-1})^t_{67}\,q_{16} + (M^{-1})^t_{77}\,q_{17}\} ,
  \{ (M^{-1})^t_{30}\,q_{23} + (M^{-1})^t_{40}\,q_{24} + (M^{-1})^t_{50}\,q_{25} + (M^{-1})^t_{60}\,q_{26} + 
    (M^{-1})^t_{70}\,q_{27},q_{21} + (M^{-1})^t_{31}\,q_{23} + (M^{-1})^t_{41}\,q_{24} + (M^{-1})^t_{51}\,q_{25} + 
    (M^{-1})^t_{61}\,q_{26} + (M^{-1})^t_{71}\,q_{27},
   q_{22} + (M^{-1})^t_{32}\,q_{23} + (M^{-1})^t_{42}\,q_{24} + (M^{-1})^t_{52}\,q_{25} + (M^{-1})^t_{62}\,q_{26} + 
    (M^{-1})^t_{72}\,q_{27},(M^{-1})^t_{33}\,q_{23} + (M^{-1})^t_{43}\,q_{24} + (M^{-1})^t_{53}\,q_{25} + 
    (M^{-1})^t_{63}\,q_{26} + (M^{-1})^t_{73}\,q_{27},
   (M^{-1})^t_{34}\,q_{23} + (M^{-1})^t_{44}\,q_{24} + (M^{-1})^t_{54}\,q_{25} + (M^{-1})^t_{64}\,q_{26} + 
    (M^{-1})^t_{74}\,q_{27},(M^{-1})^t_{35}\,q_{23} + (M^{-1})^t_{45}\,q_{24} + (M^{-1})^t_{55}\,q_{25} + 
    (M^{-1})^t_{65}\,q_{26} + (M^{-1})^t_{75}\,q_{27},
   (M^{-1})^t_{36}\,q_{23} + (M^{-1})^t_{46}\,q_{24} + (M^{-1})^t_{56}\,q_{25} + (M^{-1})^t_{66}\,q_{26} + 
    (M^{-1})^t_{76}\,q_{27},(M^{-1})^t_{37}\,q_{23} + (M^{-1})^t_{47}\,q_{24} + (M^{-1})^t_{57}\,q_{25} + 
    (M^{-1})^t_{67}\,q_{26} + (M^{-1})^t_{77}\,q_{27}\} ,
  \{ (M^{-1})^t_{40}\,q_{34} + (M^{-1})^t_{50}\,q_{35} + (M^{-1})^t_{60}\,q_{36} + (M^{-1})^t_{70}\,q_{37},
   (M^{-1})^t_{41}\,q_{34} + (M^{-1})^t_{51}\,q_{35} + (M^{-1})^t_{61}\,q_{36} + (M^{-1})^t_{71}\,q_{37},
   (M^{-1})^t_{42}\,q_{34} + (M^{-1})^t_{52}\,q_{35} + (M^{-1})^t_{62}\,q_{36} + (M^{-1})^t_{72}\,q_{37},
   (M^{-1})^t_{43}\,q_{34} + (M^{-1})^t_{53}\,q_{35} + (M^{-1})^t_{63}\,q_{36} + (M^{-1})^t_{73}\,q_{37},
   (M^{-1})^t_{44}\,q_{34} + (M^{-1})^t_{54}\,q_{35} + (M^{-1})^t_{64}\,q_{36} + (M^{-1})^t_{74}\,q_{37},
   (M^{-1})^t_{45}\,q_{34} + (M^{-1})^t_{55}\,q_{35} + (M^{-1})^t_{65}\,q_{36} + (M^{-1})^t_{75}\,q_{37},
   (M^{-1})^t_{46}\,q_{34} + (M^{-1})^t_{56}\,q_{35} + (M^{-1})^t_{66}\,q_{36} + (M^{-1})^t_{76}\,q_{37},
   (M^{-1})^t_{47}\,q_{34} + (M^{-1})^t_{57}\,q_{35} 
+ (M^{-1})^t_{67}\,q_{36} + (M^{-1})^t_{77}\,q_{37}\} ,
  \{ (M^{-1})^t_{40}\,q_{44},(M^{-1})^t_{41}\,q_{44},(M^{-1})^t_{42}\,q_{44},(M^{-1})^t_{43}\,q_{44},
   (M^{-1})^t_{44}\,q_{44},$

\noindent$
(M^{-1})^t_{45}\,q_{44},(M^{-1})^t_{46}\,q_{44},(M^{-1})^t_{47}\,q_{44}\} \}$

The above matrix equatin can then be solved for the 64-24=40 entries $(
M^{-1})^t_{ij}$ $4\leq i\leq 7,\ 0\leq j\leq 7$ to give the following result: 

$(M^{-1})^t_{30}=\frac{X_{30}}{Y_{30}}$, where

$ X_{30}= {q_{06}}\,{q_{10}}\,{q_{27}}\,{q_{35}} + {q_{16}}\,{q_{27}}\,{q_{35}} - 
            {q_{00}}\,{q_{16}}\,{q_{27}}\,{q_{35}} - {q_{05}}\,{q_{10}}\,{q_{27}}\,{q_{36}} - 
            {q_{15}}\,{q_{27}}\,{q_{36}} + {q_{00}}\,{q_{15}}\,{q_{27}}\,{q_{36}} 
( -1 + {q_{00}} ) \,{q_{17}}\,( {q_{26}}\,{q_{35}} - {q_{25}}\,{q_{36}} )  + 
            {q_{07}}\,{q_{10}}\,( -( {q_{26}}\,{q_{35}} )  + {q_{25}}\,{q_{36}} )  -$

\noindent $ {q_{06}}\,{q_{10}}\,{q_{25}}\,{q_{37}} - {q_{16}}\,{q_{25}}\,{q_{37}} + 
            {q_{00}}\,{q_{16}}\,{q_{25}}\,{q_{37}} + {q_{05}}\,{q_{10}}\,{q_{26}}\,{q_{37}} + 
            {q_{15}}\,{q_{26}}\,{q_{37}} - {q_{00}}\,{q_{15}}\,{q_{26}}\,{q_{37}}$ 

$Y_{30}=-( {q_{03}}\,
               {q_{17}}\,{q_{26}}\,{q_{35}} )  + {q_{03}}\,{q_{16}}\,{q_{27}}\,{q_{35}} - 
            {q_{05}}\,{q_{17}}\,{q_{23}}\,{q_{36}} + {q_{03}}\,{q_{17}}\,{q_{25}}\,{q_{36}} + 
            {q_{05}}\,{q_{13}}\,{q_{27}}\,{q_{36}} -
 {q_{03}}\,{q_{15}}\,{q_{27}}\,{q_{36}} + {q_{07}}\,( -( {q_{16}}\,{q_{23}}\,{q_{35}} )  
+ {q_{13}}\,{q_{26}}\,{q_{35}} + 
               {q_{15}}\,{q_{23}}\,{q_{36}} - {q_{13}}\,{q_{25}}\,{q_{36}} )  + $

\noindent$ {q_{05}}\,{q_{16}}\,{q_{23}}\,{q_{37}} - {q_{03}}\,{q_{16}}\,{q_{25}}\,{q_{37}} - 
            {q_{05}}\,{q_{13}}\,{q_{26}}\,{q_{37}} + {q_{03}}\,{q_{15}}\,{q_{26}}\,{q_{37}} +$

\noindent$   {q_{06}}\,( {q_{17}}\,{q_{23}}\,{q_{35}} - {q_{13}}\,{q_{27}}\,{q_{35}} -\, 
{q_{15}}\,{q_{23}}\,{q_{37}} + {q_{13}}\,{q_{25}}\,{q_{37}} ) , $

\vskip 0.2in $(M^{-1})^{t}_{31}=\frac{X_{31}}{Y_{31}}$ where
 
$X_{31}={q_{01}}{q_{17}}{q_{26}}{q_{35}} - {q_{01}}{q_{16}}{q_{27}}{q_{35}} + 
            {q_{05}}{q_{17}}{q_{21}}{q_{36}} - {q_{01}}{q_{17}}{q_{25}}\,{q_{36}} + 
            {q_{05}}{q_{27}}{q_{36}} - {q_{05}}{q_{11}}{q_{27}}{q_{36}} + 
            {q_{01}}{q_{15}}{q_{27}}{q_{36}} + 
            {q_{07}}( {q_{16}}{q_{21}}{q_{35}} + 
               {q_{26}}( {q_{35}} - {q_{11}}\,{q_{35}})  - $

\noindent$               ( {q_{15}}\,{q_{21}} + {q_{25}} - {q_{11}}\,{q_{25}} ) \,{q_{36}} )  - 
            {q_{05}}\,{q_{16}}\,{q_{21}}\,{q_{37}} + {q_{01}}\,{q_{16}}\,{q_{25}}\,{q_{37}} - 
            {q_{05}}\,{q_{26}}\,{q_{37}} + {q_{05}}\,{q_{11}}\,{q_{26}}\,{q_{37}} - 
            {q_{01}}\,{q_{15}}\,{q_{26}}\,{q_{37}} + 
            {q_{06}}\,( -( {q_{17}}\,{q_{21}}\,{q_{35}} )  + 
               ( -1 + {q_{11}} ) \,{q_{27}}\,{q_{35}} +$

\noindent$( {q_{15}}\,{q_{21}} + {q_{25}} - {q_{11}}\,{q_{25}} ) \,{q_{37}} )$

$Y_{31}=-( 
               {q_{03}}\,{q_{17}}\,{q_{26}}\,{q_{35}} )  + {q_{03}}\,{q_{16}}\,{q_{27}}\,{q_{35}} - 
            {q_{05}}\,{q_{17}}\,{q_{23}}\,{q_{36}} + {q_{03}}\,{q_{17}}\,{q_{25}}\,{q_{36}} + 
            {q_{05}}\,{q_{13}}\,{q_{27}}\,{q_{36}} - {q_{03}}\,{q_{15}}\,{q_{27}}\,{q_{36}} + 
            {q_{07}}\,( -( {q_{16}}\,{q_{23}}\,{q_{35}} )  + {q_{13}}\,{q_{26}}\,{q_{35}} + 
               {q_{15}}\,{q_{23}}\,{q_{36}} - {q_{13}}\,{q_{25}}\,{q_{36}} )  + 
            {q_{05}}\,{q_{16}}\,{q_{23}}\,{q_{37}} - {q_{03}}\,{q_{16}}\,{q_{25}}\,{q_{37}} - 
            {q_{05}}\,{q_{13}}\,{q_{26}}\,{q_{37}} + {q_{03}}\,{q_{15}}\,{q_{26}}\,{q_{37}} + $

\noindent${q_{06}}\,( {q_{17}}\,{q_{23}}\,{q_{35}} - {q_{13}}\,{q_{27}}\,{q_{35}} - 
               {q_{15}}\,{q_{23}}\,{q_{37}} + {q_{13}}\,{q_{25}}\,{q_{37}} ),$

\vskip 0.2in $(M^{-1})^{t}_{32}=\frac{X_{32}}{Y_{32}}$ where
 
$X_{32}= 2\,{q_{02}}\,{q_{17}}\,{q_{26}}\,{q_{35}} - 2\,{q_{02}}\,{q_{16}}\,{q_{27}}\,{q_{35}} - 
            {q_{05}}\,{q_{17}}\,{q_{36}} + 2\,{q_{05}}\,{q_{17}}\,{q_{22}}\,{q_{36}} - 
            2\,{q_{02}}\,{q_{17}}\,{q_{25}}\,{q_{36}} - 2\,{q_{05}}\,{q_{12}}\,{q_{27}}\,{q_{36}} + 
            2\,{q_{02}}\,{q_{15}}\,{q_{27}}\,{q_{36}} + 
            {q_{07}}\,( {q_{16}}\,( -1 + 2\,{q_{22}} ) \,{q_{35}} - $

\noindent              $ 2\,{q_{12}}\,{q_{26}}\,{q_{35}} + {q_{15}}\,{q_{36}} - 2\,{q_{15}}\,{q_{22}}\,{q_{36}} + 
               2\,{q_{12}}\,{q_{25}}\,{q_{36}} )  + {q_{05}}\,{q_{16}}\,{q_{37}} - 
            2\,{q_{05}}\,{q_{16}}\,{q_{22}}\,{q_{37}} + 2\,{q_{02}}\,{q_{16}}\,{q_{25}}\,{q_{37}} + 
            2\,{q_{05}}\,{q_{12}}\,{q_{26}}\,{q_{37}} - 2\,{q_{02}}\,{q_{15}}\,{q_{26}}\,{q_{37}} + 
            {q_{06}}\,( {q_{17}}\,{q_{35}} - 2\,{q_{17}}\,{q_{22}}\,{q_{35}} + 
               2\,{q_{12}}\,{q_{27}}\,{q_{35}}$

\noindent $- {q_{15}}\,{q_{37}} + 2\,{q_{15}}\,{q_{22}}\,{q_{37}} - 
               2\,{q_{12}}\,{q_{25}}\,{q_{37}} ) $

$Y_{32}=2\,
            ( -( {q_{03}}\,{q_{17}}\,{q_{26}}\,{q_{35}} )  + 
              {q_{03}}\,{q_{16}}\,{q_{27}}\,{q_{35}} - {q_{05}}\,{q_{17}}\,{q_{23}}\,{q_{36}} + 
              {q_{03}}\,{q_{17}}\,{q_{25}}\,{q_{36}}$

\noindent $+ {q_{05}}\,{q_{13}}\,{q_{27}}\,{q_{36}} - 
              {q_{03}}\,{q_{15}}\,{q_{27}}\,{q_{36}} + 
              {q_{07}}\,( -( {q_{16}}\,{q_{23}}\,{q_{35}} )  + {q_{13}}\,{q_{26}}\,{q_{35}} + 
                 {q_{15}}\,{q_{23}}\,{q_{36}} - {q_{13}}\,{q_{25}}\,{q_{36}} )  + $

\noindent${q_{05}}\,{q_{16}}\,{q_{23}}\,{q_{37}} - {q_{03}}\,{q_{16}}\,{q_{25}}\,{q_{37}} - 
              {q_{05}}\,{q_{13}}\,{q_{26}}\,{q_{37}} + {q_{03}}\,{q_{15}}\,{q_{26}}\,{q_{37}} + 
              {q_{06}}\,( {q_{17}}\,{q_{23}}\,{q_{35}} - {q_{13}}\,{q_{27}}\,{q_{35}} $

\noindent $- {q_{15}}\,{q_{23}}\,{q_{37}} + {q_{13}}\,{q_{25}}\,{q_{37}} )  )$,

\vskip 0.2in $(M^{-1})^{t}_{33}=\frac{X_{33}}{Y_{33}}$
where
 
$X_{33}=   {q_{07}}\,{q_{16}}\,{q_{25}} - {q_{06}}\,{q_{17}}\,{q_{25}} - 
            {q_{07}}\,{q_{15}}\,{q_{26}} + {q_{05}}\,{q_{17}}\,{q_{26}} + 
            {q_{06}}\,{q_{15}}\,{q_{27}} - {q_{05}}\,{q_{16}}\,{q_{27}}$

$Y_{33}=6\,
            ( -( {q_{03}}\,{q_{17}}\,{q_{26}}\,{q_{35}} )  + 
              {q_{03}}\,{q_{16}}\,{q_{27}}\,{q_{35}} - {q_{05}}\,{q_{17}}\,{q_{23}}\,{q_{36}} + 
              {q_{03}}\,{q_{17}}\,{q_{25}}\,{q_{36}} + $

\noindent${q_{05}}\,{q_{13}}\,{q_{27}}\,{q_{36}} - 
              {q_{03}}\,{q_{15}}\,{q_{27}}\,{q_{36}} + 
              {q_{07}}\,( -( {q_{16}}\,{q_{23}}\,{q_{35}} )  + {q_{13}}\,{q_{26}}\,{q_{35}} +$ 
          
\noindent $      {q_{15}}\,{q_{23}}\,{q_{36}} - {q_{13}}\,{q_{25}}\,{q_{36}} )  + 
              {q_{05}}\,{q_{16}}\,{q_{23}}\,{q_{37}} - {q_{03}}\,{q_{16}}\,{q_{25}}\,{q_{37}} - 
              {q_{05}}\,{q_{13}}\,{q_{26}}\,{q_{37}} + {q_{03}}\,{q_{15}}\,{q_{26}}\,{q_{37}} +
              {q_{06}}\,( {q_{17}}\,{q_{23}}\,{q_{35}} - {q_{13}}\,{q_{27}}\,{q_{35}} - 
                 {q_{15}}\,{q_{23}}\,{q_{37}} + {q_{13}}\,{q_{25}}\,{q_{37}} )  ),$

$
(M^{-1})^{t}_{34}=\frac{X_{34}}{Y_{34}}$ where
 
$X_{34}={q_{05}}\,{q_{17}}\,{q_{26}}\,{q_{34}} - {q_{05}}\,{q_{16}}\,{q_{27}}\,{q_{34}} - 
            {q_{04}}\,{q_{17}}\,{q_{26}}\,{q_{35}} + {q_{04}}\,{q_{16}}\,{q_{27}}\,{q_{35}} - 
            {q_{05}}\,{q_{17}}\,{q_{24}}\,{q_{36}} + {q_{04}}\,{q_{17}}\,{q_{25}}\,{q_{36}} + 
            {q_{05}}\,{q_{14}}\,{q_{27}}\,{q_{36}} - {q_{04}}\,{q_{15}}\,{q_{27}}\,{q_{36}} + 
            {q_{07}}\,( {q_{16}}\,{q_{25}}\,{q_{34}} - {q_{15}}\,{q_{26}}\,{q_{34}} - $

\noindent              $ {q_{16}}\,{q_{24}}\,{q_{35}} + {q_{14}}\,{q_{26}}\,{q_{35}} + 
               {q_{15}}\,{q_{24}}\,{q_{36}} - {q_{14}}\,{q_{25}}\,{q_{36}} )  + 
            {q_{05}}\,{q_{16}}\,{q_{24}}\,{q_{37}} - {q_{04}}\,{q_{16}}\,{q_{25}}\,{q_{37}} - 
            {q_{05}}\,{q_{14}}\,{q_{26}}\,{q_{37}} + {q_{04}}\,{q_{15}}\,{q_{26}}\,{q_{37}} + 
{q_{06}}\,( -( {q_{17}}\,{q_{25}}\,{q_{34}} )  + {q_{15}}\,{q_{27}}\,{q_{34}} + 
               {q_{17}}\,{q_{24}}\,{q_{35}} - {q_{14}}\,{q_{27}}\,{q_{35}} - $

\noindent $              {q_{15}}\,{q_{24}}\,{q_{37}} + {q_{14}}\,{q_{25}}\,{q_{37}} )$ 

$Y_{34}=24\,
            ( {q_{03}}\,{q_{17}}\,{q_{26}}\,{q_{35}} - {q_{03}}\,{q_{16}}\,{q_{27}}\,{q_{35}} + 
              {q_{05}}\,{q_{17}}\,{q_{23}}\,{q_{36}} - {q_{03}}\,{q_{17}}\,{q_{25}}\,{q_{36}} - 
              {q_{05}}\,{q_{13}}\,{q_{27}}\,{q_{36}} +$

\noindent ${q_{03}}\,{q_{15}}\,{q_{27}}\,{q_{36}} + 
              {q_{07}}\,( {q_{16}}\,{q_{23}}\,{q_{35}} - {q_{13}}\,{q_{26}}\,{q_{35}} - 
                 {q_{15}}\,{q_{23}}\,{q_{36}} + {q_{13}}\,{q_{25}}\,{q_{36}} )  - 
              {q_{05}}\,{q_{16}}\,{q_{23}}\,{q_{37}} + {q_{03}}\,{q_{16}}\,{q_{25}}\,{q_{37}} + 
              {q_{05}}\,{q_{13}}\,{q_{26}}\,{q_{37}} - {q_{03}}\,{q_{15}}\,{q_{26}}\,{q_{37}} + $

\noindent$ {q_{06}}\,( -( {q_{17}}\,{q_{23}}\,{q_{35}} )  + {q_{13}}\,{q_{27}}\,{q_{35}} + 
                 {q_{15}}\,{q_{23}}\,{q_{37}} - {q_{13}}\,{q_{25}}\,{q_{37}} )  ) \,{q_{44}},$

\vskip 0.2in $(M^{-1})^{t}_{35}=\frac{X_{35}}{Y_{35}}$ where 
       
$X_{35}=-39\,{q_{05}}\,{q_{17}}\,{q_{26}}\,{q_{34}} + 39\,{q_{05}}\,{q_{16}}\,{q_{27}}\,{q_{34}} + 
            39\,{q_{04}}\,{q_{17}}\,{q_{26}}\,{q_{35}} - 39\,{q_{04}}\,{q_{16}}\,{q_{27}}\,{q_{35}} + 
            39\,{q_{05}}\,{q_{17}}\,{q_{24}}\,{q_{36}} - 39\,{q_{04}}\,{q_{17}}\,{q_{25}}\,{q_{36}} - 
            39\,{q_{05}}\,{q_{14}}\,{q_{27}}\,{q_{36}} + 39\,{q_{04}}\,{q_{15}}\,{q_{27}}\,{q_{36}} - 
            39\,{q_{05}}\,{q_{16}}\,{q_{24}}\,{q_{37}} + 39\,{q_{04}}\,{q_{16}}\,{q_{25}}\,{q_{37}} + 
            39\,{q_{05}}\,{q_{14}}\,{q_{26}}\,{q_{37}} - 39\,{q_{04}}\,{q_{15}}\,{q_{26}}\,{q_{37}} + 
            460\,{q_{05}}\,{q_{17}}\,{q_{26}}\,{q_{44}} - 460\,{q_{05}}\,{q_{16}}\,{q_{27}}\,{q_{44}} - 
            1175799\,{q_{17}}\,{q_{26}}\,{q_{35}}\,{q_{44}} + 
            1175799\,{q_{16}}\,{q_{27}}\,{q_{35}}\,{q_{44}} - 
            5190\,{q_{05}}\,{q_{17}}\,{q_{36}}\,{q_{44}} + 
            1175799\,{q_{17}}\,{q_{25}}\,{q_{36}}\,{q_{44}} + 
            35826\,{q_{05}}\,{q_{27}}\,{q_{36}}\,{q_{44}} - 
            1175799\,{q_{15}}\,{q_{27}}\,{q_{36}}\,{q_{44}} + 
            5190\,{q_{05}}\,{q_{16}}\,{q_{37}}\,{q_{44}} - 
            1175799\,{q_{16}}\,{q_{25}}\,{q_{37}}\,{q_{44}} - 
            35826\,{q_{05}}\,{q_{26}}\,{q_{37}}\,{q_{44}} + 
            1175799\,{q_{15}}\,{q_{26}}\,{q_{37}}\,{q_{44}} + 
            {q_{07}}\,( -3\,( {q_{26}}\,{q_{35}} - {q_{25}}\,{q_{36}} ) \,
                ( 13\,{q_{14}} - 11942\,{q_{44}} )  + $

\noindent$
               {q_{16}}\,( -39\,{q_{25}}\,{q_{34}} + 39\,{q_{24}}\,{q_{35}} + 460\,{q_{25}}\,{q_{44}} - 
                  5190\,{q_{35}}\,{q_{44}} )  + $

\noindent$
               {q_{15}}\,( 39\,{q_{26}}\,{q_{34}} - 39\,{q_{24}}\,{q_{36}} - 460\,{q_{26}}\,{q_{44}} + 
                  5190\,{q_{36}}\,{q_{44}} )  )  + $

\noindent$ {q_{06}}\,( 3\,( {q_{27}}\,{q_{35}} - {q_{25}}\,{q_{37}} ) \,
                ( 13\,{q_{14}} - 11942\,{q_{44}} )  + 
               {q_{17}}\,( 39\,{q_{25}}\,{q_{34}} - 39\,{q_{24}}\,{q_{35}} - 460\,{q_{25}}\,{q_{44}} + $

\noindent$
                  5190\,{q_{35}}\,{q_{44}} )  + 
               {q_{15}}\,( -39\,{q_{27}}\,{q_{34}} + 39\,{q_{24}}\,{q_{37}} + 460\,{q_{27}}\,{q_{44}} - 
                  5190\,{q_{37}}\,{q_{44}} )  )$

$Y_{35}=96\,
            ( {q_{03}}\,{q_{17}}\,{q_{26}}\,{q_{35}} - {q_{03}}\,{q_{16}}\,{q_{27}}\,{q_{35}} + 
              {q_{05}}\,{q_{17}}\,{q_{23}}\,{q_{36}} - {q_{03}}\,{q_{17}}\,{q_{25}}\,{q_{36}} - 
              {q_{05}}\,{q_{13}}\,{q_{27}}\,{q_{36}}$

\noindent $+ {q_{03}}\,{q_{15}}\,{q_{27}}\,{q_{36}} + 
              {q_{07}}\,( {q_{16}}\,{q_{23}}\,{q_{35}} - {q_{13}}\,{q_{26}}\,{q_{35}} - 
                 {q_{15}}\,{q_{23}}\,{q_{36}} + {q_{13}}\,{q_{25}}\,{q_{36}} )  - 
              {q_{05}}\,{q_{16}}\,{q_{23}}\,{q_{37}} + {q_{03}}\,{q_{16}}\,{q_{25}}\,{q_{37}} + 
              {q_{05}}\,{q_{13}}\,{q_{26}}\,{q_{37}} - {q_{03}}\,{q_{15}}\,{q_{26}}\,{q_{37}} + 
              {q_{06}}\,( -( {q_{17}}\,{q_{23}}\,{q_{35}} )  + {q_{13}}\,{q_{27}}\,{q_{35}}$

\noindent$ + 
                 {q_{15}}\,{q_{23}}\,{q_{37}} - {q_{13}}\,{q_{25}}\,{q_{37}} )  ) \,{q_{44}},$

$
(M^{-1})^{t}_{36}=\frac{X_{36}}{Y_{36}}$ where
 
       $ X_{36}=3\,( 6\,{q_{05}}\,{q_{17}}\,{q_{26}}\,{q_{34}} - 
              6\,{q_{05}}\,{q_{16}}\,{q_{27}}\,{q_{34}} - 6\,{q_{04}}\,{q_{17}}\,{q_{26}}\,{q_{35}} + 
              6\,{q_{04}}\,{q_{16}}\,{q_{27}}\,{q_{35}} -$

\noindent$ 6\,{q_{05}}\,{q_{17}}\,{q_{24}}\,{q_{36}} + 
              6\,{q_{04}}\,{q_{17}}\,{q_{25}}\,{q_{36}} + 6\,{q_{05}}\,{q_{14}}\,{q_{27}}\,{q_{36}} - 
              6\,{q_{04}}\,{q_{15}}\,{q_{27}}\,{q_{36}} + 6\,{q_{05}}\,{q_{16}}\,{q_{24}}\,{q_{37}} - 
              6\,{q_{04}}\,{q_{16}}\,{q_{25}}\,{q_{37}} - 6\,{q_{05}}\,{q_{14}}\,{q_{26}}\,{q_{37}} + 
              6\,{q_{04}}\,{q_{15}}\,{q_{26}}\,{q_{37}} - 140\,{q_{05}}\,{q_{17}}\,{q_{26}}\,{q_{44}} + 
              140\,{q_{05}}\,{q_{16}}\,{q_{27}}\,{q_{44}} + 43359\,{q_{17}}\,{q_{26}}\,{q_{35}}\,{q_{44}} - 
              43359\,{q_{16}}\,{q_{27}}\,{q_{35}}\,{q_{44}} + 
              1758\,{q_{05}}\,{q_{17}}\,{q_{36}}\,{q_{44}} - 
              43359\,{q_{17}}\,{q_{25}}\,{q_{36}}\,{q_{44}} - 
              12810\,{q_{05}}\,{q_{27}}\,{q_{36}}\,{q_{44}} + 
              43359\,{q_{15}}\,{q_{27}}\,{q_{36}}\,{q_{44}} - 
              1758\,{q_{05}}\,{q_{16}}\,{q_{37}}\,{q_{44}} + 
              43359\,{q_{16}}\,{q_{25}}\,{q_{37}}\,{q_{44}} + 
              12810\,{q_{05}}\,{q_{26}}\,{q_{37}}\,{q_{44}} - 
              43359\,{q_{15}}\,{q_{26}}\,{q_{37}}\,{q_{44}} + 
              2\,{q_{07}}\,( 3\,( {q_{26}}\,{q_{35}} - {q_{25}}\,{q_{36}} ) \,
                  ( {q_{14}} - 2135\,{q_{44}} )  + 
                 {q_{16}}\,( 3\,{q_{25}}\,{q_{34}} - 3\,{q_{24}}\,{q_{35}} - 70\,{q_{25}}\,{q_{44}} + 
                    879\,{q_{35}}\,{q_{44}} )  + $

\noindent$ {q_{15}}\,( -3\,{q_{26}}\,{q_{34}} + 3\,{q_{24}}\,{q_{36}} + 70\,{q_{26}}\,{q_{44}} - 
                    879\,{q_{36}}\,{q_{44}} )  )  + $

\noindent$ 2\,{q_{06}}\,( -3\,( {q_{27}}\,{q_{35}} - {q_{25}}\,{q_{37}} ) \,
                  ( {q_{14}} - 2135\,{q_{44}} )  + $

\noindent$ {q_{17}}\,( -3\,{q_{25}}\,{q_{34}} + 3\,{q_{24}}\,{q_{35}} + 70\,{q_{25}}\,{q_{44}} - 
                    879\,{q_{35}}\,{q_{44}} )  + 
                 {q_{15}}\,( 3\,{q_{27}}\,{q_{34}} - 3\,{q_{24}}\,{q_{37}}$

\noindent$ - 70\,{q_{27}}\,{q_{44}} + 
                    879\,{q_{37}}\,{q_{44}} )  )  )$

$Y_{36}=32\,
            ( {q_{03}}\,{q_{17}}\,{q_{26}}\,{q_{35}} - {q_{03}}\,{q_{16}}\,{q_{27}}\,{q_{35}} + 
              {q_{05}}\,{q_{17}}\,{q_{23}}\,{q_{36}} - {q_{03}}\,{q_{17}}\,{q_{25}}\,{q_{36}} - 
              {q_{05}}\,{q_{13}}\,{q_{27}}\,{q_{36}} +$

\noindent ${q_{03}}\,{q_{15}}\,{q_{27}}\,{q_{36}} + 
              {q_{07}}\,( {q_{16}}\,{q_{23}}\,{q_{35}} - {q_{13}}\,{q_{26}}\,{q_{35}} - 
                 {q_{15}}\,{q_{23}}\,{q_{36}} + {q_{13}}\,{q_{25}}\,{q_{36}} )  - 
              {q_{05}}\,{q_{16}}\,{q_{23}}\,{q_{37}} + {q_{03}}\,{q_{16}}\,{q_{25}}\,{q_{37}} + 
              {q_{05}}\,{q_{13}}\,{q_{26}}\,{q_{37}} - {q_{03}}\,{q_{15}}\,{q_{26}}\,{q_{37}} + $

\noindent$ {q_{06}}\,( -( {q_{17}}\,{q_{23}}\,{q_{35}} )  + {q_{13}}\,{q_{27}}\,{q_{35}} + 
                 {q_{15}}\,{q_{23}}\,{q_{37}} - {q_{13}}\,{q_{25}}\,{q_{37}} )  ) \,{q_{44}},$

\vskip 0.2in $(M^{-1})^{t}_{37}=\frac{X_{37}}{Y_{37}}$ where 

$X_{37}=-3\,( 2\,{q_{05}}\,{q_{17}}\,{q_{26}}\,{q_{34}} - 
              2\,{q_{05}}\,{q_{16}}\,{q_{27}}\,{q_{34}} - 2\,{q_{04}}\,{q_{17}}\,{q_{26}}\,{q_{35}} + 
              2\,{q_{04}}\,{q_{16}}\,{q_{27}}\,{q_{35}} - $

\noindent$ 2\,{q_{05}}\,{q_{17}}\,{q_{24}}\,{q_{36}} + 
              2\,{q_{04}}\,{q_{17}}\,{q_{25}}\,{q_{36}} + 2\,{q_{05}}\,{q_{14}}\,{q_{27}}\,{q_{36}} - 
              2\,{q_{04}}\,{q_{15}}\,{q_{27}}\,{q_{36}} + 2\,{q_{05}}\,{q_{16}}\,{q_{24}}\,{q_{37}} - 
              2\,{q_{04}}\,{q_{16}}\,{q_{25}}\,{q_{37}} - 2\,{q_{05}}\,{q_{14}}\,{q_{26}}\,{q_{37}} + 
              2\,{q_{04}}\,{q_{15}}\,{q_{26}}\,{q_{37}} - 52\,{q_{05}}\,{q_{17}}\,{q_{26}}\,{q_{44}} + 
              52\,{q_{05}}\,{q_{16}}\,{q_{27}}\,{q_{44}} + 17913\,{q_{17}}\,{q_{26}}\,{q_{35}}\,{q_{44}} - 
              17913\,{q_{16}}\,{q_{27}}\,{q_{35}}\,{q_{44}} + 690\,{q_{05}}\,{q_{17}}\,{q_{36}}\,{q_{44}} - 
              17913\,{q_{17}}\,{q_{25}}\,{q_{36}}\,{q_{44}} - 
              5190\,{q_{05}}\,{q_{27}}\,{q_{36}}\,{q_{44}} + 
              17913\,{q_{15}}\,{q_{27}}\,{q_{36}}\,{q_{44}} - 690\,{q_{05}}\,{q_{16}}\,{q_{37}}\,{q_{44}} + 
              17913\,{q_{16}}\,{q_{25}}\,{q_{37}}\,{q_{44}} + 
              5190\,{q_{05}}\,{q_{26}}\,{q_{37}}\,{q_{44}} - 
              17913\,{q_{15}}\,{q_{26}}\,{q_{37}}\,{q_{44}} + 
              2\,{q_{07}}\,( ( {q_{26}}\,{q_{35}} - {q_{25}}\,{q_{36}} ) \,
                  ( {q_{14}} - 2595\,{q_{44}} )  + 
                 {q_{16}}\,( {q_{25}}\,{q_{34}} - {q_{24}}\,{q_{35}} - 26\,{q_{25}}\,{q_{44}} + 
                    345\,{q_{35}}\,{q_{44}} )  + $

\noindent$ {q_{15}}\,( -( {q_{26}}\,{q_{34}} )  + {q_{24}}\,{q_{36}} + 
                    26\,{q_{26}}\,{q_{44}} - 345\,{q_{36}}\,{q_{44}} )  )  + 
              2\,{q_{06}}\,( -( ( {q_{27}}\,{q_{35}} - {q_{25}}\,{q_{37}} ) \,
                    ( {q_{14}} - 2595\,{q_{44}} )  )   $

\noindent$
                 +{q_{17}}\,( -( {q_{25}}\,{q_{34}} )  + {q_{24}}\,{q_{35}} + 
                    26\,{q_{25}}\,{q_{44}} - 345\,{q_{35}}\,{q_{44}} )  +$

\noindent $ {q_{15}}\,( {q_{27}}\,{q_{34}} - {q_{24}}\,{q_{37}} - 26\,{q_{27}}\,{q_{44}} + 
                    345\,{q_{37}}\,{q_{44}} )  )  )$ 

$Y_{37}= 16\,
            ( {q_{03}}\,{q_{17}}\,{q_{26}}\,{q_{35}} - {q_{03}}\,{q_{16}}\,{q_{27}}\,{q_{35}} + 
              {q_{05}}\,{q_{17}}\,{q_{23}}\,{q_{36}} - {q_{03}}\,{q_{17}}\,{q_{25}}\,{q_{36}} - 
              {q_{05}}\,{q_{13}}\,{q_{27}}\,{q_{36}} +$

\noindent ${q_{03}}\,{q_{15}}\,{q_{27}}\,{q_{36}} + 
              {q_{07}}\,( {q_{16}}\,{q_{23}}\,{q_{35}} - {q_{13}}\,{q_{26}}\,{q_{35}} - 
                 {q_{15}}\,{q_{23}}\,{q_{36}} + {q_{13}}\,{q_{25}}\,{q_{36}} )  - 
              {q_{05}}\,{q_{16}}\,{q_{23}}\,{q_{37}} + {q_{03}}\,{q_{16}}\,{q_{25}}\,{q_{37}} + 
              {q_{05}}\,{q_{13}}\,{q_{26}}\,{q_{37}} - {q_{03}}\,{q_{15}}\,{q_{26}}\,{q_{37}} + $

\noindent ${q_{06}}\,( -( {q_{17}}\,{q_{23}}\,{q_{35}} )  + {q_{13}}\,{q_{27}}\,{q_{35}} + 
                 {q_{15}}\,{q_{23}}\,{q_{37}} - {q_{13}}\,{q_{25}}\,{q_{37}} )  ) \,{q_{44}},$

$
(M^{-1})^{t}_{50}=\frac{X_{50}}{Y_{50}}$ where
 
$        X_{50}=-( {q_{07}}\,{q_{10}}\,{q_{23}}\,{q_{36}} )  + 
            ( -1 + {q_{00}} ) \,{q_{17}}\,{q_{23}}\,{q_{36}} + 
            {q_{03}}\,{q_{10}}\,{q_{27}}\,{q_{36}} + {q_{13}}\,{q_{27}}\,{q_{36}} - 
            {q_{00}}\,{q_{13}}\,{q_{27}}\,{q_{36}} + {q_{06}}\,{q_{10}}\,{q_{23}}\,{q_{37}} + 
            {q_{16}}\,{q_{23}}\,{q_{37}} - {q_{00}}\,{q_{16}}\,{q_{23}}\,{q_{37}} - 
            {q_{03}}\,{q_{10}}\,{q_{26}}\,{q_{37}} - {q_{13}}\,{q_{26}}\,{q_{37}} + 
            {q_{00}}\,{q_{13}}\,{q_{26}}\,{q_{37}}$

$Y_{50}=-( {q_{03}}\,{q_{17}}\,{q_{26}}\,{q_{35}} )
                + {q_{03}}\,{q_{16}}\,{q_{27}}\,{q_{35}} - {q_{05}}\,{q_{17}}\,{q_{23}}\,{q_{36}} + 
            {q_{03}}\,{q_{17}}\,{q_{25}}\,{q_{36}} + {q_{05}}\,{q_{13}}\,{q_{27}}\,{q_{36}} - 
            {q_{03}}\,{q_{15}}\,{q_{27}}\,{q_{36}} + 
            {q_{07}}\,( -( {q_{16}}\,{q_{23}}\,{q_{35}} )  + {q_{13}}\,{q_{26}}\,{q_{35}} + 
               {q_{15}}\,{q_{23}}\,{q_{36}} - {q_{13}}\,{q_{25}}\,{q_{36}} )  + 
            {q_{05}}\,{q_{16}}\,{q_{23}}\,{q_{37}} - {q_{03}}\,{q_{16}}\,{q_{25}}\,{q_{37}} - 
            {q_{05}}\,{q_{13}}\,{q_{26}}\,{q_{37}} + {q_{03}}\,{q_{15}}\,{q_{26}}\,{q_{37}} + 
            {q_{06}}\,( {q_{17}}\,{q_{23}}\,{q_{35}} - {q_{13}}\,{q_{27}}\,{q_{35}} - $

\noindent$ {q_{15}}\,{q_{23}}\,{q_{37}} + {q_{13}}\,{q_{25}}\,{q_{37}} ),$

\vskip 0.2in $(M^{-1})^{t}_{51}=\frac{X_{51}}{Y_{51}}$ where

        $X_{51}=
{q_{01}}\,{q_{17}}\,{q_{23}}\,{q_{36}} + 
            {q_{07}}\,( {q_{13}}\,{q_{21}} + {q_{23}} - {q_{11}}\,{q_{23}} ) \,{q_{36}} - 
            {q_{01}}\,{q_{13}}\,{q_{27}}\,{q_{36}} - {q_{06}}\,{q_{13}}\,{q_{21}}\,{q_{37}} - 
            {q_{06}}\,{q_{23}}\,{q_{37}} + {q_{06}}\,{q_{11}}\,{q_{23}}\,{q_{37}} - 
            {q_{01}}\,{q_{16}}\,{q_{23}}\,{q_{37}} + {q_{01}}\,{q_{13}}\,{q_{26}}\,{q_{37}} + 
            {q_{03}}\,( -( {q_{17}}\,{q_{21}}\,{q_{36}} )  + 
               ( -1 + {q_{11}} ) \,{q_{27}}\,{q_{36}} + 
               ( {q_{16}}\,{q_{21}} + {q_{26}} - {q_{11}}\,{q_{26}} ) \,{q_{37}} )$

$ Y_{51}=-( 
               {q_{03}}\,{q_{17}}\,{q_{26}}\,{q_{35}} )  + {q_{03}}\,{q_{16}}\,{q_{27}}\,{q_{35}} - 
            {q_{05}}\,{q_{17}}\,{q_{23}}\,{q_{36}} + {q_{03}}\,{q_{17}}\,{q_{25}}\,{q_{36}} + 
            {q_{05}}\,{q_{13}}\,{q_{27}}\,{q_{36}} - {q_{03}}\,{q_{15}}\,{q_{27}}\,{q_{36}} + 
            {q_{07}}\,( -( {q_{16}}\,{q_{23}}\,{q_{35}} )  + {q_{13}}\,{q_{26}}\,{q_{35}} + 
               {q_{15}}\,{q_{23}}\,{q_{36}} - {q_{13}}\,{q_{25}}\,{q_{36}} )  + 
            {q_{05}}\,{q_{16}}\,{q_{23}}\,{q_{37}} - {q_{03}}\,{q_{16}}\,{q_{25}}\,{q_{37}} - 
            {q_{05}}\,{q_{13}}\,{q_{26}}\,{q_{37}} + {q_{03}}\,{q_{15}}\,{q_{26}}\,{q_{37}} + 
            {q_{06}}\,( {q_{17}}\,{q_{23}}\,{q_{35}} - {q_{13}}\,{q_{27}}\,{q_{35}} - $

\noindent$
               {q_{15}}\,{q_{23}}\,{q_{37}} + {q_{13}}\,{q_{25}}\,{q_{37}} ),$

\vskip 0.2in $(M^{-1})^{t}_{52}=\frac{X_{52}}{Y_{52}}$ where
     
$X_{52}=-2\,{q_{02}}\,{q_{17}}\,{q_{23}}\,{q_{36}} + 
            {q_{07}}\,( {q_{13}} - 2\,{q_{13}}\,{q_{22}} + 2\,{q_{12}}\,{q_{23}} ) \,{q_{36}} + 
            2\,{q_{02}}\,{q_{13}}\,{q_{27}}\,{q_{36}} - {q_{06}}\,{q_{13}}\,{q_{37}} + 
            2\,{q_{06}}\,{q_{13}}\,{q_{22}}\,{q_{37}} - 2\,{q_{06}}\,{q_{12}}\,{q_{23}}\,{q_{37}} + 
            2\,{q_{02}}\,{q_{16}}\,{q_{23}}\,{q_{37}} - 2\,{q_{02}}\,{q_{13}}\,{q_{26}}\,{q_{37}} + 
            {q_{03}}\,( {q_{17}}\,( -1 + 2\,{q_{22}} ) \,{q_{36}} - 
               2\,{q_{12}}\,{q_{27}}\,{q_{36}} + {q_{16}}\,{q_{37}} - 2\,{q_{16}}\,{q_{22}}\,{q_{37}} + 
               2\,{q_{12}}\,{q_{26}}\,{q_{37}} )$
$Y_{52}= 2\,
            ( {q_{03}}\,{q_{17}}\,{q_{26}}\,{q_{35}} - {q_{03}}\,{q_{16}}\,{q_{27}}\,{q_{35}} + 
              {q_{05}}\,{q_{17}}\,{q_{23}}\,{q_{36}} - {q_{03}}\,{q_{17}}\,{q_{25}}\,{q_{36}} - 
              {q_{05}}\,{q_{13}}\,{q_{27}}\,{q_{36}} +$

\noindent$
 {q_{03}}\,{q_{15}}\,{q_{27}}\,{q_{36}} + 
              {q_{07}}\,( {q_{16}}\,{q_{23}}\,{q_{35}} - {q_{13}}\,{q_{26}}\,{q_{35}} - 
                 {q_{15}}\,{q_{23}}\,{q_{36}} + {q_{13}}\,{q_{25}}\,{q_{36}} )  - 
              {q_{05}}\,{q_{16}}\,{q_{23}}\,{q_{37}} + {q_{03}}\,{q_{16}}\,{q_{25}}\,{q_{37}} + $

\noindent$
              {q_{05}}\,{q_{13}}\,{q_{26}}\,{q_{37}} - {q_{03}}\,{q_{15}}\,{q_{26}}\,{q_{37}} + 
              {q_{06}}\,( -( {q_{17}}\,{q_{23}}\,{q_{35}} )  + {q_{13}}\,{q_{27}}\,{q_{35}} + 
                 {q_{15}}\,{q_{23}}\,{q_{37}} - {q_{13}}\,{q_{25}}\,{q_{37}} )  ),$

\vskip 0.2in $(M^{-1})^t_{53}=\frac{X_{53}}{Y_{53}}$ where
 
$X_{53}=-( {q_{07}}\,{q_{16}}\,{q_{23}} )  + {q_{06}}\,{q_{17}}\,{q_{23}} + 
            {q_{07}}\,{q_{13}}\,{q_{26}} - {q_{03}}\,{q_{17}}\,{q_{26}} - 
            {q_{06}}\,{q_{13}}\,{q_{27}} + {q_{03}}\,{q_{16}}\,{q_{27}}$

$Y_{53}=6\,
            ( -( {q_{03}}\,{q_{17}}\,{q_{26}}\,{q_{35}} )  + 
              {q_{03}}\,{q_{16}}\,{q_{27}}\,{q_{35}} - {q_{05}}\,{q_{17}}\,{q_{23}}\,{q_{36}} + 
              {q_{03}}\,{q_{17}}\,{q_{25}}\,{q_{36}} + {q_{05}}\,{q_{13}}\,{q_{27}}\,{q_{36}} $

\noindent$- 
              {q_{03}}\,{q_{15}}\,{q_{27}}\,{q_{36}} + 
              {q_{07}}\,( -( {q_{16}}\,{q_{23}}\,{q_{35}} )  + {q_{13}}\,{q_{26}}\,{q_{35}} + 
                 {q_{15}}\,{q_{23}}\,{q_{36}} - {q_{13}}\,{q_{25}}\,{q_{36}} )  + 
              {q_{05}}\,{q_{16}}\,{q_{23}}\,{q_{37}} - {q_{03}}\,{q_{16}}\,{q_{25}}\,{q_{37}} - 
              {q_{05}}\,{q_{13}}\,{q_{26}}\,{q_{37}} + {q_{03}}\,{q_{15}}\,{q_{26}}\,{q_{37}} + 
              {q_{06}}\,( {q_{17}}\,{q_{23}}\,{q_{35}} - {q_{13}}\,{q_{27}}\,{q_{35}} - $

\noindent $ {q_{15}}\,{q_{23}}\,{q_{37}} + {q_{13}}\,{q_{25}}\,{q_{37}} )  ),$

\vskip 0.2in $(M^{-1})^{t}_{54}=\frac{X_{54}}{Y_{54}}$ where
 
$X_{54}=-( {q_{03}}\,{q_{17}}\,{q_{26}}\,{q_{34}} )  + 
            {q_{03}}\,{q_{16}}\,{q_{27}}\,{q_{34}} - {q_{04}}\,{q_{17}}\,{q_{23}}\,{q_{36}} + 
            {q_{03}}\,{q_{17}}\,{q_{24}}\,{q_{36}} + {q_{04}}\,{q_{13}}\,{q_{27}}\,{q_{36}} - 
            {q_{03}}\,{q_{14}}\,{q_{27}}\,{q_{36}} + 
            {q_{07}}\,( -( {q_{16}}\,{q_{23}}\,{q_{34}} )  + {q_{13}}\,{q_{26}}\,{q_{34}} + 
               {q_{14}}\,{q_{23}}\,{q_{36}} - {q_{13}}\,{q_{24}}\,{q_{36}} )  + 
            {q_{04}}\,{q_{16}}\,{q_{23}}\,{q_{37}} - {q_{03}}\,{q_{16}}\,{q_{24}}\,{q_{37}} - 
            {q_{04}}\,{q_{13}}\,{q_{26}}\,{q_{37}} + {q_{03}}\,{q_{14}}\,{q_{26}}\,{q_{37}} + 
            {q_{06}}\,( {q_{17}}\,{q_{23}}\,{q_{34}} - {q_{13}}\,{q_{27}}\,{q_{34}} - $

\noindent$ {q_{14}}\,{q_{23}}\,{q_{37}} + {q_{13}}\,{q_{24}}\,{q_{37}} ) $

$Y_{54}=24\,
            ( {q_{03}}\,{q_{17}}\,{q_{26}}\,{q_{35}} - {q_{03}}\,{q_{16}}\,{q_{27}}\,{q_{35}} + 
              {q_{05}}\,{q_{17}}\,{q_{23}}\,{q_{36}} - {q_{03}}\,{q_{17}}\,{q_{25}}\,{q_{36}} - 
              {q_{05}}\,{q_{13}}\,{q_{27}}\,{q_{36}}$

\noindent$ + {q_{03}}\,{q_{15}}\,{q_{27}}\,{q_{36}} + 
{q_{07}}\,( {q_{16}}\,{q_{23}}\,{q_{35}} - {q_{13}}\,{q_{26}}\,{q_{35}} - 
                 {q_{15}}\,{q_{23}}\,{q_{36}} + {q_{13}}\,{q_{25}}\,{q_{36}} )  - 
              {q_{05}}\,{q_{16}}\,{q_{23}}\,{q_{37}} + {q_{03}}\,{q_{16}}\,{q_{25}}\,{q_{37}} + 
              {q_{05}}\,{q_{13}}\,{q_{26}}\,{q_{37}} - {q_{03}}\,{q_{15}}\,{q_{26}}\,{q_{37}} + $

\noindent$ {q_{06}}\,( -( {q_{17}}\,{q_{23}}\,{q_{35}} )  + {q_{13}}\,{q_{27}}\,{q_{35}} + 
                 {q_{15}}\,{q_{23}}\,{q_{37}} - {q_{13}}\,{q_{25}}\,{q_{37}} )  ) \,{q_{44}},$

\vskip 0.2in $(M^{-1})^{t}_{55}=\frac{X_{55}}{Y_{55}}$ where
 
$X_{55}=39\,{q_{03}}\,{q_{17}}\,{q_{26}}\,{q_{34}} - 39\,{q_{03}}\,{q_{16}}\,{q_{27}}\,{q_{34}} + 
            39\,{q_{04}}\,{q_{17}}\,{q_{23}}\,{q_{36}} - 39\,{q_{03}}\,{q_{17}}\,{q_{24}}\,{q_{36}} - 
            39\,{q_{04}}\,{q_{13}}\,{q_{27}}\,{q_{36}} + 39\,{q_{03}}\,{q_{14}}\,{q_{27}}\,{q_{36}} - 
            39\,{q_{04}}\,{q_{16}}\,{q_{23}}\,{q_{37}} + 39\,{q_{03}}\,{q_{16}}\,{q_{24}}\,{q_{37}} + 
            39\,{q_{04}}\,{q_{13}}\,{q_{26}}\,{q_{37}} - 39\,{q_{03}}\,{q_{14}}\,{q_{26}}\,{q_{37}} - 
            460\,{q_{03}}\,{q_{17}}\,{q_{26}}\,{q_{44}} + 460\,{q_{03}}\,{q_{16}}\,{q_{27}}\,{q_{44}} + 
            5190\,{q_{03}}\,{q_{17}}\,{q_{36}}\,{q_{44}} - 
            1175799\,{q_{17}}\,{q_{23}}\,{q_{36}}\,{q_{44}} - 
            35826\,{q_{03}}\,{q_{27}}\,{q_{36}}\,{q_{44}} + 
            1175799\,{q_{13}}\,{q_{27}}\,{q_{36}}\,{q_{44}} - 
            5190\,{q_{03}}\,{q_{16}}\,{q_{37}}\,{q_{44}} + 
            1175799\,{q_{16}}\,{q_{23}}\,{q_{37}}\,{q_{44}} + 
            35826\,{q_{03}}\,{q_{26}}\,{q_{37}}\,{q_{44}} - 
            1175799\,{q_{13}}\,{q_{26}}\,{q_{37}}\,{q_{44}} + 
            {q_{07}}\,( -39\,{q_{14}}\,{q_{23}}\,{q_{36}} + 
               {q_{16}}\,{q_{23}}\,( 39\,{q_{34}} - 460\,{q_{44}} )  + 
               35826\,{q_{23}}\,{q_{36}}\,{q_{44}} + $

\noindent $ {q_{13}}\,( -39\,{q_{26}}\,{q_{34}} + 39\,{q_{24}}\,{q_{36}} + 460\,{q_{26}}\,{q_{44}} - 
                  5190\,{q_{36}}\,{q_{44}} )  )  + 
            {q_{06}}\,( 3\,{q_{23}}\,{q_{37}}\,( 13\,{q_{14}} - 11942\,{q_{44}} )  + $

\noindent$ {q_{17}}\,{q_{23}}\,( -39\,{q_{34}} + 460\,{q_{44}} )  + 
               {q_{13}}\,( 39\,{q_{27}}\,{q_{34}} - 39\,{q_{24}}\,{q_{37}} - 460\,{q_{27}}\,{q_{44}} + 
                  5190\,{q_{37}}\,{q_{44}} )  ) $

$Y_{55}=96\,
            ( {q_{03}}\,{q_{17}}\,{q_{26}}\,{q_{35}} - {q_{03}}\,{q_{16}}\,{q_{27}}\,{q_{35}} + 
              {q_{05}}\,{q_{17}}\,{q_{23}}\,{q_{36}} - {q_{03}}\,{q_{17}}\,{q_{25}}\,{q_{36}} - 
              {q_{05}}\,{q_{13}}\,{q_{27}}\,{q_{36}} + $

\noindent ${q_{03}}\,{q_{15}}\,{q_{27}}\,{q_{36}} + 
{q_{07}}\,( {q_{16}}\,{q_{23}}\,{q_{35}} - {q_{13}}\,{q_{26}}\,{q_{35}} - 
                 {q_{15}}\,{q_{23}}\,{q_{36}} + {q_{13}}\,{q_{25}}\,{q_{36}} )  - 
              {q_{05}}\,{q_{16}}\,{q_{23}}\,{q_{37}} + {q_{03}}\,{q_{16}}\,{q_{25}}\,{q_{37}} + 
              {q_{05}}\,{q_{13}}\,{q_{26}}\,{q_{37}} - {q_{03}}\,{q_{15}}\,{q_{26}}\,{q_{37}} + $

\noindent$ {q_{06}}\,( -( {q_{17}}\,{q_{23}}\,{q_{35}} )  + {q_{13}}\,{q_{27}}\,{q_{35}} + 
                 {q_{15}}\,{q_{23}}\,{q_{37}} - {q_{13}}\,{q_{25}}\,{q_{37}} )  ) \,{q_{44}},$

\vskip 0.2in $(M^{-1})^{t}_{56}=\frac{X_{56}}{Y_{56}}$ where
 
$X_{56}=3\,( -6\,{q_{03}}\,{q_{17}}\,{q_{26}}\,{q_{34}} + 
              6\,{q_{03}}\,{q_{16}}\,{q_{27}}\,{q_{34}} - 6\,{q_{04}}\,{q_{17}}\,{q_{23}}\,{q_{36}} + 
              6\,{q_{03}}\,{q_{17}}\,{q_{24}}\,{q_{36}}$

\noindent$ + 6\,{q_{04}}\,{q_{13}}\,{q_{27}}\,{q_{36}} - 
              6\,{q_{03}}\,{q_{14}}\,{q_{27}}\,{q_{36}} + 6\,{q_{04}}\,{q_{16}}\,{q_{23}}\,{q_{37}} - 
              6\,{q_{03}}\,{q_{16}}\,{q_{24}}\,{q_{37}} - 6\,{q_{04}}\,{q_{13}}\,{q_{26}}\,{q_{37}} + 
              6\,{q_{03}}\,{q_{14}}\,{q_{26}}\,{q_{37}} + 140\,{q_{03}}\,{q_{17}}\,{q_{26}}\,{q_{44}} - 
              140\,{q_{03}}\,{q_{16}}\,{q_{27}}\,{q_{44}} - 1758\,{q_{03}}\,{q_{17}}\,{q_{36}}\,{q_{44}} + 
              43359\,{q_{17}}\,{q_{23}}\,{q_{36}}\,{q_{44}} + 
              12810\,{q_{03}}\,{q_{27}}\,{q_{36}}\,{q_{44}} - 
              43359\,{q_{13}}\,{q_{27}}\,{q_{36}}\,{q_{44}} + 
              1758\,{q_{03}}\,{q_{16}}\,{q_{37}}\,{q_{44}} - 
              43359\,{q_{16}}\,{q_{23}}\,{q_{37}}\,{q_{44}} - 
              12810\,{q_{03}}\,{q_{26}}\,{q_{37}}\,{q_{44}} + 
              43359\,{q_{13}}\,{q_{26}}\,{q_{37}}\,{q_{44}} + 
              2\,{q_{07}}\,( 3\,{q_{23}}\,{q_{36}}\,( {q_{14}} - 2135\,{q_{44}} )  + 
                 {q_{16}}\,{q_{23}}\,( -3\,{q_{34}} + 70\,{q_{44}} )  + $

\noindent$ {q_{13}}\,( 3\,{q_{26}}\,{q_{34}} - 3\,{q_{24}}\,{q_{36}} - 70\,{q_{26}}\,{q_{44}} + 
                    879\,{q_{36}}\,{q_{44}} )  )  + $

\noindent$ 2\,{q_{06}}\,( -3\,{q_{23}}\,{q_{37}}\,( {q_{14}} - 2135\,{q_{44}} )  + 
                 {q_{17}}\,{q_{23}}\,( 3\,{q_{34}} - 70\,{q_{44}} )  + $

\noindent$ {q_{13}}\,( -3\,{q_{27}}\,{q_{34}} + 3\,{q_{24}}\,{q_{37}} + 70\,{q_{27}}\,{q_{44}} - 
                    879\,{q_{37}}\,{q_{44}} )  )  )$

$Y_{56}= 32\,
            ( {q_{03}}\,{q_{17}}\,{q_{26}}\,{q_{35}} - {q_{03}}\,{q_{16}}\,{q_{27}}\,{q_{35}} + 
              {q_{05}}\,{q_{17}}\,{q_{23}}\,{q_{36}} - {q_{03}}\,{q_{17}}\,{q_{25}}\,{q_{36}} - $

\noindent$ {q_{05}}\,{q_{13}}\,{q_{27}}\,{q_{36}} + {q_{03}}\,{q_{15}}\,{q_{27}}\,{q_{36}} + 
              {q_{07}}\,( {q_{16}}\,{q_{23}}\,{q_{35}} - {q_{13}}\,{q_{26}}\,{q_{35}} - 
                 {q_{15}}\,{q_{23}}\,{q_{36}} + {q_{13}}\,{q_{25}}\,{q_{36}} )  - 
              {q_{05}}\,{q_{16}}\,{q_{23}}\,{q_{37}} + {q_{03}}\,{q_{16}}\,{q_{25}}\,{q_{37}} + 
              {q_{05}}\,{q_{13}}\,{q_{26}}\,{q_{37}} - {q_{03}}\,{q_{15}}\,{q_{26}}\,{q_{37}} + $

\noindent$
              {q_{06}}\,( -( {q_{17}}\,{q_{23}}\,{q_{35}} )  + {q_{13}}\,{q_{27}}\,{q_{35}} + 
                 {q_{15}}\,{q_{23}}\,{q_{37}} - {q_{13}}\,{q_{25}}\,{q_{37}} )  ) \,{q_{44}},$

\vskip 0.2in $(M^{-1})^{t}_{57}=\frac{X_{57}}{Y_{57}}$ where
 
$X_{57}=-3\,( -2\,{q_{03}}\,{q_{17}}\,{q_{26}}\,{q_{34}} + 
              2\,{q_{03}}\,{q_{16}}\,{q_{27}}\,{q_{34}} - 2\,{q_{04}}\,{q_{17}}\,{q_{23}}\,{q_{36}} + 
              2\,{q_{03}}\,{q_{17}}\,{q_{24}}\,{q_{36}} +$

\noindent$ 2\,{q_{04}}\,{q_{13}}\,{q_{27}}\,{q_{36}} - 
              2\,{q_{03}}\,{q_{14}}\,{q_{27}}\,{q_{36}} + 2\,{q_{04}}\,{q_{16}}\,{q_{23}}\,{q_{37}} - 
              2\,{q_{03}}\,{q_{16}}\,{q_{24}}\,{q_{37}} - 2\,{q_{04}}\,{q_{13}}\,{q_{26}}\,{q_{37}} + 
              2\,{q_{03}}\,{q_{14}}\,{q_{26}}\,{q_{37}} + 52\,{q_{03}}\,{q_{17}}\,{q_{26}}\,{q_{44}} - 
              52\,{q_{03}}\,{q_{16}}\,{q_{27}}\,{q_{44}} - 690\,{q_{03}}\,{q_{17}}\,{q_{36}}\,{q_{44}} + 
              17913\,{q_{17}}\,{q_{23}}\,{q_{36}}\,{q_{44}} + 
              5190\,{q_{03}}\,{q_{27}}\,{q_{36}}\,{q_{44}} - 
              17913\,{q_{13}}\,{q_{27}}\,{q_{36}}\,{q_{44}} + 690\,{q_{03}}\,{q_{16}}\,{q_{37}}\,{q_{44}} - 
              17913\,{q_{16}}\,{q_{23}}\,{q_{37}}\,{q_{44}} - 
              5190\,{q_{03}}\,{q_{26}}\,{q_{37}}\,{q_{44}} + 
              17913\,{q_{13}}\,{q_{26}}\,{q_{37}}\,{q_{44}} + 
              2\,{q_{07}}\,( {q_{23}}\,{q_{36}}\,( {q_{14}} - 2595\,{q_{44}} )  - 
                 {q_{16}}\,{q_{23}}\,( {q_{34}} - 26\,{q_{44}} )  + $

\noindent$ {q_{13}}\,( {q_{26}}\,{q_{34}} - {q_{24}}\,{q_{36}} - 26\,{q_{26}}\,{q_{44}} + 
                    345\,{q_{36}}\,{q_{44}} )  )  + $

\noindent$ 2\,{q_{06}}\,( -( {q_{23}}\,{q_{37}}\,( {q_{14}} - 2595\,{q_{44}} )  )  + 
                 {q_{17}}\,{q_{23}}\,( {q_{34}} - 26\,{q_{44}} )  + $

\noindent$ {q_{13}}\,( -( {q_{27}}\,{q_{34}} )  + {q_{24}}\,{q_{37}} + 
                    26\,{q_{27}}\,{q_{44}} - 345\,{q_{37}}\,{q_{44}} )  )  )$

$Y_{57}=16\,
            ( {q_{03}}\,{q_{17}}\,{q_{26}}\,{q_{35}} - {q_{03}}\,{q_{16}}\,{q_{27}}\,{q_{35}} + 
              {q_{05}}\,{q_{17}}\,{q_{23}}\,{q_{36}} - {q_{03}}\,{q_{17}}\,{q_{25}}\,{q_{36}} - 
              {q_{05}}\,{q_{13}}\,{q_{27}}\,{q_{36}} $

\noindent $+ {q_{03}}\,{q_{15}}\,{q_{27}}\,{q_{36}} + 
              {q_{07}}\,( {q_{16}}\,{q_{23}}\,{q_{35}} - {q_{13}}\,{q_{26}}\,{q_{35}} - 
                 {q_{15}}\,{q_{23}}\,{q_{36}} + {q_{13}}\,{q_{25}}\,{q_{36}} )  - 
              {q_{05}}\,{q_{16}}\,{q_{23}}\,{q_{37}} + {q_{03}}\,{q_{16}}\,{q_{25}}\,{q_{37}} + 
              {q_{05}}\,{q_{13}}\,{q_{26}}\,{q_{37}} - {q_{03}}\,{q_{15}}\,{q_{26}}\,{q_{37}} + $

\noindent$ {q_{06}}\,( -( {q_{17}}\,{q_{23}}\,{q_{35}} )  + {q_{13}}\,{q_{27}}\,{q_{35}} + 
                 {q_{15}}\,{q_{23}}\,{q_{37}} - {q_{13}}\,{q_{25}}\,{q_{37}} )  ) \,{q_{44}},$

\vskip 0.2in $(M^{-1})^{t}_{60}=\frac{X_{60}}{Y_{60}}$ where
 
$X_{60}=-( {q_{07}}\,{q_{10}}\,{q_{23}}\,{q_{35}} )  + 
            ( -1 + {q_{00}} ) \,{q_{17}}\,{q_{23}}\,{q_{35}} + 
            {q_{03}}\,{q_{10}}\,{q_{27}}\,{q_{35}} + {q_{13}}\,{q_{27}}\,{q_{35}} - 
            {q_{00}}\,{q_{13}}\,{q_{27}}\,{q_{35}} + {q_{05}}\,{q_{10}}\,{q_{23}}\,{q_{37}} + 
            {q_{15}}\,{q_{23}}\,{q_{37}} - {q_{00}}\,{q_{15}}\,{q_{23}}\,{q_{37}} - 
            {q_{03}}\,{q_{10}}\,{q_{25}}\,{q_{37}} - {q_{13}}\,{q_{25}}\,{q_{37}} + 
            {q_{00}}\,{q_{13}}\,{q_{25}}\,{q_{37}}$

$Y_{60}={q_{03}}\,{q_{17}}\,{q_{26}}\,{q_{35}} - 
            {q_{03}}\,{q_{16}}\,{q_{27}}\,{q_{35}} + {q_{05}}\,{q_{17}}\,{q_{23}}\,{q_{36}} - 
            {q_{03}}\,{q_{17}}\,{q_{25}}\,{q_{36}} - {q_{05}}\,{q_{13}}\,{q_{27}}\,{q_{36}} + 
            {q_{03}}\,{q_{15}}\,{q_{27}}\,{q_{36}} + 
            {q_{07}}\,( {q_{16}}\,{q_{23}}\,{q_{35}} - {q_{13}}\,{q_{26}}\,{q_{35}} - 
               {q_{15}}\,{q_{23}}\,{q_{36}} + {q_{13}}\,{q_{25}}\,{q_{36}} )  - 
            {q_{05}}\,{q_{16}}\,{q_{23}}\,{q_{37}} + {q_{03}}\,{q_{16}}\,{q_{25}}\,{q_{37}} + 
            {q_{05}}\,{q_{13}}\,{q_{26}}\,{q_{37}} - {q_{03}}\,{q_{15}}\,{q_{26}}\,{q_{37}} + $

\noindent $ {q_{06}}\,( -( {q_{17}}\,{q_{23}}\,{q_{35}} )  + {q_{13}}\,{q_{27}}\,{q_{35}} + 
               {q_{15}}\,{q_{23}}\,{q_{37}} - {q_{13}}\,{q_{25}}\,{q_{37}} ),$

\vskip 0.2in $(M^{-1})^{t}_{61}=\frac{X_{61}}{Y_{61}}$ where
 
$X_{61}={q_{01}}\,{q_{17}}\,{q_{23}}\,{q_{35}} + 
            {q_{07}}\,( {q_{13}}\,{q_{21}} + {q_{23}} - {q_{11}}\,{q_{23}} ) \,{q_{35}} - 
            {q_{01}}\,{q_{13}}\,{q_{27}}\,{q_{35}} - {q_{05}}\,{q_{13}}\,{q_{21}}\,{q_{37}} - 
            {q_{05}}\,{q_{23}}\,{q_{37}} + {q_{05}}\,{q_{11}}\,{q_{23}}\,{q_{37}} - 
            {q_{01}}\,{q_{15}}\,{q_{23}}\,{q_{37}} + {q_{01}}\,{q_{13}}\,{q_{25}}\,{q_{37}} + 
            {q_{03}}\,( -( {q_{17}}\,{q_{21}}\,{q_{35}} )  + 
               ( -1 + {q_{11}} ) \,{q_{27}}\,{q_{35}} + 
               ( {q_{15}}\,{q_{21}} + {q_{25}} - {q_{11}}\,{q_{25}} ) \,{q_{37}} )$

$Y_{61}={q_{03}}\,
             {q_{17}}\,{q_{26}}\,{q_{35}} - {q_{03}}\,{q_{16}}\,{q_{27}}\,{q_{35}} + 
            {q_{05}}\,{q_{17}}\,{q_{23}}\,{q_{36}} - {q_{03}}\,{q_{17}}\,{q_{25}}\,{q_{36}} - 
            {q_{05}}\,{q_{13}}\,{q_{27}}\,{q_{36}} + {q_{03}}\,{q_{15}}\,{q_{27}}\,{q_{36}} + 
            {q_{07}}\,( {q_{16}}\,{q_{23}}\,{q_{35}} - {q_{13}}\,{q_{26}}\,{q_{35}} - 
               {q_{15}}\,{q_{23}}\,{q_{36}} + {q_{13}}\,{q_{25}}\,{q_{36}} )  - 
            {q_{05}}\,{q_{16}}\,{q_{23}}\,{q_{37}} + {q_{03}}\,{q_{16}}\,{q_{25}}\,{q_{37}} + 
            {q_{05}}\,{q_{13}}\,{q_{26}}\,{q_{37}} - {q_{03}}\,{q_{15}}\,{q_{26}}\,{q_{37}} + $

\noindent $ {q_{06}}\,( -( {q_{17}}\,{q_{23}}\,{q_{35}} )  + {q_{13}}\,{q_{27}}\,{q_{35}} + 
               {q_{15}}\,{q_{23}}\,{q_{37}} - {q_{13}}\,{q_{25}}\,{q_{37}} ),$

\vskip 0.2in $(M^{-1})^{t}_{62}=\frac{X_{62}}{Y_{62}}$ where
 
$X_{62}=-2\,{q_{02}}\,{q_{17}}\,{q_{23}}\,{q_{35}} + 
            {q_{07}}\,( {q_{13}} - 2\,{q_{13}}\,{q_{22}} + 2\,{q_{12}}\,{q_{23}} ) \,{q_{35}} + 
            2\,{q_{02}}\,{q_{13}}\,{q_{27}}\,{q_{35}} - {q_{05}}\,{q_{13}}\,{q_{37}} + 
            2\,{q_{05}}\,{q_{13}}\,{q_{22}}\,{q_{37}} - 2\,{q_{05}}\,{q_{12}}\,{q_{23}}\,{q_{37}} + 
            2\,{q_{02}}\,{q_{15}}\,{q_{23}}\,{q_{37}} - 2\,{q_{02}}\,{q_{13}}\,{q_{25}}\,{q_{37}} + 
            {q_{03}}\,( {q_{17}}\,( -1 + 2\,{q_{22}} ) \,{q_{35}} - 
               2\,{q_{12}}\,{q_{27}}\,{q_{35}} + {q_{15}}\,{q_{37}} - 2\,{q_{15}}\,{q_{22}}\,{q_{37}} + 
               2\,{q_{12}}\,{q_{25}}\,{q_{37}} )$

$Y_{62}= 2\,
            ( -( {q_{03}}\,{q_{17}}\,{q_{26}}\,{q_{35}} )  + 
              {q_{03}}\,{q_{16}}\,{q_{27}}\,{q_{35}} - {q_{05}}\,{q_{17}}\,{q_{23}}\,{q_{36}} + 
              {q_{03}}\,{q_{17}}\,{q_{25}}\,{q_{36}} $

\noindent $ + {q_{05}}\,{q_{13}}\,{q_{27}}\,{q_{36}} - 
              {q_{03}}\,{q_{15}}\,{q_{27}}\,{q_{36}} + 
              {q_{07}}\,( -( {q_{16}}\,{q_{23}}\,{q_{35}} )  + {q_{13}}\,{q_{26}}\,{q_{35}} + 
                 {q_{15}}\,{q_{23}}\,{q_{36}} - {q_{13}}\,{q_{25}}\,{q_{36}} )  + 
              {q_{05}}\,{q_{16}}\,{q_{23}}\,{q_{37}} - {q_{03}}\,{q_{16}}\,{q_{25}}\,{q_{37}} - 
              {q_{05}}\,{q_{13}}\,{q_{26}}\,{q_{37}} + {q_{03}}\,{q_{15}}\,{q_{26}}\,{q_{37}} + $

\noindent$ {q_{06}}\,( {q_{17}}\,{q_{23}}\,{q_{35}} - {q_{13}}\,{q_{27}}\,{q_{35}} - 
                 {q_{15}}\,{q_{23}}\,{q_{37}} + {q_{13}}\,{q_{25}}\,{q_{37}} )  ),$

\vskip 0.2in $(M^{-1})^{t}_{63}=\frac{X_{63}}{Y_{63}}$ where
 
$X_{63}={q_{07}}\,{q_{15}}\,{q_{23}} - {q_{05}}\,{q_{17}}\,{q_{23}} - 
            {q_{07}}\,{q_{13}}\,{q_{25}} + {q_{03}}\,{q_{17}}\,{q_{25}} + 
{q_{05}}\,{q_{13}}\,{q_{27}} - {q_{03}}\,{q_{15}}\,{q_{27}}$

$Y_{63}=6\,
            ( -( {q_{03}}\,{q_{17}}\,{q_{26}}\,{q_{35}} )  + 
              {q_{03}}\,{q_{16}}\,{q_{27}}\,{q_{35}} - {q_{05}}\,{q_{17}}\,{q_{23}}\,{q_{36}} + 
              {q_{03}}\,{q_{17}}\,{q_{25}}\,{q_{36}} +$

\noindent $ {q_{05}}\,{q_{13}}\,{q_{27}}\,{q_{36}} - 
              {q_{03}}\,{q_{15}}\,{q_{27}}\,{q_{36}} + 
              {q_{07}}\,( -( {q_{16}}\,{q_{23}}\,{q_{35}} )  + {q_{13}}\,{q_{26}}\,{q_{35}} + 
                 {q_{15}}\,{q_{23}}\,{q_{36}} - {q_{13}}\,{q_{25}}\,{q_{36}} )  + 
              {q_{05}}\,{q_{16}}\,{q_{23}}\,{q_{37}} - {q_{03}}\,{q_{16}}\,{q_{25}}\,{q_{37}} - 
              {q_{05}}\,{q_{13}}\,{q_{26}}\,{q_{37}} + {q_{03}}\,{q_{15}}\,{q_{26}}\,{q_{37}} + $

\noindent$ {q_{06}}\,( {q_{17}}\,{q_{23}}\,{q_{35}} - {q_{13}}\,{q_{27}}\,{q_{35}} - 
                 {q_{15}}\,{q_{23}}\,{q_{37}} + {q_{13}}\,{q_{25}}\,{q_{37}} )  ),$

\vskip 0.2in $(M^{-1})^{t}_{64}=\frac{X_{64}}{Y_{64}}$ where
 
$X_{64}=-( {q_{03}}\,{q_{17}}\,{q_{25}}\,{q_{34}} )  + 
            {q_{03}}\,{q_{15}}\,{q_{27}}\,{q_{34}} - {q_{04}}\,{q_{17}}\,{q_{23}}\,{q_{35}} + 
            {q_{03}}\,{q_{17}}\,{q_{24}}\,{q_{35}} + {q_{04}}\,{q_{13}}\,{q_{27}}\,{q_{35}} - 
            {q_{03}}\,{q_{14}}\,{q_{27}}\,{q_{35}} + 
            {q_{07}}\,( -( {q_{15}}\,{q_{23}}\,{q_{34}} )  + {q_{13}}\,{q_{25}}\,{q_{34}} + 
               {q_{14}}\,{q_{23}}\,{q_{35}} - {q_{13}}\,{q_{24}}\,{q_{35}} )  + 
            {q_{04}}\,{q_{15}}\,{q_{23}}\,{q_{37}} - {q_{03}}\,{q_{15}}\,{q_{24}}\,{q_{37}} - 
            {q_{04}}\,{q_{13}}\,{q_{25}}\,{q_{37}} + {q_{03}}\,{q_{14}}\,{q_{25}}\,{q_{37}} + $

\noindent$ {q_{05}}\,( {q_{17}}\,{q_{23}}\,{q_{34}} - {q_{13}}\,{q_{27}}\,{q_{34}} - 
               {q_{14}}\,{q_{23}}\,{q_{37}} + {q_{13}}\,{q_{24}}\,{q_{37}} )$

$Y_{64}=24\,
            ( -( {q_{03}}\,{q_{17}}\,{q_{26}}\,{q_{35}} )  + 
              {q_{03}}\,{q_{16}}\,{q_{27}}\,{q_{35}} - {q_{05}}\,{q_{17}}\,{q_{23}}\,{q_{36}} + 
              {q_{03}}\,{q_{17}}\,{q_{25}}\,{q_{36}} $

\noindent
$+ {q_{05}}\,{q_{13}}\,{q_{27}}\,{q_{36}} - 
              {q_{03}}\,{q_{15}}\,{q_{27}}\,{q_{36}} + 
              {q_{07}}\,( -( {q_{16}}\,{q_{23}}\,{q_{35}} )  + {q_{13}}\,{q_{26}}\,{q_{35}} + 
                 {q_{15}}\,{q_{23}}\,{q_{36}} - {q_{13}}\,{q_{25}}\,{q_{36}} )  + 
              {q_{05}}\,{q_{16}}\,{q_{23}}\,{q_{37}} - {q_{03}}\,{q_{16}}\,{q_{25}}\,{q_{37}} - 
              {q_{05}}\,{q_{13}}\,{q_{26}}\,{q_{37}} + {q_{03}}\,{q_{15}}\,{q_{26}}\,{q_{37}} + $

\noindent$ {q_{06}}\,( {q_{17}}\,{q_{23}}\,{q_{35}} - {q_{13}}\,{q_{27}}\,{q_{35}} - 
                 {q_{15}}\,{q_{23}}\,{q_{37}} + {q_{13}}\,{q_{25}}\,{q_{37}} )  ) \,{q_{44}},$

\vskip 0.2in $(M^{-1})^{t}_{65}=\frac{X_{65}}{Y_{65}}$ where 

$X_{65}=-39\,{q_{03}}\,{q_{17}}\,{q_{25}}\,{q_{34}} + 39\,{q_{03}}\,{q_{15}}\,{q_{27}}\,{q_{34}} - 
            39\,{q_{04}}\,{q_{17}}\,{q_{23}}\,{q_{35}} + 39\,{q_{03}}\,{q_{17}}\,{q_{24}}\,{q_{35}} + 
            39\,{q_{04}}\,{q_{13}}\,{q_{27}}\,{q_{35}} - 39\,{q_{03}}\,{q_{14}}\,{q_{27}}\,{q_{35}} + 
            39\,{q_{04}}\,{q_{15}}\,{q_{23}}\,{q_{37}} - 39\,{q_{03}}\,{q_{15}}\,{q_{24}}\,{q_{37}} - 
            39\,{q_{04}}\,{q_{13}}\,{q_{25}}\,{q_{37}} + 39\,{q_{03}}\,{q_{14}}\,{q_{25}}\,{q_{37}} + 
            460\,{q_{03}}\,{q_{17}}\,{q_{25}}\,{q_{44}} - 460\,{q_{03}}\,{q_{15}}\,{q_{27}}\,{q_{44}} - 
            5190\,{q_{03}}\,{q_{17}}\,{q_{35}}\,{q_{44}} + 
            1175799\,{q_{17}}\,{q_{23}}\,{q_{35}}\,{q_{44}} + 
            35826\,{q_{03}}\,{q_{27}}\,{q_{35}}\,{q_{44}} - 
            1175799\,{q_{13}}\,{q_{27}}\,{q_{35}}\,{q_{44}} + 
            5190\,{q_{03}}\,{q_{15}}\,{q_{37}}\,{q_{44}} - 
            1175799\,{q_{15}}\,{q_{23}}\,{q_{37}}\,{q_{44}} - 
            35826\,{q_{03}}\,{q_{25}}\,{q_{37}}\,{q_{44}} + 
            1175799\,{q_{13}}\,{q_{25}}\,{q_{37}}\,{q_{44}} + 
            {q_{07}}\,( 3\,{q_{23}}\,{q_{35}}\,( 13\,{q_{14}} - 11942\,{q_{44}} )  + $

\noindent$ {q_{15}}\,{q_{23}}\,( -39\,{q_{34}} + 460\,{q_{44}} )  + 
               {q_{13}}\,( 39\,{q_{25}}\,{q_{34}} - 39\,{q_{24}}\,{q_{35}} - 460\,{q_{25}}\,{q_{44}} + 
                  5190\,{q_{35}}\,{q_{44}} )  )  + 
            {q_{05}}\,( -39\,{q_{14}}\,{q_{23}}\,{q_{37}} + 
               {q_{17}}\,{q_{23}}\,( 39\,{q_{34}} - 460\,{q_{44}} )  + 
               35826\,{q_{23}}\,{q_{37}}\,{q_{44}} + $

\noindent$ {q_{13}}\,( -39\,{q_{27}}\,{q_{34}} + 39\,{q_{24}}\,{q_{37}} + 460\,{q_{27}}\,{q_{44}} - 
                  5190\,{q_{37}}\,{q_{44}} )  )$ 

$Y_{65}=96\,
            ( {q_{03}}\,{q_{17}}\,{q_{26}}\,{q_{35}} - {q_{03}}\,{q_{16}}\,{q_{27}}\,{q_{35}} + 
              {q_{05}}\,{q_{17}}\,{q_{23}}\,{q_{36}} - {q_{03}}\,{q_{17}}\,{q_{25}}\,{q_{36}} - 
              {q_{05}}\,{q_{13}}\,{q_{27}}\,{q_{36}} + $

\noindent$ {q_{03}}\,{q_{15}}\,{q_{27}}\,{q_{36}} + 
              {q_{07}}\,( {q_{16}}\,{q_{23}}\,{q_{35}} - {q_{13}}\,{q_{26}}\,{q_{35}} - 
                 {q_{15}}\,{q_{23}}\,{q_{36}} + {q_{13}}\,{q_{25}}\,{q_{36}} )  - 
              {q_{05}}\,{q_{16}}\,{q_{23}}\,{q_{37}} + {q_{03}}\,{q_{16}}\,{q_{25}}\,{q_{37}} + 
              {q_{05}}\,{q_{13}}\,{q_{26}}\,{q_{37}} - {q_{03}}\,{q_{15}}\,{q_{26}}\,{q_{37}} + $

\noindent$ {q_{06}}\,( -( {q_{17}}\,{q_{23}}\,{q_{35}} )  + {q_{13}}\,{q_{27}}\,{q_{35}} + 
                 {q_{15}}\,{q_{23}}\,{q_{37}} - {q_{13}}\,{q_{25}}\,{q_{37}} )  ) \,{q_{44}},$

$
(M^{-1})^{t}_{66}=\frac{X_{66}}{Y_{66}}$ where 

$X_{66}=3\,( 6\,{q_{03}}\,{q_{17}}\,{q_{25}}\,{q_{34}} - 
              6\,{q_{03}}\,{q_{15}}\,{q_{27}}\,{q_{34}} + 6\,{q_{04}}\,{q_{17}}\,{q_{23}}\,{q_{35}} - 
              6\,{q_{03}}\,{q_{17}}\,{q_{24}}\,{q_{35}} - $

\noindent$
6\,{q_{04}}\,{q_{13}}\,{q_{27}}\,{q_{35}} + 
              6\,{q_{03}}\,{q_{14}}\,{q_{27}}\,{q_{35}} - 6\,{q_{04}}\,{q_{15}}\,{q_{23}}\,{q_{37}} + 
              6\,{q_{03}}\,{q_{15}}\,{q_{24}}\,{q_{37}} + 6\,{q_{04}}\,{q_{13}}\,{q_{25}}\,{q_{37}} - 
              6\,{q_{03}}\,{q_{14}}\,{q_{25}}\,{q_{37}} - 140\,{q_{03}}\,{q_{17}}\,{q_{25}}\,{q_{44}} + 
              140\,{q_{03}}\,{q_{15}}\,{q_{27}}\,{q_{44}} + 1758\,{q_{03}}\,{q_{17}}\,{q_{35}}\,{q_{44}} - 
              43359\,{q_{17}}\,{q_{23}}\,{q_{35}}\,{q_{44}} - 
              12810\,{q_{03}}\,{q_{27}}\,{q_{35}}\,{q_{44}} + 
              43359\,{q_{13}}\,{q_{27}}\,{q_{35}}\,{q_{44}} - 
              1758\,{q_{03}}\,{q_{15}}\,{q_{37}}\,{q_{44}} + 
              43359\,{q_{15}}\,{q_{23}}\,{q_{37}}\,{q_{44}} + 
              12810\,{q_{03}}\,{q_{25}}\,{q_{37}}\,{q_{44}} - 
              43359\,{q_{13}}\,{q_{25}}\,{q_{37}}\,{q_{44}} + 
              2\,{q_{07}}\,( -3\,{q_{23}}\,{q_{35}}\,( {q_{14}} - 2135\,{q_{44}} )  + 
                 {q_{15}}\,{q_{23}}\,( 3\,{q_{34}} - 70\,{q_{44}} )  + $

\noindent$
                 {q_{13}}\,( -3\,{q_{25}}\,{q_{34}} + 3\,{q_{24}}\,{q_{35}} + 70\,{q_{25}}\,{q_{44}} - 
                    879\,{q_{35}}\,{q_{44}} )  )  +  
2\,{q_{05}}\,( 3\,{q_{23}}\,{q_{37}}\,( {q_{14}} - 2135\,{q_{44}} )  + $

\noindent$ {q_{17}}\,{q_{23}}\,( -3\,{q_{34}} + 70\,{q_{44}} )  + 
{q_{13}}\,( 3\,{q_{27}}\,{q_{34}} - 3\,{q_{24}}\,{q_{37}} - 70\,{q_{27}}\,{q_{44}} + 
                    879\,{q_{37}}\,{q_{44}} )  )  )$ 

$Y_{66}=32\,
            ( {q_{03}}\,{q_{17}}\,{q_{26}}\,{q_{35}} - {q_{03}}\,{q_{16}}\,{q_{27}}\,{q_{35}} + 
              {q_{05}}\,{q_{17}}\,{q_{23}}\,{q_{36}} - {q_{03}}\,{q_{17}}\,{q_{25}}\,{q_{36}} - 
              {q_{05}}\,{q_{13}}\,{q_{27}}\,{q_{36}} +$

\noindent$ {q_{03}}\,{q_{15}}\,{q_{27}}\,{q_{36}} + 
{q_{07}}\,( {q_{16}}\,{q_{23}}\,{q_{35}} - {q_{13}}\,{q_{26}}\,{q_{35}} - 
                 {q_{15}}\,{q_{23}}\,{q_{36}} + {q_{13}}\,{q_{25}}\,{q_{36}} )  - 
              {q_{05}}\,{q_{16}}\,{q_{23}}\,{q_{37}} + 
{q_{03}}\,{q_{16}}\,{q_{25}}\,{q_{37}} + 
              {q_{05}}\,{q_{13}}\,{q_{26}}\,{q_{37}} - {q_{03}}\,{q_{15}}\,{q_{26}}\,{q_{37}} + 
              {q_{06}}\,( -( {q_{17}}\,{q_{23}}\,{q_{35}} )  + {q_{13}}\,{q_{27}}\,{q_{35}} + $

\noindent$ {q_{15}}\,{q_{23}}\,{q_{37}} - {q_{13}}\,{q_{25}}\,{q_{37}} )  ) \,{q_{44}},$
      
\vskip 0.2in $(M^{-1})^{t}_{67}=\frac{X_{67}}{Y_{67}}$ where
 
$X_{67}=3\,( -2\,{q_{03}}\,{q_{17}}\,{q_{25}}\,{q_{34}} + 
              2\,{q_{03}}\,{q_{15}}\,{q_{27}}\,{q_{34}} - 2\,{q_{04}}\,{q_{17}}\,{q_{23}}\,{q_{35}} + 
              2\,{q_{03}}\,{q_{17}}\,{q_{24}}\,{q_{35}} + $

\noindent$
2\,{q_{04}}\,{q_{13}}\,{q_{27}}\,{q_{35}} - 
              2\,{q_{03}}\,{q_{14}}\,{q_{27}}\,{q_{35}} + 2\,{q_{04}}\,{q_{15}}\,{q_{23}}\,{q_{37}} - 
              2\,{q_{03}}\,{q_{15}}\,{q_{24}}\,{q_{37}} - 2\,{q_{04}}\,{q_{13}}\,{q_{25}}\,{q_{37}} + 
              2\,{q_{03}}\,{q_{14}}\,{q_{25}}\,{q_{37}} + 52\,{q_{03}}\,{q_{17}}\,{q_{25}}\,{q_{44}} - 
              52\,{q_{03}}\,{q_{15}}\,{q_{27}}\,{q_{44}} - 690\,{q_{03}}\,{q_{17}}\,{q_{35}}\,{q_{44}} + 
              17913\,{q_{17}}\,{q_{23}}\,{q_{35}}\,{q_{44}} + 
              5190\,{q_{03}}\,{q_{27}}\,{q_{35}}\,{q_{44}} - 
              17913\,{q_{13}}\,{q_{27}}\,{q_{35}}\,{q_{44}} + 690\,{q_{03}}\,{q_{15}}\,{q_{37}}\,{q_{44}} - 
              17913\,{q_{15}}\,{q_{23}}\,{q_{37}}\,{q_{44}} - 
              5190\,{q_{03}}\,{q_{25}}\,{q_{37}}\,{q_{44}} + 
              17913\,{q_{13}}\,{q_{25}}\,{q_{37}}\,{q_{44}} + 
              2\,{q_{07}}\,( {q_{23}}\,{q_{35}}\,( {q_{14}} - 2595\,{q_{44}} )  - 
                 {q_{15}}\,{q_{23}}\,( {q_{34}} - 26\,{q_{44}} )  + $

\noindent${q_{13}}\,( {q_{25}}\,{q_{34}} - {q_{24}}\,{q_{35}} - 26\,{q_{25}}\,{q_{44}} + 
                    345\,{q_{35}}\,{q_{44}} )  )  + 
              2\,{q_{05}}\,( -( {q_{23}}\,{q_{37}}\,( {q_{14}} - 2595\,{q_{44}} )  )  + $

\noindent$ {q_{17}}\,{q_{23}}\,( {q_{34}} - 26\,{q_{44}} )  + 
                 {q_{13}}\,( -( {q_{27}}\,{q_{34}} )  + {q_{24}}\,{q_{37}} + 
                    26\,{q_{27}}\,{q_{44}} - 345\,{q_{37}}\,{q_{44}} )  )  )$

$Y_{67}= 16\,
            ( {q_{03}}\,{q_{17}}\,{q_{26}}\,{q_{35}} - {q_{03}}\,{q_{16}}\,{q_{27}}\,{q_{35}} + 
              {q_{05}}\,{q_{17}}\,{q_{23}}\,{q_{36}} - {q_{03}}\,{q_{17}}\,{q_{25}}\,{q_{36}} - 
              {q_{05}}\,{q_{13}}\,{q_{27}}\,{q_{36}} $

\noindent$
+ {q_{03}}\,{q_{15}}\,{q_{27}}\,{q_{36}} + 
              {q_{07}}\,( {q_{16}}\,{q_{23}}\,{q_{35}} - {q_{13}}\,{q_{26}}\,{q_{35}} - 
                 {q_{15}}\,{q_{23}}\,{q_{36}} + {q_{13}}\,{q_{25}}\,{q_{36}} )  - 
              {q_{05}}\,{q_{16}}\,{q_{23}}\,{q_{37}} + {q_{03}}\,{q_{16}}\,{q_{25}}\,{q_{37}} + 
              {q_{05}}\,{q_{13}}\,{q_{26}}\,{q_{37}} - {q_{03}}\,{q_{15}}\,{q_{26}}\,{q_{37}} + $

\noindent$ {q_{06}}\,( -( {q_{17}}\,{q_{23}}\,{q_{35}} )  + {q_{13}}\,{q_{27}}\,{q_{35}} + 
                 {q_{15}}\,{q_{23}}\,{q_{37}} - {q_{13}}\,{q_{25}}\,{q_{37}} )  ) \,{q_{44}},$

\vskip 0.2in $(M^{-1})^{t}_{70}=\frac{X_{70}}{Y_{70}}$ where
 
$X_{70}=-( {q_{23}}\,{q_{35}}\,( -( {q_{10}}\,
                   ( {q_{06}}\,{q_{23}}\,{q_{35}} - {q_{03}}\,{q_{26}}\,{q_{35}} - 
                     {q_{05}}\,{q_{23}}\,{q_{36}} + {q_{03}}\,{q_{25}}\,{q_{36}} )  )  + $

\noindent$ ( -1 + {q_{00}} ) \,( {q_{16}}\,{q_{23}}\,{q_{35}} - 
                   {q_{13}}\,{q_{26}}\,{q_{35}} - {q_{15}}\,{q_{23}}\,{q_{36}} + 
                   {q_{13}}\,{q_{25}}\,{q_{36}} )  )$

$Y_{70}= ( {q_{16}}\,{q_{23}}\,{q_{35}} - 
                 {q_{13}}\,{q_{26}}\,{q_{35}} - {q_{15}}\,{q_{23}}\,{q_{36}} + 
                 {q_{13}}\,{q_{25}}\,{q_{36}} ) \, $

\noindent$\times ( {q_{07}}\,{q_{23}}\,{q_{35}} - {q_{03}}\,{q_{27}}\,{q_{35}} - 
                 {q_{05}}\,{q_{23}}\,{q_{37}} + {q_{03}}\,{q_{25}}\,{q_{37}} )  - $

\noindent$ ( {q_{06}}\,{q_{23}}\,{q_{35}} - {q_{03}}\,{q_{26}}\,{q_{35}} - 
                 {q_{05}}\,{q_{23}}\,{q_{36}} + {q_{03}}\,{q_{25}}\,{q_{36}} ) \,$

\noindent$\times ( {q_{17}}\,{q_{23}}\,{q_{35}} - {q_{13}}\,{q_{27}}\,{q_{35}} - 
                 {q_{15}}\,{q_{23}}\,{q_{37}} + {q_{13}}\,{q_{25}}\,{q_{37}} )  )$,

\vskip 0.2in $(M^{-1})^{t}_{71}=\frac{X_{71}}{Y_{71}}$ where
 
$X_{71}=-(-( ( {q_{13}}\,{q_{21}} + {q_{23}} - {q_{11}}\,{q_{23}} ) \,{q_{35}}\,
                 ( {q_{06}}\,{q_{23}}\,{q_{35}} - {q_{03}}\,{q_{26}}\,{q_{35}} - 
                   {q_{05}}\,{q_{23}}\,{q_{36}} + {q_{03}}\,{q_{25}}\,{q_{36}} )  )   $

\noindent$+ ( {q_{03}}\,{q_{21}} - {q_{01}}\,{q_{23}} ) \,{q_{35}}\,
               ( {q_{16}}\,{q_{23}}\,{q_{35}} - {q_{13}}\,{q_{26}}\,{q_{35}} - 
                 {q_{15}}\,{q_{23}}\,{q_{36}} + {q_{13}}\,{q_{25}}\,{q_{36}} )$

$Y_{71}= -( 
                 ( {q_{16}}\,{q_{23}}\,{q_{35}} - {q_{13}}\,{q_{26}}\,{q_{35}} - 
                   {q_{15}}\,{q_{23}}\,{q_{36}} + {q_{13}}\,{q_{25}}\,{q_{36}} ) \,$

\noindent$\times ( {q_{07}}\,{q_{23}}\,{q_{35}} - {q_{03}}\,{q_{27}}\,{q_{35}} - 
                   {q_{05}}\,{q_{23}}\,{q_{37}} + {q_{03}}\,{q_{25}}\,{q_{37}} )  )  + $

$\noindent ( {q_{06}}\,{q_{23}}\,{q_{35}} - {q_{03}}\,{q_{26}}\,{q_{35}} - 
                 {q_{05}}\,{q_{23}}\,{q_{36}} + {q_{03}}\,{q_{25}}\,{q_{36}} ) \,$

\noindent$\times
               ( {q_{17}}\,{q_{23}}\,{q_{35}} - {q_{13}}\,{q_{27}}\,{q_{35}} - 
                 {q_{15}}\,{q_{23}}\,{q_{37}} + {q_{13}}\,{q_{25}}\,{q_{37}} )  ),$

\vskip 0.2in $(M^{-1})^{t}_{72}=\frac{X_{72}}{Y_{72}}$ where

$X_{72}= -( {q_{35}}\,( -( ( {q_{13}}\,( -1 + 2\,{q_{22}} )  - 2\,{q_{12}}\,{q_{23}} )
                     \,( {q_{06}}\,{q_{23}}\,{q_{35}} - {q_{03}}\,{q_{26}}\,{q_{35}} - 
                     {q_{05}}\,{q_{23}}\,{q_{36}} +$

\noindent$ {q_{03}}\,{q_{25}}\,{q_{36}} )  )  + 
                ( {q_{03}}\,( -1 + 2\,{q_{22}} )  - 2\,{q_{02}}\,{q_{23}} ) \,
                 ( {q_{16}}\,{q_{23}}\,{q_{35}} - {q_{13}}\,{q_{26}}\,{q_{35}} - 
                   {q_{15}}\,{q_{23}}\,{q_{36}} + {q_{13}}\,{q_{25}}\,{q_{36}} )  )  )$

$Y_{72}=2\, ( -( ( {q_{16}}\,{q_{23}}\,{q_{35}} - {q_{13}}\,{q_{26}}\,{q_{35}} - 
                   {q_{15}}\,{q_{23}}\,{q_{36}} + {q_{13}}\,{q_{25}}\,{q_{36}} ) \,
                 ( {q_{07}}\,{q_{23}}\,{q_{35}} - {q_{03}}\,{q_{27}}\,{q_{35}} - $

\noindent$ {q_{05}}\,{q_{23}}\,{q_{37}} + {q_{03}}\,{q_{25}}\,{q_{37}} )  )  + 
              ( {q_{06}}\,{q_{23}}\,{q_{35}} - {q_{03}}\,{q_{26}}\,{q_{35}} - 
                 {q_{05}}\,{q_{23}}\,{q_{36}} + {q_{03}}\,{q_{25}}\,{q_{36}} ) \,$

\noindent$\times
               ( {q_{17}}\,{q_{23}}\,{q_{35}} - {q_{13}}\,{q_{27}}\,{q_{35}} - 
                 {q_{15}}\,{q_{23}}\,{q_{37}} + {q_{13}}\,{q_{25}}\,{q_{37}} )  ),$

\vskip 0.2in $(M^{-1})^{t}_{73}=\frac{X_{73}}{Y_{73}}$ where
 
$X_{73}=-( {q_{06}}\,{q_{15}}\,{q_{23}} )  + {q_{05}}\,{q_{16}}\,{q_{23}} + 
            {q_{06}}\,{q_{13}}\,{q_{25}} - {q_{03}}\,{q_{16}}\,{q_{25}} - 
            {q_{05}}\,{q_{13}}\,{q_{26}} + {q_{03}}\,{q_{15}}\,{q_{26}}$

$Y_{73}=6\,
            ( -( {q_{03}}\,{q_{17}}\,{q_{26}}\,{q_{35}} )  + 
              {q_{03}}\,{q_{16}}\,{q_{27}}\,{q_{35}} - {q_{05}}\,{q_{17}}\,{q_{23}}\,{q_{36}} + 
              {q_{03}}\,{q_{17}}\,{q_{25}}\,{q_{36}} + $

\noindent${q_{05}}\,{q_{13}}\,{q_{27}}\,{q_{36}} - 
              {q_{03}}\,{q_{15}}\,{q_{27}}\,{q_{36}} + 
              {q_{07}}\,( -( {q_{16}}\,{q_{23}}\,{q_{35}} )  + {q_{13}}\,{q_{26}}\,{q_{35}} + 
                 {q_{15}}\,{q_{23}}\,{q_{36}} - {q_{13}}\,{q_{25}}\,{q_{36}} )  + 
              {q_{05}}\,{q_{16}}\,{q_{23}}\,{q_{37}} - {q_{03}}\,{q_{16}}\,{q_{25}}\,{q_{37}} - 
              {q_{05}}\,{q_{13}}\,{q_{26}}\,{q_{37}} + {q_{03}}\,{q_{15}}\,{q_{26}}\,{q_{37}} + $

\noindent$
              {q_{06}}\,( {q_{17}}\,{q_{23}}\,{q_{35}} - {q_{13}}\,{q_{27}}\,{q_{35}} - 
                 {q_{15}}\,{q_{23}}\,{q_{37}} + {q_{13}}\,{q_{25}}\,{q_{37}} )  ),$

\vskip 0.2in $(M^{-1})^{t}_{74}=\frac{X_{74}}{Y_{74}}$ where
        
$X_{74}=-( -( ( {q_{15}}\,{q_{23}}\,{q_{34}} - {q_{13}}\,{q_{25}}\,{q_{34}} - 
                   {q_{14}}\,{q_{23}}\,{q_{35}} + {q_{13}}\,{q_{24}}\,{q_{35}} ) \,$

\noindent$\times ( {q_{06}}\,{q_{23}}\,{q_{35}} - {q_{03}}\,{q_{26}}\,{q_{35}} - 
                   {q_{05}}\,{q_{23}}\,{q_{36}} + {q_{03}}\,{q_{25}}\,{q_{36}} )  )  + $

\noindent$ ( {q_{05}}\,{q_{23}}\,{q_{34}} - {q_{03}}\,{q_{25}}\,{q_{34}} - 
                 {q_{04}}\,{q_{23}}\,{q_{35}} + {q_{03}}\,{q_{24}}\,{q_{35}} ) \,$

\noindent$\times
               ( {q_{16}}\,{q_{23}}\,{q_{35}} - {q_{13}}\,{q_{26}}\,{q_{35}} - 
                 {q_{15}}\,{q_{23}}\,{q_{36}} + {q_{13}}\,{q_{25}}\,{q_{36}} )  )$

$Y_{74}=24\,
            ( -( ( {q_{16}}\,{q_{23}}\,{q_{35}} - {q_{13}}\,{q_{26}}\,{q_{35}} - 
                   {q_{15}}\,{q_{23}}\,{q_{36}} + {q_{13}}\,{q_{25}}\,{q_{36}} ) \,$

\noindent$\times
                 ( {q_{07}}\,{q_{23}}\,{q_{35}} - {q_{03}}\,{q_{27}}\,{q_{35}} - 
                   {q_{05}}\,{q_{23}}\,{q_{37}} + {q_{03}}\,{q_{25}}\,{q_{37}} )  )  + $

\noindent$
              ( {q_{06}}\,{q_{23}}\,{q_{35}} - {q_{03}}\,{q_{26}}\,{q_{35}} - 
                 {q_{05}}\,{q_{23}}\,{q_{36}} + {q_{03}}\,{q_{25}}\,{q_{36}} ) \,$

\noindent$\times
               ( {q_{17}}\,{q_{23}}\,{q_{35}} - {q_{13}}\,{q_{27}}\,{q_{35}} - 
                 {q_{15}}\,{q_{23}}\,{q_{37}} + {q_{13}}\,{q_{25}}\,{q_{37}} )  ) \,{q_{44}},$

\vskip 0.2in $(M^{-1})^{t}_{75}=\frac{X_{75}}{Y_{75}}$ where
 
$X_{75}=39\,{q_{03}}\,{q_{16}}\,{q_{25}}\,{q_{34}} - 39\,{q_{03}}\,{q_{15}}\,{q_{26}}\,{q_{34}} + 
            39\,{q_{04}}\,{q_{16}}\,{q_{23}}\,{q_{35}} - 39\,{q_{03}}\,{q_{16}}\,{q_{24}}\,{q_{35}} - 
            39\,{q_{04}}\,{q_{13}}\,{q_{26}}\,{q_{35}} + 39\,{q_{03}}\,{q_{14}}\,{q_{26}}\,{q_{35}} - 
            39\,{q_{04}}\,{q_{15}}\,{q_{23}}\,{q_{36}} + 39\,{q_{03}}\,{q_{15}}\,{q_{24}}\,{q_{36}} + 
            39\,{q_{04}}\,{q_{13}}\,{q_{25}}\,{q_{36}} - 39\,{q_{03}}\,{q_{14}}\,{q_{25}}\,{q_{36}} - 
            460\,{q_{03}}\,{q_{16}}\,{q_{25}}\,{q_{44}} + 460\,{q_{03}}\,{q_{15}}\,{q_{26}}\,{q_{44}} + 
            5190\,{q_{03}}\,{q_{16}}\,{q_{35}}\,{q_{44}} - 
            1175799\,{q_{16}}\,{q_{23}}\,{q_{35}}\,{q_{44}} - 
            35826\,{q_{03}}\,{q_{26}}\,{q_{35}}\,{q_{44}} + 
            1175799\,{q_{13}}\,{q_{26}}\,{q_{35}}\,{q_{44}} - 
            5190\,{q_{03}}\,{q_{15}}\,{q_{36}}\,{q_{44}} + 
            1175799\,{q_{15}}\,{q_{23}}\,{q_{36}}\,{q_{44}} + 
            35826\,{q_{03}}\,{q_{25}}\,{q_{36}}\,{q_{44}} - 
            1175799\,{q_{13}}\,{q_{25}}\,{q_{36}}\,{q_{44}} + 
            {q_{06}}\,( -39\,{q_{14}}\,{q_{23}}\,{q_{35}} + 
               {q_{15}}\,{q_{23}}\,( 39\,{q_{34}} - 460\,{q_{44}} )  + 
               35826\,{q_{23}}\,{q_{35}}\,{q_{44}} + $

\noindent$
               {q_{13}}\,( -39\,{q_{25}}\,{q_{34}} + 39\,{q_{24}}\,{q_{35}} + 460\,{q_{25}}\,{q_{44}} - 
                  5190\,{q_{35}}\,{q_{44}} )  )  + $

\noindent$ {q_{05}}\,( 3\,{q_{23}}\,{q_{36}}\,( 13\,{q_{14}} - 11942\,{q_{44}} )  + 
               {q_{16}}\,{q_{23}}\,( -39\,{q_{34}} + 460\,{q_{44}} )  + $

\noindent$ {q_{13}}\,( 39\,{q_{26}}\,{q_{34}} - 39\,{q_{24}}\,{q_{36}} - 460\,{q_{26}}\,{q_{44}} + 
                  5190\,{q_{36}}\,{q_{44}} )  )$

$Y_{75}=96\,
            ( {q_{03}}\,{q_{17}}\,{q_{26}}\,{q_{35}} - {q_{03}}\,{q_{16}}\,{q_{27}}\,{q_{35}} + 
              {q_{05}}\,{q_{17}}\,{q_{23}}\,{q_{36}} - {q_{03}}\,{q_{17}}\,{q_{25}}\,{q_{36}} - $

\noindent$ {q_{05}}\,{q_{13}}\,{q_{27}}\,{q_{36}} + {q_{03}}\,{q_{15}}\,{q_{27}}\,{q_{36}} + 
              {q_{07}}\,( {q_{16}}\,{q_{23}}\,{q_{35}} - {q_{13}}\,{q_{26}}\,{q_{35}} - 
                 {q_{15}}\,{q_{23}}\,{q_{36}} + {q_{13}}\,{q_{25}}\,{q_{36}} )  - 
              {q_{05}}\,{q_{16}}\,{q_{23}}\,{q_{37}} + {q_{03}}\,{q_{16}}\,{q_{25}}\,{q_{37}} + 
              {q_{05}}\,{q_{13}}\,{q_{26}}\,{q_{37}} - {q_{03}}\,{q_{15}}\,{q_{26}}\,{q_{37}} + $

\noindent
     $         {q_{06}}\,( -( {q_{17}}\,{q_{23}}\,{q_{35}} )  + {q_{13}}\,{q_{27}}\,{q_{35}} + 
                 {q_{15}}\,{q_{23}}\,{q_{37}} - {q_{13}}\,{q_{25}}\,{q_{37}} )  ) \,{q_{44}},$

\vskip 0.2in $(M^{-1})^{t}_{76}=\frac{X_{76}}{Y_{76}}$ where
 
$X_{76}=3\,( -6\,{q_{03}}\,{q_{16}}\,{q_{25}}\,{q_{34}} + 
6\,{q_{03}}\,{q_{15}}\,{q_{26}}\,{q_{34}} - 6\,{q_{04}}\,{q_{16}}\,{q_{23}}\,{q_{35}} + 
              6\,{q_{03}}\,{q_{16}}\,{q_{24}}\,{q_{35}}$

\noindent$ + 6\,{q_{04}}\,{q_{13}}\,{q_{26}}\,{q_{35}} - 
              6\,{q_{03}}\,{q_{14}}\,{q_{26}}\,{q_{35}} + 6\,{q_{04}}\,{q_{15}}\,{q_{23}}\,{q_{36}} - 
              6\,{q_{03}}\,{q_{15}}\,{q_{24}}\,{q_{36}} - 6\,{q_{04}}\,{q_{13}}\,{q_{25}}\,{q_{36}} + 
              6\,{q_{03}}\,{q_{14}}\,{q_{25}}\,{q_{36}} + 140\,{q_{03}}\,{q_{16}}\,{q_{25}}\,{q_{44}} - 
              140\,{q_{03}}\,{q_{15}}\,{q_{26}}\,{q_{44}} - 1758\,{q_{03}}\,{q_{16}}\,{q_{35}}\,{q_{44}} + 
              43359\,{q_{16}}\,{q_{23}}\,{q_{35}}\,{q_{44}} + 
              12810\,{q_{03}}\,{q_{26}}\,{q_{35}}\,{q_{44}} - 
              43359\,{q_{13}}\,{q_{26}}\,{q_{35}}\,{q_{44}} + 
              1758\,{q_{03}}\,{q_{15}}\,{q_{36}}\,{q_{44}} - 
              43359\,{q_{15}}\,{q_{23}}\,{q_{36}}\,{q_{44}} - 
              12810\,{q_{03}}\,{q_{25}}\,{q_{36}}\,{q_{44}} + 
              43359\,{q_{13}}\,{q_{25}}\,{q_{36}}\,{q_{44}} + 
              2\,{q_{06}}\,( 3\,{q_{23}}\,{q_{35}}\,( {q_{14}} - 2135\,{q_{44}} )  + 
                 {q_{15}}\,{q_{23}}\,( -3\,{q_{34}} + 70\,{q_{44}} )  + $

\noindent$ {q_{13}}\,( 3\,{q_{25}}\,{q_{34}} - 3\,{q_{24}}\,{q_{35}} - 70\,{q_{25}}\,{q_{44}} + 
                    879\,{q_{35}}\,{q_{44}} )  )  + $

\noindent$ 2\,{q_{05}}\,( -3\,{q_{23}}\,{q_{36}}\,( {q_{14}} - 2135\,{q_{44}} )  + 
                 {q_{16}}\,{q_{23}}\,( 3\,{q_{34}} - 70\,{q_{44}} )  + $

\noindent$ {q_{13}}\,( -3\,{q_{26}}\,{q_{34}} + 3\,{q_{24}}\,{q_{36}} + 70\,{q_{26}}\,{q_{44}} - 
                    879\,{q_{36}}\,{q_{44}} )  )  )$ 

$Y_{76}=32\,
            ( {q_{03}}\,{q_{17}}\,{q_{26}}\,{q_{35}} - {q_{03}}\,{q_{16}}\,{q_{27}}\,{q_{35}} + 
              {q_{05}}\,{q_{17}}\,{q_{23}}\,{q_{36}} - {q_{03}}\,{q_{17}}\,{q_{25}}\,{q_{36}} - $

\noindent$
              {q_{05}}\,{q_{13}}\,{q_{27}}\,{q_{36}} + {q_{03}}\,{q_{15}}\,{q_{27}}\,{q_{36}} + 
              {q_{07}}\,( {q_{16}}\,{q_{23}}\,{q_{35}} - {q_{13}}\,{q_{26}}\,{q_{35}} - 
                 {q_{15}}\,{q_{23}}\,{q_{36}} + {q_{13}}\,{q_{25}}\,{q_{36}} )  - 
              {q_{05}}\,{q_{16}}\,{q_{23}}\,{q_{37}} + {q_{03}}\,{q_{16}}\,{q_{25}}\,{q_{37}} + 
              {q_{05}}\,{q_{13}}\,{q_{26}}\,{q_{37}} - {q_{03}}\,{q_{15}}\,{q_{26}}\,{q_{37}} + $

\noindent$ {q_{06}}\,( -( {q_{17}}\,{q_{23}}\,{q_{35}} )  + {q_{13}}\,{q_{27}}\,{q_{35}} + 
                 {q_{15}}\,{q_{23}}\,{q_{37}} - {q_{13}}\,{q_{25}}\,{q_{37}} )  ) \,{q_{44}},$

\vskip 0.2in $(M^{-1})^{t}_{77}=\frac{X_{77}}{Y_{77}}$ where
 
$X_{77}=-3\,( -2\,{q_{03}}\,{q_{16}}\,{q_{25}}\,{q_{34}} + 
              2\,{q_{03}}\,{q_{15}}\,{q_{26}}\,{q_{34}} - 2\,{q_{04}}\,{q_{16}}\,{q_{23}}\,{q_{35}} + 
              2\,{q_{03}}\,{q_{16}}\,{q_{24}}\,{q_{35}} +$

\noindent$ 2\,{q_{04}}\,{q_{13}}\,{q_{26}}\,{q_{35}} - 
              2\,{q_{03}}\,{q_{14}}\,{q_{26}}\,{q_{35}} + 2\,{q_{04}}\,{q_{15}}\,{q_{23}}\,{q_{36}} - 
              2\,{q_{03}}\,{q_{15}}\,{q_{24}}\,{q_{36}} - 2\,{q_{04}}\,{q_{13}}\,{q_{25}}\,{q_{36}} + 
              2\,{q_{03}}\,{q_{14}}\,{q_{25}}\,{q_{36}} + 52\,{q_{03}}\,{q_{16}}\,{q_{25}}\,{q_{44}} - 
              52\,{q_{03}}\,{q_{15}}\,{q_{26}}\,{q_{44}} - 690\,{q_{03}}\,{q_{16}}\,{q_{35}}\,{q_{44}} + 
              17913\,{q_{16}}\,{q_{23}}\,{q_{35}}\,{q_{44}} + 
              5190\,{q_{03}}\,{q_{26}}\,{q_{35}}\,{q_{44}} - 
              17913\,{q_{13}}\,{q_{26}}\,{q_{35}}\,{q_{44}} + 690\,{q_{03}}\,{q_{15}}\,{q_{36}}\,{q_{44}} - 
              17913\,{q_{15}}\,{q_{23}}\,{q_{36}}\,{q_{44}} - 
              5190\,{q_{03}}\,{q_{25}}\,{q_{36}}\,{q_{44}} + 
              17913\,{q_{13}}\,{q_{25}}\,{q_{36}}\,{q_{44}} + 
              2\,{q_{06}}\,( {q_{23}}\,{q_{35}}\,( {q_{14}} - 2595\,{q_{44}} )  - 
                 {q_{15}}\,{q_{23}}\,( {q_{34}} - 26\,{q_{44}} )  + $

\noindent$ {q_{13}}\,( {q_{25}}\,{q_{34}} - {q_{24}}\,{q_{35}} - 26\,{q_{25}}\,{q_{44}} + 
345\,{q_{35}}\,{q_{44}} )  )  + 
              2\,{q_{05}}\,( -( {q_{23}}\,{q_{36}}\,( {q_{14}} - 2595\,{q_{44}} )  
)  + $
                 
\noindent$ {q_{16}}\,{q_{23}}\,( {q_{34}} - 26\,{q_{44}} )  + 
                 {q_{13}}\,( -( {q_{26}}\,{q_{34}} )  + {q_{24}}\,{q_{36}} + 
                    26\,{q_{26}}\,{q_{44}} - 345\,{q_{36}}\,{q_{44}} )  )  )$

$Y_{77}=16\,
            ( {q_{03}}\,{q_{17}}\,{q_{26}}\,{q_{35}} - {q_{03}}\,{q_{16}}\,{q_{27}}\,{q_{35}} + 
              {q_{05}}\,{q_{17}}\,{q_{23}}\,{q_{36}} - {q_{03}}\,{q_{17}}\,{q_{25}}\,{q_{36}} - $

\noindent$ {q_{05}}\,{q_{13}}\,{q_{27}}\,{q_{36}} + {q_{03}}\,{q_{15}}\,{q_{27}}\,{q_{36}} + 
              {q_{07}}\,( {q_{16}}\,{q_{23}}\,{q_{35}} - {q_{13}}\,{q_{26}}\,{q_{35}} - 
                 {q_{15}}\,{q_{23}}\,{q_{36}} + {q_{13}}\,{q_{25}}\,{q_{36}} )  - 
              {q_{05}}\,{q_{16}}\,{q_{23}}\,{q_{37}} + {q_{03}}\,{q_{16}}\,{q_{25}}\,{q_{37}} + 
              {q_{05}}\,{q_{13}}\,{q_{26}}\,{q_{37}} - {q_{03}}\,{q_{15}}\,{q_{26}}\,{q_{37}} + $

\noindent$ {q_{06}}\,( -( {q_{17}}\,{q_{23}}\,{q_{35}} )  + {q_{13}}\,{q_{27}}\,{q_{35}} + 
                 {q_{15}}\,{q_{23}}\,{q_{37}} - {q_{13}}\,{q_{25}}\,{q_{37}} )  ) \,{q_{44}},$

\vskip 0.2in $(M^{-1})^{t}_{40}=(M^{-1})^{t}_{41}=(M^{-1})^{t}_{42}=(M^{-1})^{t}_{43}=0,
(M^{-1})^{t}_{44}=\frac{1}{24\,{q_{44}}},$

\noindent$
(M^{-1})^{t}_{45}=\frac{-13}{32\,{q_{44}}},
(M^{-1})^{t}_{46}=\frac{9}{16\,{q_{44}}},(M^{-1})^{t}_{47}=\frac{-3}{8\,{q_{44}}}$
\vskip 0.2in
In the  above expressions for  $(M^{-1})^t$, the non-zero $q_{ij}'s,\ 0\leq i\leq 4,\ 
0\leq j\leq 7$,
are given by:
\begin{eqnarray}
\label{eq:qijdieufs}
& & q_{00}={8\pi^3\over3}
\biggl[ln\biggl({2^2.3^3\over6^6}\biggr)+i\pi\biggr]\nonumber\\
& & q_{10}={8\pi^3\over3}\nonumber\\
& & q_{01}=2\pi^2\biggl[ln\biggl({2^2.3^3\over6^6}\biggr)+{14\over3}\pi^2\biggr]\nonumber\\
& & q_{11}=4\pi^2ln\biggl({2^2.3^3\over6^6}\biggr)\nonumber\\
& & 
q_{02}={4\pi^2\over\sqrt{3}} 
\biggl(\biggl[ln\biggl({2^2.3^3\over6^6}\biggr)+{\pi\over\sqrt{3}}\biggr]^2+5\pi^2\biggr)
\nonumber\\
& & q_{12}=
{8\pi^2\over\sqrt{3}} 
\biggl[ln\biggl({2^2.3^3\over6^6}\biggr)+{\pi\over\sqrt{3}}\biggr]\nonumber\\
& & q_{22}={4\pi^2\over\sqrt{3}}\nonumber\\
& & q_{03}=
\biggl(\biggl[ln\biggl({2^2.3^3\over6^6}\biggr)-{\pi\over\sqrt{3}}\biggr]^2+5\pi^2\biggr)
\nonumber\\
& & q_{13}= 
{8\pi^2\over\sqrt{3}} 
\biggl[ln\biggl({2^2.3^3\over6^6}\biggr)-{\pi\over\sqrt{3}}\biggr]\nonumber\\
& & q_{23}={4\pi62\over\sqrt{3}}\nonumber\\
& & q_{04}=4\sqrt{2}\pi\Biggl(
\biggl[ln\biggl({2^2.3^3\over 6^6}\biggr)+i\pi\biggr]^4
+{65\over3}\biggl[ln\biggl({2^2.3^3\over 6^6}\biggr)+i\pi\biggr]^2+{169\pi^4\over3}\nonumber\\
& & -1440\biggl[ln\biggl({2^2.3^3\over 6^6}\biggr)+i\pi\biggr]+402\zeta(4)\Biggr]\nonumber\\
& & q_{14}=4\sqrt{2}\pi\Biggl[4 \biggl[ln\biggl({2^2.3^3\over 6^6}\biggr)+i\pi\biggr]^3
+{65\over3}-1440\zeta(3)\Biggr]\nonumber\\
& & q_{24}
=4\sqrt{2}\pi\biggl(6\biggl[ln\biggl({2^2.3^3\over 6^6}\biggr)+i\pi\biggr]+{65\over3}
\nonumber\\
& & q_{34}=16\sqrt{2}\pi\biggl[ln\biggl({2^2.3^3\over6^6}\biggr)+i\pi\biggr]\nonumber\\
& & q_{44}=4\sqrt{2}\pi\nonumber\\
& & q_{05}=2\pi^2\Biggl(
\biggl[ln\biggl({2^2.33^3\over6^6}\biggr)+i\pi\biggr)^3
+15\pi^2ln\biggl({2^2.33^3\over6^6}\biggr)-356\zeta(3)\Biggr)\nonumber\\
& & q_{15}=2\pi^2\Biggl(
\biggl[ln\biggl({2^2.3^3\over6^6}\biggr)+i\pi\biggr)^2+15\pi^2\biggr]\Biggr)\nonumber\\
& & q_{25}=6\pi^2\biggl[ln\biggl({2^2.3^3\over6^6}\biggr)+i\pi\biggr]\Biggr)\nonumber\\
& & q_{35}=2\pi^2\nonumber\\
& & q_{06}=
{4\pi^2\over\sqrt{3}}\Biggl(
\biggl[ln\biggl({2^2.3^3\over 6^6}+i\pi +{\pi\over\sqrt{3}}\biggr]^3
+16\pi^2 \biggl[ln\biggl({2^2.3^3\over 6^6}+i\pi +{\pi\over\sqrt{3}}\biggr]\nonumber\\
& & -360\zeta(3)-{8\pi^3\over3\sqrt{3}}\Biggr)\nonumber\\
& & q_{16}={4\pi^2\over\sqrt{3}}\Biggl(
3\biggl[ln\biggl({2^2.3^3\over 6^6}+i\pi +{\pi\over\sqrt{3}}\biggr]^2+16\pi^2\Biggr)
\nonumber\\
& & q_{26}={4\pi^2\over\sqrt{3}}\biggl[ln\biggl({2^2.3^3\over6^6}\biggr)+i\pi\biggr)
\nonumber\\
& & q_{36}={4\pi^2\over\sqrt{3}}\nonumber\\
& & q_{07}=
{4\pi^2\over\sqrt{3}}\Biggl(
\biggl[ln\biggl({2^2.3^3\over 6^6}+i\pi -{\pi\over\sqrt{3}}\biggr]^3
+16\pi^2 \biggl[ln\biggl({2^2.3^3\over 6^6}+i\pi -{\pi\over\sqrt{3}}\biggr]\nonumber\\
& & -360\zeta(3)+{8\pi^3\over3\sqrt{3}}\Biggr)\nonumber\\
& & q_{17}={4\pi^2\over\sqrt{3}}\Biggl(
3\biggl[ln\biggl({2^2.3^3\over 6^6}+i\pi -{\pi\over\sqrt{3}}\biggr]^2+16\pi^2\Biggr)
\nonumber\\
& & q_{27}={4\pi^2\over\sqrt{3}}\biggl[ln\biggl({2^2.3^3\over6^6}\biggr)-i\pi\biggr)
\nonumber\\
& & q_{36}={4\pi^2\over\sqrt{3}}
\end{eqnarray}

The matrix $(M^{-1})^t$ is non-singular as the determinant is non-zero.
\footnote{The determinant is given by:
$\frac{P}{Q}$ where $P=\frac{10613001429\,i  }{131072}$ and
$Q= \sqrt{2}\,{\pi }^7\,( 8\,( -7\,i   + 325\,{\sqrt{3}} ) \,{\pi }^7 + 
      104\,i  \,( 87\,i   + 35\,{\sqrt{3}} ) \,{\pi }^6\,\log (432) - 
      2\,{\pi }^5\,( 3423 - 889\,i  \,{\sqrt{3}} + 62208\,( -i   + {\sqrt{3}} )
 \,\log \,\log (432) -$

\noindent$ 6\,( -742\,i   + 393\,{\sqrt{3}} ) \,{\log (432)}^2 )  - 108\,i  \,{\log (432)}^2\,\zeta(3) + 
      6\,{\pi }^4\,( ( -1351\,i   + 53\,{\sqrt{3}} ) \,\log (432) + 
         4478976\,i  \,( 3\,i   + {\sqrt{3}} ) \,{\log }^2\,\log (432) 
- 20736\,i  \,{\sqrt{3}}\,\log \,{\log (432)}^2 + $

\noindent$
         24\,i  \,( 9\,i   + 28\,{\sqrt{3}} ) \,{\log (432)}^3 + 16\,( -3 + 7\,i  \,{\sqrt{3}} ) \,\zeta(3) )  + 
      9\,i  \,\pi \,\log (432)\,( 3\,{\sqrt{3}}\,{\log (432)}^3 - 8\,( -3\,i   + {\sqrt{3}} ) \,\zeta(3) )  + 
      3\,{\pi }^3\,\log (432)\,( ( -495 + 2\,i  \,{\sqrt{3}} ) \,\log (432) - 26873856\,i  \,{\log }^2\,\log (432) - 
         180\,i  \,{\log (432)}^3 + 96\,( -3\,i   + {\sqrt{3}} ) \,\zeta(3) )  + 
      36\,{\pi }^2\,( 3\,( -5\,i   + {\sqrt{3}} ) \,{\log (432)}^3 - 2\,( 23\,i   + {\sqrt{3}} ) \,\zeta(3) + 
         20\,i  \,{\sqrt{3}}\,{\log (432)}^2\,\zeta(3) )  ) $}.
From the above the matrix $M$ is found to be as given in {\bf 2}.
 
\setcounter{equation}{0}
\section{Monodromy around $z=\infty$}
\seceqbb
In this appendix we give the details on the evaluation of  the monodromy matrix around
$z\rightarrow\infty$.
The non-zero entries of the $8\times 5$ matrix $A_{ai}(\infty)$ 
with $a=0,...,7\ i=1,...,5$ for $z\rightarrow\infty$, are given
below:

\begin{eqnarray}
\label{eq:Anonzero}
& & A_{01}(\infty)=-{36\prod_{j=1}^5\Gamma({j\over6})\Gamma({2\over3})\Gamma({1\over2})
\over\Gamma({7\over6})(\Gamma({5\over6}))^2}=
-(2\pi)^{{5\over2}}6^{{3\over2}}{\sqrt{\pi}\Gamma({2\over3})\over\Gamma(
{7\over6})(\Gamma({5\over6})^2}\nonumber\\
& & A_{02}(\infty)=54\sqrt{3}\pi^{{3\over2}}{\Gamma({5\over6})\over\Gamma({4\over3})\Gamma
({2\over3})}\nonumber\\
& & A_{03}(\infty)=-{4\Gamma({5\over6}\Gamma({7\over6}\Gamma(-{1\over3})\Gamma(-{1\over6})
\Gamma({1\over6})\Gamma({1\over3})\over\Gamma({3\over2})(\Gamma({1\over2})^2}=
-288
\sqrt{{\pi\over3}}\nonumber\\
& & A_{04}(\infty)=-{9\over4}{\Gamma({7\over6})\Gamma({3\over2})\Gamma({4\over3})
\Gamma(-{1\over2})\Gamma(-{1\over3})\Gamma(-{1\over6})\Gamma({1\over6})\over\Gamma({5\over3})
(\Gamma({1\over3}))^2}=-3\pi^2{\Gamma({7\over6})\Gamma(-{1\over3})\over\Gamma({5\over3})
\Gamma({1\over3})}\nonumber\\
& & A_{05}(\infty)=-{36\over25}{\Gamma({4\over3})\Gamma({7\over6})\Gamma({3\over2})
\Gamma(-{2\over3})\Gamma(-{1\over2})\Gamma(-{1\over6})\Gamma(-{1\over3})
\over\Gamma({11\over6})(\Gamma({1\over6})^2}={12\pi\over25\sqrt{3}}{\Gamma(-{2\over3})
\Gamma(-{1\over6})\over\Gamma({11\over6})\Gamma({1\over6})};\nonumber\\
& & A_{11}(\infty)={216(\Gamma({2\over3})^2\Gamma({1\over3})\over(\Gamma({5\over6}))^2}
\nonumber\\
& & A_{13}(\infty)={8\Gamma(-{1\over3}\Gamma({1\over3})\over
(\Gamma({1\over2})^2}=-{16\over\sqrt{3}\pi}\nonumber\\
& & A_{15}(\infty)={216\over125}{\Gamma({4\over3}\Gamma(-{2\over3})\Gamma(-{1\over3}
\over(\Gamma({1\over6})^2}=-{144\sqrt{3}\pi\over125}{\Gamma(-{2\over3})\over
(\Gamma({1\over6}))^2};\nonumber\\
& & A_{21}(\infty)={216(\Gamma({1\over2}))^2\Gamma({2\over3})\over(\Gamma({5\over6}))^2}
={216\pi\Gamma({2\over3})\over(\Gamma({5\over6}))^2}\nonumber\\
& & A_{24}(\infty)={27\Gamma(-{1\over2}\Gamma({1\over6})\over8(\Gamma({1\over3})^2}\nonumber\\
& & A_{25}(\infty)={216\over125}{\Gamma({7\over6})\Gamma(-{2\over3})\Gamma(-{1\over6})\over
\Gamma({1\over6})}={36\over125}\Gamma(-{2\over3})\Gamma(-{1\over6})\nonumber\\
& & A_{31}(\infty)={216\Gamma({1\over6})\Gamma({2\over3})\over\Gamma({5\over6})}\nonumber\\
& & A_{32}(\infty)={27\Gamma(-{1\over6})\Gamma(-{1\over2})\over(\Gamma({2\over3})^2}
={27\sqrt{\pi}\Gamma(-{1\over6})\over(\Gamma({2\over3}))^2}\nonumber\\
& & A_{32}(\infty)={27\over8}{\Gamma({3\over2})\Gamma(-{2\over3})\Gamma(-{1\over2})
\over(\Gamma({1\over6}))^2}=-{27\pi\over8}{\Gamma(-{2\over3})\over(\Gamma({1\over6}))^2};
\nonumber\\
& & A_{41}(\infty)=-216(\Gamma({1\over6}))^2\Gamma({2\over3})\nonumber\\
& & A_{45}(\infty)=-{216\over125}(\Gamma({5\over6}))^2\Gamma(-{2\over3});\nonumber\\
& & A_{51}(\infty)=-216{\Gamma({1\over6})\Gamma({1\over3})\Gamma({2\over3})\over
\Gamma({5\over6})}=-{432\pi\Gamma({1\over6})\over\sqrt{3}\Gamma({5\over6})}\nonumber\\
& & A_{53}(\infty)=-8\Gamma(-{1\over3})\Gamma({1\over3})=2\sqrt{3}\pi\nonumber\\
& & A_{55}(\infty)=-{216\over125}{\Gamma({5\over6})\Gamma(-{2\over3})\Gamma(-{1\over3})
\over\Gamma({1\over6})};\nonumber\\
& & A_{61}(\infty)={216\Gamma({1\over6})(\Gamma({1\over2}))^2\Gamma({2\over3})
\over\Gamma({5\over6})}={216\pi\Gamma({1\over6})\Gamma({2\over3})\over\Gamma({5\over6})}
\nonumber\\
& & A_{64}(\infty)={27\over8}{\Gamma({2\over3}\Gamma({1\over6})\Gamma(-{1\over2})\over
\Gamma({1\over3})}=-{27\sqrt{\pi}\over4}{\Gamma({2\over3})\Gamma({1\over6})\over
\Gamma({1\over3})}\nonumber\\
& & A_{65}(\infty)={216\over125}{\Gamma({5\over6})\Gamma({7\over6}\Gamma(-{2\over3})\Gamma
(-{1\over6})\over\Gamma({1\over6})};\nonumber\\
& & A_{71}(\infty)=216(\Gamma({1\over6}))^2\Gamma({2\over3})\nonumber\\
& & A_{72}(\infty)=27\Gamma({1\over3})\Gamma(-{1\over6})\Gamma({1\over2})
=27\sqrt{\pi}\Gamma({1\over3})\Gamma(-{1\over6})\nonumber\\
& & A_{75}(\infty)={216\over125}{\Gamma({5\over6}\Gamma({3\over2})\Gamma(-{2\over3})
\Gamma(-{1\over2})\over\Gamma({1\over6})}=-{216\over125}{\Gamma({5\over6})\Gamma(-{2\over3})
\over\Gamma({1\over6})}.
\end{eqnarray}
Given that $(T_u)_{ij}=e^{-{\sqrt{-1}\pi i\over3}}\delta_{ij}$, no sum over $i$, 
one sees that the equation
(\ref{eq:infty3}) becomes:
\begin{equation}
\label{eq:infty4}
e^{-{\sqrt{-1}\pi j\over3}}A_{aj}(\infty)=T_{ab}(\infty)A_{bj}(\infty),
\end{equation}
(no sum over $j$) which needs to be solved  for $T_{ab}(\infty)$. MATHEMATICA is unable
to perform the required computation - however, it is in principle, doable.

\end{document}